\documentclass[fleqn,usenatbib,twocolumn]{mnras}
\usepackage{geometry}
\usepackage{stackengine}
\usepackage{newtxtext,newtxmath}
\usepackage[T1]{fontenc}
\usepackage[normalem]{ulem}
\usepackage{tikz}
\def\angle{\tikz[baseline=0mm,scale=0.4]{\draw(30:1)--(0,0)--(1,0);\draw[->](-10:0.6)arc(-10:55:.6);}}

\DeclareRobustCommand{\VAN}[3]{#2}
\let\VANthebibliography\thebibliography
\def\thebibliography{\DeclareRobustCommand{\VAN}[3]{##3}\VANthebibliography}

\usepackage{graphicx}	
\usepackage{amsmath}	
\usepackage{float}
\usepackage{appendix}
\usepackage{xcolor}
\usepackage{multirow}
\usepackage{caption}
\definecolor{G01}{rgb}{0.0000,0.4470,0.7410}
\definecolor{G02}{rgb}{0.9290,0.6940,0.1250}
\definecolor{G07}{rgb}{0.4660,0.6740,0.1880}
\definecolor{G08}{rgb}{0.6350,0.0780,0.1840}
\definecolor{G28}{rgb}{0.8500,0.3250,0.0980}
\definecolor{G29}{rgb}{0.4940,0.1840,0.5560}

\title[Ionosphere of Ganymede]{Ionosphere of Ganymede: Galileo observations versus test particle simulation}

\author[A. Beth et al.]{
A. Beth,$^{1}$\thanks{E-mail: arnaud.beth@gmail.com}
M. Galand,$^{1}$
R. Modolo,$^{2}$
X. Jia,$^{3}$
F. Leblanc,$^{2}$
H.~L.~F. Huybrighs$^{4}$
\\
$^{1}$Department of Physics, Imperial College London, London, UK\\
$^{2}$LATMOS/CNRS, Sorbonne Université, UVSQ, Paris, France\\
$^{3}$Department of Climate and Space Sciences and Engineering, University of Michigan ‐ Ann Arbor, Ann Arbor, MI, USA\\
$^{4}$Astronomy \& Astrophysics Section, School of Cosmic Physics, Dublin Institute for Advanced Studies, DIAS Dunsink Observatory, Dublin D15 XR2R, Ireland.
}

\date{Accepted XXX. Received YYY; in original form ZZZ}

\pubyear{\the\year{}}
\graphicspath{{figures/}}

\begin{document}
\label{firstpage}
\pagerange{\pageref{firstpage}--\pageref{lastpage}}
\maketitle

\begin{abstract}
In this paper, we model the plasma environment of Ganymede by means of a collisionless test particle simulation. By coupling the outputs from a DSMC simulation of Ganymede's exosphere (i.e. number density profiles of neutral species such as $\mathrm{H}$, $\mathrm{H_2}$, $\mathrm{O}$, $\mathrm{HO}$, $\mathrm{H_2O}$, $\mathrm{O_2}$ for which we provide parametrisation) with those of a MagnetoHydroDynamic simulation of the interaction between Ganymede and the Jovian plasma (i.e. electric and magnetic fields), we perform a comparison between simulated ion plasma densities and ion energy spectra with those observed in-situ during 6 close flybys of Ganymede by the Galileo spacecraft. We find that not only our test particle simulation sometimes can well reproduce the in-situ ion number density measurement, but also the dominant ion species during these flybys are $\mathrm{H_2^+}$, $\mathrm{O_2^+}$, and occasionally $\mathrm{H_2O^+}$. Although the observed ion energy spectra cannot be reproduced exactly, the simulated ion energy spectra exhibit similar trends to those observed near the closest approach and near the magnetopause crossings but at lower energies. We show that the neutral exosphere plays an important role in supplying plasma to Ganymede's magnetised environment and that additional mechanisms may be at play to energise/accelerate newborn ions from the neutral exosphere.
\end{abstract}




\section{Introduction}
Ganymede is the largest moon of the Solar System, even exceeding the size of Mercury. Since its observation by Galileo in 1610, and maybe even earlier by Gan De, Ganymede has always been of huge interest amongst astronomers. Notwithstanding its size, Ganymede is also known as the only moon with an intrinsic magnetic field \citep{Kivelson1996}
 as well as likely hosting a subsurface ocean \citep{Kivelson2002,Saur2015}.
These peculiar characteristics rely on both remote sensing (e.g. with the Hubble Space Telescope) and in situ observations done by Galileo mission launched in 1989 to explore the Jovian system \citep{Oneil1983,Johnson1992,Young1998} and its magnetosphere, the largest one within the Solar system.

One of the scientific objectives of the Galileo mission regarding the exploration of Jovian satellites was to study their atmospheres, their ionospheres, and their interactions with the Jovian magnetosphere. Thanks to a series of instruments, including the plasma science instrument \citep[PLS,][]{Frank1992}, the plasma wave instrument \citep[PWS,][]{Gurnett1992}, the energetic particles detector \citep[EPD,][]{Williams1992}, the fluxgate magnetometer \citep[MAG,][]{Kivelson1992}, and the Ultraviolet Spectrometer \citep[UVS,][]{Hord1992}, Galileo had the opportunity to probe the close plasma environment of Ganymede during six flybys from June 1996 to May 2000. During these flybys, not only the fact that Ganymede bore a magnetic field came out as a surprise, but also the presence of an ionosphere was revealed by PWS \citep{Eviatar2001} and PLS \citep{Frank1992}. Amongst these flybys, the second one, namely G02, has been the most investigated as it went the closest to the surface down to 260~km altitude. During the G02 flyby, within the Jovian plasma sheet, prior to the magnetopause (referred to as MP hereafter) inbound crossing, PLS detected two distinct peaks undoubtedly attributed to H$^+$ and O$^+$ as their positions were located at ions streaming at the plasma corotating velocity and a mass-to-charge ratio $m/z$ of 16 (if singly ionised). However, within Ganymede's magnetosphere, only one peak was detected and assumptions had to be made on the ion mass-to-charge ratio near Ganymede's surface leading to the wrong conclusion of a proton outflow from Ganymede \citep{Frank1997} as shown by \citet{Vasyliunas2000} and \citet{Eviatar2001} who proposed another interpretation. They suggested that the ion flow from Ganymede was caused by O$^+$ at a much lower speed in Ganymede's vicinity to reconcile the electron number density from PWS with the derived ion number density from PLS and the ion convection speed reported by \citet{Williams1992}. Indeed, to derive moments from PLS measurements, one has to fix and assume a mass-to-charge ratio for the observed ions. For instance, the derived PLS ion number density scales linearly with the assumed mass \citep[see e.g. ][for G01]{Collinson2018} as PLS measured the ion flux. O$^+$ appeared to be a better candidate for making the most of Ganymede's ionosphere during G02. However, more recently, on 7 June 2021, Juno spacecraft performed a flyby of Ganymede, confirming the unambiguous presence of several ion species such as H$^+$, H$_2^+$, O$_2^+$, and unexpectedly H$_3^+$ \citep{Valek2022,Waite2024}. While ions were also detected around a mass-to-charge ratio $m/z\approx 16$~u\,q$^{-1}$, the ion mass spectrometer cannot distinguish O$^+$ from HO$^+$ or H$_2$O$^+$ as ions are primarily identified according to their mass. Hence, one has to be cautious in the identification process.

Ganymede's ionospheric composition and source remained puzzling, where observed, and uncertain over the full range of latitude, longitude, and altitude. Around any planetary bodies, the ionosphere is produced from the ionisation by energetic solar photons and particles (both typically above $\sim$10~eV) of the neutral layer that surrounds it, either its upper atmosphere or, for airless bodies, its exosphere. In the case of Ganymede, the neutral environment of Ganymede is primarily exospheric (i.e. collisions between neutral molecules/atoms are scarce) and poorly constrained. During G01, \citet{Barth1997} reported, based on UVS observation made onboard Galileo, the presence of a neutral hydrogen corona suspected to be supplied by photodissociation of water vapour, photodesorption of water ice, and/or ion sputtering of water ice. Remote sensing observations from the Earth with the Hubble Space Telescope (HST) revealed strong emissions of oxygen lines, O I 1304 and O I] 1356 \citep{Hall1998} mainly confined in the polar regions. Their ratio indicated that the main cause of emission was electron dissociative excitation of O$_2$. This was supported by laboratory experiments that showed ion sputtering of water ices may supply the atmosphere with H$_2$O, O$_2$, H$_2$, and H \citep{BarNun1985,Johnson1996}. As these observations are always remote-sensing in essence, only the column density can be derived with large uncertainties. The emissions depend on the electron distribution function that is poorly constrained, and there has been so far no direct in-situ measurement of Ganymede's neutral exosphere. To address this question, it is necessary to develop adequate models for the neutral environment. As detailed by \citet{Marconi2007}, Ganymede is quite challenging from a modelling perspective. Ganymede's neutral environment over the few hundred kilometres above the surface is not fully collisional with a spatially-dependent transition from a collisional to a collisionless regime: its exobase is mainly confined near the surface, and its altitude depends on the subsolar longitude. \citet{Marconi2007} demonstrated the need to apply Direct Simulation Monte Carlo (DSMC) approach \citep{Bird1994} for Ganymede, to capture the transition between continuous and free molecular flows appropriately. \citet{Marconi2007} was the first to use the DSMC method applied in 2D at Ganymede. He showed that the composition of the exosphere was dominated by H$_2$O, O$_2$, and H$_2$. These results are consistent with those of \citet{Shematovich2016}. Other authors try to investigate different sources and mechanisms that yield Ganymede's exosphere. \citet{Plainaki2015} and \citet{Vorburger2022} focused on the ion sputtering yield from Jovian ions while \citet{Vorburger2024} did on electron precipitation. The initial approach by \citet{Marconi2007} has been extended in 3D years later by \citet{Turc2014}. \citet{Leblanc2017} improved the latter by including, amongst other processes, Jupiter's gravitational influence that said mainly Ganymede rotating around Jupiter. These processes are important for species with long lifetimes within Ganymede's exosphere, such as O$_2$: the timescale to form the O$_2$ exosphere is of the order of Ganymede's period of revolution. It also caused a shift between the subsolar point and where H$_2$O number density peaks which is associated with the thermal inertia of the ice. Ganymede's exosphere is not in a steady state due to its revolution and because of the varying inclination of the plasma sheet with respect to Ganymede as Jupiter rotates. As the magnetic axis of Jupiter is not perpendicular to the orbit of Ganymede, Ganymede goes above and below the plasma sheet, passing through it every 5-6 hours.

In order to address the mystery of Ganymede's ionospheric composition in relation to its neutral exosphere, \citet{Carnielli2019} developed the first 3D collisionless test-particle model to quantify the contribution of the ionised neutral exosphere to the plasma surrounding the moon and understand how the exosphere once ionised supplies the plasma around Ganymede. Ions are produced from the exospheric model of \citet{Leblanc2017} following ionisation of neutral species; they then drift around Ganymede's magnetosphere due to the presence of electric and magnetic fields obtained from the MHD model developed by \citet{Jia2008,Jia2009}. \citet{Carnielli2019} modelled ion number density, velocity, and ion distribution for the configuration of the G02 flyby comparing with the in-situ observation of moments derived from PWS (electron number density) and PLS (ion bulk velocity, ion energy spectra): they showed that the ionosphere was dominated by O$_2^+$ but the total ion number density was still an order of magnitude lower than those observed \citet{Carnielli2019} and boosting the neutral exosphere may help reconcile observations with modelling for G02 \citet{Carnielli2020a}. \citet{Galand2025} provide a full review of the previous modelling of Ganymede's ionosphere.

The exospheric model by \citet{Leblanc2017} was considered as ``dry'' with a low sublimation rate, and the H$_2$O/O$_2$ ratio was inconsistent with those derived from the recent HST observations made by \citet{Roth2021}. To better agree with the latter, \citet{Leblanc2023} reviewed the exospheric model by increasing the sublimation rate. Nevertheless, unlike \citet{Leblanc2017}’s version, the exospheric model of \citet{Leblanc2023} did not include collisions between neutral species as it significantly lowers the computational time and may only have a limited impact \citep[e.g. by increasing O$_2$ number density at higher altitude, see][and Appendix~\ref{AppExo}]{Leblanc2017}.

In this paper, we extend the original approach by \citet{Carnielli2019} applied to G02 to the remaining flybys not analysed so far in terms of the ionospheric number density and composition, namely G01, G07, G08, G28, and G29. The Juno flyby has been excluded as it has a different set of plasma instruments and would require a dedicated study for itself; this will be the scope of a future study. We are using an updated version of the ionospheric model (including higher spatial resolution), combined with an updated version of Ganymede's exosphere from \citet{Leblanc2023} and of the electromagnetic fields from \citet{Jia2008,Jia2009}. In Section~\ref{section2}, we detail the main changes and updates regarding the different models compared with \citet{Carnielli2019}. In Section~\ref{section3}, we present our results for the six flybys as well as a parameterisation of the neutral exosphere (detailed in Appendix~\ref{AppExo}, more suitable and appropriate for further more straightforward use).  In Sections~\ref{section4} and \ref{section5}, we finally discuss our results and conclude. 

\section{Method}\label{section2}
	
\subsection{An improved test particle code}\label{section21}

\citet{Carnielli2019} were the first to model Ganymede's ionosphere in 3D. They developed a collisionless test particle code applied to ions produced by ionising exospheric neutrals. First, $N_\text{stat}$ particles are produced from each exospheric cell: the associated weight depends on the local number density, volume of the exospheric cell, and local ionisation rate (photoionisation and electron-impact). Secondly, macroparticles are initialised randomly within the cell in terms of position and velocity, though the latter is set at $\vec{0}$ as it does not change the outcome of the simulations. They are transported using a particle pusher \citep{Boris1970} within Ganymede's surrounding by means of electric and magnetic fields based on plasma simulation. Finally, ion moments (e.g. number density, velocity, kinetic energy) are derived on an ionospheric grid from the time spent $\Delta t$ (and associated quantities such as $\Delta t$, $\varv_x\Delta t$, $\varv_y\Delta t$, $\varv_z\Delta t$, etc.) by each macroparticle in each cell of this grid. The ions (or macroparticles) are assumed to not collide either between them or with the neutral background species.

We have applied the same approach as \citet{Carnielli2019} in the present study, but the code we have developed has been optimised regarding computing time and statistics. The code is fully parallelised as particles are sent independently from
each exospheric cell. This is justified because we ignore ion-ion collisions and assume that the ions have no feedback on the electromagnetic fields (electric and magnetic fields are kept fixed during every test-particle simulation). The contribution of each exospheric cell to the ion moments is calculated separately and put together once all macroparticles have been sent from every exospheric cell.

The ion moments $\mathbf{M}^{s'}_{k_\text{grid}}$, associated with the ion species $s'$ in the exospheric cell $k_\text{grid}$ are derived as follows:
\begin{align*}
\mathbf{M}^{s'}_{k_\text{grid}}=&\sum_{j_\text{exo}}^{N_\text{exo}}\mathbf{M}^{s'}_{j_\text{exo}\rightarrow k_\text{grid}}
\end{align*}
where
    \begin{align*}
    \mathbf{M}^{s'}_{j_\text{exo}\rightarrow k_\text{grid}}=&\overbrace{\dfrac{\sum_{s} \nu_{j_\text{exo}}^{\text{ioni},s\rightarrow s'} n_{j_\text{exo}}^{s\rightarrow s'} }{N_{j_\text{exo},\text{stat}}}\dfrac{V_{j_\text{exo}}}{V_{k_\text{grid}}}}^{\substack{\text{independent of the traj.,}\\ \text{intrinsic to a macropart.}}}\times\\
    &\left[\sum_{\text{macropart.}}^{N_\text{stat}}\mathcal{M}_\text{macropart.} \Delta t\right]_{j_\text{exo}\rightarrow k_\text{grid}}\\
\end{align*}
where $\sum_s\nu_{j_\text{exo}}^{\text{ioni},s\rightarrow s'} n_{j_\text{exo}}^{s\rightarrow s'} V_{j_\text{exo}}$ is the total number of neutral species $s$ turned into ions $s'$ per second in the exospheric cell $j_\text{exo}$, $N_{j_\text{exo},\text{stat}}$ is the number of macroparticles sent from the exospheric cell $j_\text{exo}$, and  $V_{k_\text{grid}}$ is the volume of the grid cell $k_\text{grid}$. $\mathbf{M}$ is the ion moment of interest: $\mathbf{M}=n$, the number density, $\mathbf{M}=n\vec{u}$, the flux where $\vec{u}$ is the bulk velocity, etc.,  $\mathbf{M}^{s'}_{j_\text{exo}\rightarrow k_\text{grid}}$ is the contribution of the exospheric cell $j_\text{exo}$ to the ionospheric grid cell $k_\text{grid}$ for the moment of the species $s'$, and $\mathcal{M}_\text{macropart.}\Delta t$ the associated quantity that must be saved for a given macroparticle at each time step $\Delta t$ (e.g. $\mathcal{M}=1$ for $\mathbf{M}=n$, and $\mathcal{M}=\vec{\varv}$ for $\mathbf{M}=n\vec{u}$). 

Though we must prevent macroparticles from passing through several cells during one timestep, the timestep $\Delta t$ is primarily constrained by the local gyroperiod, hence the magnetic field strength and ion mass. More iterations are performed either where the magnetic field strength is higher near Ganymede's surface (around $\sim$1440\,nT near the magnetic pole) and within the dipole field, or for a light ion. H$^+$ requires 32 iterations more than O$_2^+$ for the same field strength to simulate the same physical time. In the simulations by \citet{Carnielli2019}, $\Delta t$ was updated at each iteration to be $0.05\,T_\text{gyro}$ based on the interpolated magnetic field. However, that is not optimum as the Boris scheme loses its stability properties. To be more accurate, the timestep $\Delta t$ should be kept constant during the simulation of a given macroparticle. That is, however, not possible as the magnetic field strength changes by 1 order of magnitude around Ganymede. To ensure the Boris scheme conserves part of its stability properties as a leapfrog scheme \citep{Skeel1992}, $\Delta t$ is kept constant depending on where ions are. We fixed $\Delta t = 10^{24}m_{s'}$ for ions where $m_{s'}$ corresponds to the ion mass in kg.  $\Delta t = 10^{24}m_{s'}$ is such that the timestep is at least 1/20 of the gyroperiod over the North Pole of Ganymede, where the magnetic field is the strongest. 

While the original version developed by \citet{Carnielli2019} adopted $N_\text{stat}=25$ particles, we upgraded to $N_\text{stat}=10000$ to improve statistics. This means that we simulate a total of $N_\text{exo}\times N_\text{stat}=9.6\times10^{8}$ macroparticles for a given ion species (that said a total of 7.68 billion per simulation/flyby), where $N_\text{exo}$ is the total number of exospheric cells (see Section~\ref{section22}). In addition, while the maximum number of iterations for each particle was originally limited to 10$^7$, we have set no limit, expecting the particles to impact Ganymede's surface or to leave the simulation box. Setting a limit for iterations may have an influence on the moment depending on the species: light species need more iteration than heavy ones as their gyroperiod is shorter; hence, the timestep must be scaled down accordingly to simulate the same physical time. Simulations for O$_2^+$ are the fastest to perform. We have also improved the spatial resolution from $\sim$150~km on a spherical grid \citep{Carnielli2019} to $\sim$65~km on a cartesian grid as we have increased $N_\text{stat}$. Indeed, the main limitation in increasing the spatial resolution is ensuring there are still enough macroparticles to derive moments in each grid cell. This allows us, combined with a better resolved MHD simulation (see Section\,\ref{section23}), to capture narrower structures such as the magnetopause. 

Regarding the ionisation frequency of neutral species, we use the same values as published by \citet{Carnielli2019} for photoionisation and electron-impact ionisation. However, the electron-impact ionisation frequency is kept constant everywhere around Ganymede to ease interpretation and not have another free parameter to adjust, as it cannot be clearly constrained. This is discussed in more detail in Section~\ref{section4}. 

We have also simulated Jovian ions. The latter are simulated by launching ions from the faces of a cube ($[-4R_G;4R_G]^3$) centred on Ganymede assuming similar input parameters to those of \citet{Jia2008,Jia2009}: a slightly higher density of 4~cm$^{-3}$ \citep[except 8 for G8, 13\% H$^+$, 87\% O$^+$,][]{Gurnett1996,Kivelson2004chap}, an initial velocity of 140~km\,s$^{-1}$ \citep[except 130 for G07,][]{Jia2008} in GPhiO coordinates (where $X$ points in the direction of the corotating flow, $Y$ towards Jupiter, $Z$ completing the orthonormal frame), and an ion temperature of 360~eV. However, charge exchange between the Jovian ions and the exospheric neutrals as a source of ionospheric ions is negligible compared with other ionisation mechanisms (photo- and electron-impact), hence ignored here. Jovian O$^+$ and H$^+$ have limited access to Ganymede's magnetosphere (cf. Sections~\ref{section311} and \ref{section321}), efficiently shielded by its own dipole field \citep[see][and Section~\ref{section3}]{Carnielli2019, Fatemi2022} where the exosphere is dense. However, the Jovian plasma can still access the surface to some extent depending on its energy (e.g. the polar regions), sputter the surface, and contribute to the neutral exosphere.

\subsection{A revisited exosphere}\label{section22}

The test-particle model is driven by exospheric density profiles from a collisional exospheric Direct Simulation Monte Carlo model \citep{Leblanc2023}. These new simulations are motivated by new remote sensing auroral emissions observed with HST \citep{Roth2021,Roth2023}. The simulation box of the exospheric model is set up in spherical coordinates $(r,\theta,\phi)$ and divided into $100\times24\times40$~cells; respectively, the cells are quadratically binned along $r$ from the surface to $5\,R_G$ such that $\Delta r\propto r$, are binned along $\theta$ such that $\Delta\cos\theta=2/24$, and are evenly spaced in longitude such that $\Delta\phi=2\pi/40$. All neutral species ($\mathrm{H}$, $\mathrm{H_2}$, $\mathrm{O}$, $\mathrm{HO}$, $\mathrm{H_2O}$, $\mathrm{O_2}$) were affected in terms of their number densities, their scale height, and their spatial distribution. Compared with previous versions \citep[i.e.][]{Leblanc2017}, $\mathrm{O_2}$ number density is slightly higher at the surface, more spatially homogeneous, and less dependent on the illumination, whereas its scale height is smaller, being even more confined near the surface. $\mathrm{H_2O}$ number density relies more on sublimation near the subsolar point than sputtering over the polar regions. $\mathrm{H_2}$ number density is slightly higher, more spatially homogeneous, and exhibits lower diurnal variations.

Analysis of the simulated exospheric density profiles in altitude from \citet{Leblanc2023} has revealed that they (in particular for light species, such as $\mathrm{H_2}$) are inconsistent, that said departing significantly, with a hydrostatic equilibrium in plane parallel approximation \citep[prescribed in plasma and auroral studies, such as ][]{Duling2022,Stahl2023} following:
\begin{equation}
    n_n(z)=n_n(z_0)\exp\left(-\dfrac{z-z_0}{H_n}\right)
\end{equation}
where $z_0$ is the altitude of reference where the number density $n_n(z_0)$ is set for the neutral species $n$, $z=r-r_G$ is the altitude, $H_n=\frac{k_BT_\text{exo}}{m_ng}$ is the scale height defined as a function of $T_\text{exo}$ the exospheric temperature , $m_n$ the mass of the neutral species, $g$ Ganymede's surface gravity, and $k_B$ the Boltzmann constant. The exospheric temperature and gravity are assumed constant throughout the exosphere. The neutral density profiles derived from \citet{Leblanc2023} are however more consistent with a hydrostatic equilibrium in spherical symmetry defined as:
\begin{equation}
    n_n(r)=n_n(r_c)\exp(\lambda_n(r)-\lambda_{n,c})
    \label{Eq2}
\end{equation}
where $r_c$ corresponds to the critical radius and usually refers to the exobase radial distance from the centre of Ganymede. The Jeans' parameter $\lambda_{n}(r)=\dfrac{GMm_n}{k_BT_\text{exo}r}=\dfrac{\varv_\text{esc}^2(r)}{\varv_{n,\text{th}}^2}$ where $\varv_\text{esc}=\sqrt{2GM/r}$ is the escape speed and $\varv_{n,\text{th}}=\sqrt{2k_BT_\text{exo}/m_n}$ is the thermal speed, and $\lambda_{n,c}=\lambda_n(r_c)$ \citep{Jeans1925,Chamberlain1963}. The latter is often referred as the Jeans' parameter. The second hydrostatic equilibrium model takes into account the change in gravity (decreasing in $1/r$ and neglecting Jupiter's influence). Hence, even in the case of an isothermal atmosphere, the scale height varies with $r$ and is found to be:
\begin{equation}
H_n(r)=(\mathrm{d}\log n_n(r)/\mathrm{d}r)^{-1}=r/\lambda_n(r)=H_n(r_c)(r/r_c)^2
\label{Eq2p}
\end{equation}
It can be interpreted as the maximum altitude reached by a species launched vertically at $\varv_{n,\text{th}}$ from $r$. This is treated in more detail in Appendix~\ref{AppExo}. As inputs of our simulations, we use the neutral exosphere simulated with a phase angle similar to those of each Galileo flyby (cf. Table~\ref{table1}, \ref{tableA1}, and \ref{tableA2}).

\subsection{A more refined MHD field}\label{section23}

The simulations for the electromagnetic fields are those from \citet{Jia2008,Jia2009}. These MHD simulations were performed over a staggered grid, that is, each component ($E_x$, $E_y$, $E_z$, $B_x$, $B_y$, $B_z$) is saved at different locations within the same cell. In \citet{Carnielli2019}, the electromagnetic fields were extrapolated on a low-resolution cartesian grid of $0.05\,R_G$ resolution, losing the capability to capture sharp transitions in the electromagnetic fields such as the magnetopause. Here, we still use the same initial fields from \citet{Jia2008,Jia2009} on an interpolated grid albeit spherical with an improved spatial resolution especially near the surface closer to the resolution of the original simulations by \citet{Jia2008,Jia2009}. However, the calculation of the electromagnetic fields is only undertaken down to $\sim1.05\,R_G$. It is critical to set up electromagnetic fields down to the surface as some ions species, such as O$_2^+$ ions from ionisation of O$_2$, are produced near the surface where O$_2$ is confined (neutral species with the lowest scale height, see Section~\ref{section22}). To overcome this issue, we assume a dipole field to extrapolate the magnetic field down to the surface of Ganymede to fill the gap between $1R_G$ and $\sim1.05\,R_G$. To extrapolate the electric field down to the surface, as the MHD model assumes ideal MHD (i.e. $\vec{E}=-\vec{\varv}\times \vec{B}$), we evaluate $\vec{\varv}$ at the lowest boundary of the MHD simulation around 1.05 $R_G$ such that $\vec{\varv}=\vec{E}\times \vec{B}/B^2$ (neglecting the component along $\vec{B}$) and then use it into Ohm's law using the dipole field at the surface. This approach has the benefit to ensure ideal MHD down to the surface. However, the overall MHD model has neither a Hall term nor an ambipolar electric field that could accelerate ions along magnetic field lines. The latter must be kept in mind for the interpretation of the simulation and comparison with the observations. 

\begin{figure*}
\centering
\includegraphics[width=2\columnwidth]{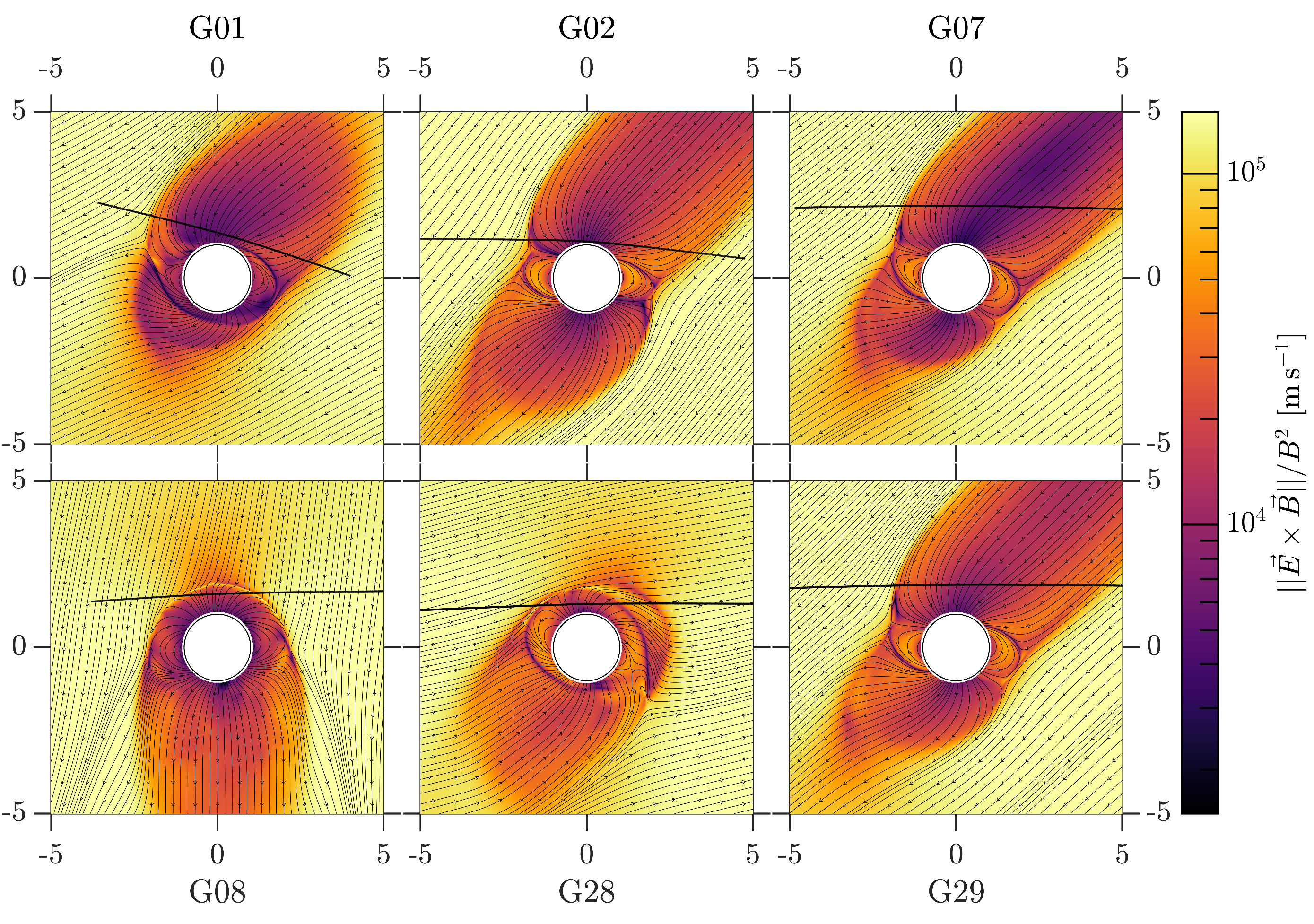}
\caption{Overview of MHD simulations for the 6 Galileo's flybys under study showing $||\vec{\varv}_{\vec{E}\times\vec{B}}||$, represented from $1.5\times 10^3$ to $1.5\times 10^5$\,m\,s$^{-1}$. These are cuts through the 3D simulation domain over a plane containing Galileo's orbit, that is, through a plane that is perpendicular to the average angular momentum of the Galileo spacecraft. Though Galileo has a $\Delta V$ during the flyby, the Galileo's angular momentum direction is almost kept constant. Galileo travelled from left to right. The centred black circle is Ganymede. Streamlines correspond to the projected magnetic field lines in these planes with arrows indicating the direction.} \label{Fig1}
\end{figure*}

Fig.~\ref{Fig1} shows a cut through the MHD simulation for each of the selected, close Galileo flybys, in a plane containing the spacecraft trajectory. It shows$||\vec{\varv}_{\vec{E}\times\vec{B}}||=||\vec{E}\times\vec{B}||/B^2=||\vec{E}_\perp||/B$ as it exhibits  most clearly the different plasma regions. The maximum value is reached outside Ganymede's magnetosphere within the corotating Jovian thermal plasma flowing at around $1.5\times 10^5$\,m\,s$^{-1}$. The lowest drift speed is reached at the magnetopause where ions are heavily decelerated. Another minimum is also found along the centre of the Alfvén wings. These plots do not take into account the aforementioned extrapolation to the surface for the electric and magnetic fields: this allows to visualise how far the lower boundary is from Ganymede's surface. G08 and G28 crossed the closed field line regions, while G01, G02, G07, and G29 passed across the Alfvén wings. In particular, G01 and G07 reached closer to the centre of the wings, where $||\vec{\varv}_{\vec{E}\times\vec{B}}||$ is much lower, than G02 and G29.

\subsection{Data and observations}
\begin{figure}
\centering
\textcolor{G01}{\large\textbf{G01}} \textcolor{G02}{\large\textbf{G02}} \textcolor{G07}{\large\textbf{G07}} \textcolor{G08}{\large\textbf{G08}} \textcolor{G28}{\large\textbf{G28}} \textcolor{G29}{\large\textbf{G29}}\\

\includegraphics[width=.35\linewidth,angle=90,trim=1cm 1cm 1cm 1cm, clip]{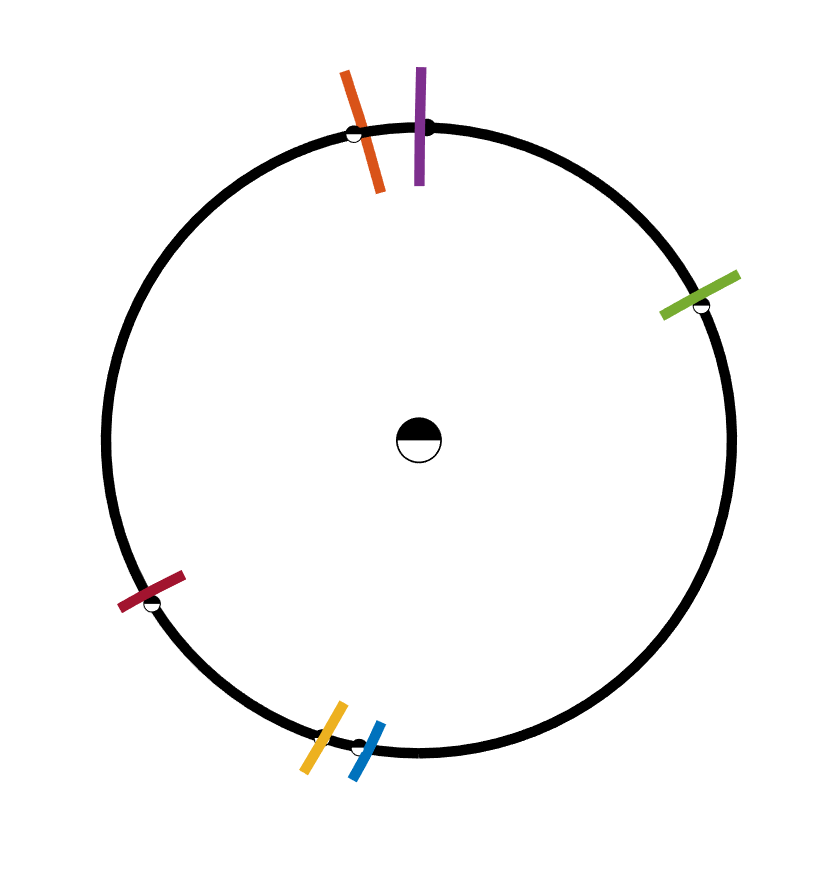}
\includegraphics[width=.55\linewidth]{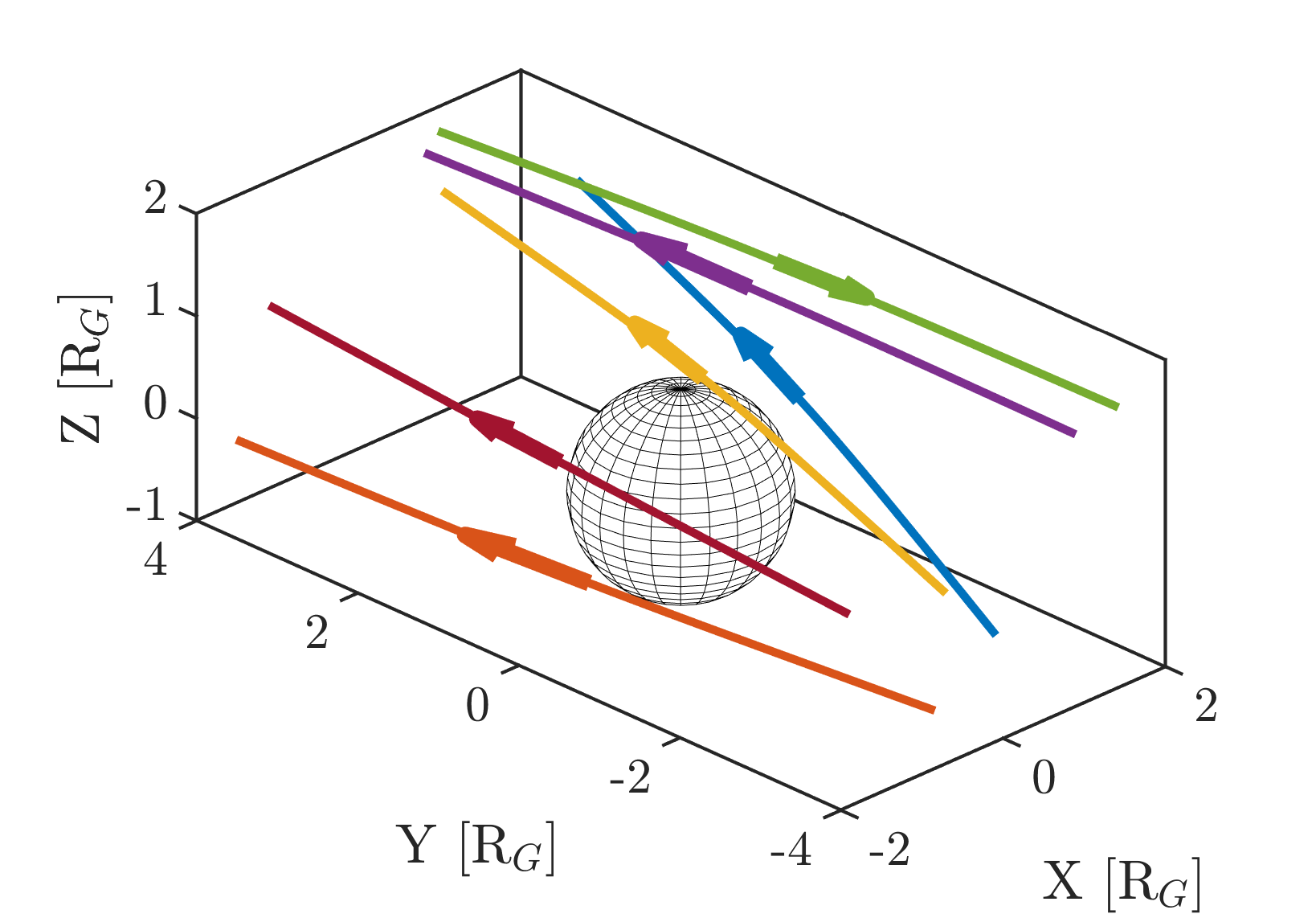}
\caption{ (Left panel) Ganymede's phase around Jupiter for the different flybys and their associated trajectories. Ganymede and flybys are not at scale. The Sun is on the right. G29 occurred during Jovian eclipse. (Right panel) Galileo flyby trajectories around Ganymede in GPhiO coordinates. Arrows indicate spacecraft direction. The $X$-axis points towards the corotating Jovian plasma flow and the $Y$-axis points towards Jupiter. The base of the arrow corresponds to the closest approach.}\label{Fig2}
\end{figure}

Our simulations are compared against particle and plasma data acquired by PLS and PWS during 6 Galileo's flybys of Ganymede. A summary of their flyby's characteristics is presented in Table~\ref{table1}. Fig.~\ref{Fig2} shows the local time of each flyby with respect to Ganymede's position around Jupiter as well as Galileo's trajectories in GPhiO frame.

\begin{table*}
    \centering
    \caption{Characteristics of the 6 Galileo's flybys: acronym, position with respect to the plasma sheet ($\uparrow$ above, $=$ within, $\downarrow$ below), to Jupiter (the angle is counted positive and increases as Ganymede revolves around Jupiter), to Ganymede, crossed plasma regions (e.g. Alfvén wings, closed field line regions (CFL hereafter)), and coordinates of closest approach (CA) in the GPhiO frame. The last three columns correspond to the mean Galileo velocity vector in the GPhiO frame we assumed during the flybys to compute ion energy spectra in the spacecraft frame. Note that for G29, Ganymede is in Jupiter's shadow.}
    \begin{tabular}{cccccccrccc}
    \hline
        Flyby   & \multicolumn{7}{c}{Location}& \multicolumn{3}{c}{$<\vec{\varv}_\text{SC}>$}\\
        \cline{2-8}\cline{9-11}
        & Rel. to& $\angle$ Sun-Jup.-Gan.&Rel. to&Plasma&&CA&&\multicolumn{1}{c}{$<\varv_{\text{SC},x}>$}&\multicolumn{1}{c}{$<\varv_{\text{SC},y}>$}&\multicolumn{1}{c}{$<\varv_{\text{SC},z}>$}\\
        \cline{6-8}
        & the PS&&Ganymede&regions&$r\, [R_G]$&\hphantom{-}lat.&long.&\multicolumn{3}{c}{[km\,s$^{-1}$]}\\
        \hline
        \textcolor{G01}{\textbf{G01}} &$\uparrow$&  349$^{\circ}$&central wake &Alfvén wing&$1.32$&$\hphantom{-}30.6^{\circ}$&$-21.1^{\circ}$&$\hphantom{-}2.08$&$\hphantom{30}7.30\hphantom{7}$&$\hphantom{-}1.12$\\
        \textcolor{G02}{\textbf{G02}} & $\uparrow$&342$^{\circ}$&polar&Alfvén wing&$1.10$&$\hphantom{-}79.5^{\circ}$&$-31.5^{\circ}$&$\hphantom{-}1.63$&$\hphantom{54}7.54\hphantom{7}\,$&$\hphantom{-}0.46$\\
        \textcolor{G07}{\textbf{G07}} & $\downarrow$& 115.5$^{\circ}$&mid-lat. downstr.&Alfvén wing&$2.18$&$\hphantom{-}55.6^{\circ}$&$2.9^{\circ}$&$\hphantom{-}0.46$&$\hphantom{.43}{-8.43}\hphantom{-8.}$&$-0.04$\\
        \textcolor{G29}{\textbf{G29}} & $\uparrow$& 178.5$^{\circ}$&mid-lat. downstr. &Alfvén wing&$1.89$&$\hphantom{-}62.4^{\circ}$&$1.4^{\circ}$&$-0.12$&$\hphantom{45}10.45\hphantom{10}$&$-0.07$\\
        \hline
        \textcolor{G08}{\textbf{G08}} & = & 301.5$^{\circ}$&low-lat. upstr.&CFL&$1.61$&$\hphantom{-}28.1^{\circ}$&$177.2^{\circ}$&$\hphantom{-}0.55$&$\hphantom{53}8.53\hphantom{8}$&$\hphantom{-}0.14$\\
        \textcolor{G28}{\textbf{G28}} &  $\downarrow$ & 192$^{\circ}$&low-lat. upstr. &CFL&$1.31$&$-19.2^{\circ}$&$-175.3^{\circ}$&$-0.94$&$\hphantom{11}11.11\hphantom{11}$&$\hphantom{-}0.07$\\
    \end{tabular}
    \label{table1}
\end{table*}

\begin{table*}
    \centering
    \caption{Timing of the different boundaries crossed by Galileo based on the MHD simulations from \citet{Jia2008,Jia2009} and Galileo's position. We determine the field line going through Galileo and check its connectivity: whether it is connected to Jupiter only, Jupiter and Ganymede, or Ganymede only. MP: magnetopause crossing, OFL: open field line region, CFL: closed field line region, CA: closest approach.\label{table2}}
    \begin{tabular}{ccccccc}
    \hline
        Flyby & Date  & \multicolumn{5}{c}{Timing [hh:mm]}\\
        \cline{3-7}
        &[YYYY-MM-DD]& MP inbound & OFL $\longrightarrow$ CFL&CA&CFL $\longrightarrow$ OFL&MP outbound\\
        \hline
        \textcolor{G01}{\textbf{G01}} & 1996-06-27&06:19&N/A&06:29&N/A&N/A\\
        \textcolor{G02}{\textbf{G02}} &
        1996-09-06&18:50&N/A&19:00&N/A&19:22\\
        \textcolor{G07}{\textbf{G07}}& 1996-04-05&07:02&N/A&07:10&N/A&07:37\\
        \textcolor{G29}{\textbf{G29}} &2000-12-28&08:18&N/A&08:25&N/A&08:45\\
        \hline
        \textcolor{G08}{\textbf{G08}} &1997-05-07&15:50&15:52&15:56&16:01&16:03\\
        \textcolor{G28}{\textbf{G28}}&
        2000-05-20&10:04&10:05&10:10&10:14&10:19\\
    \end{tabular}   
\end{table*}

\begin{figure}
\centering
\includegraphics[width=\linewidth]{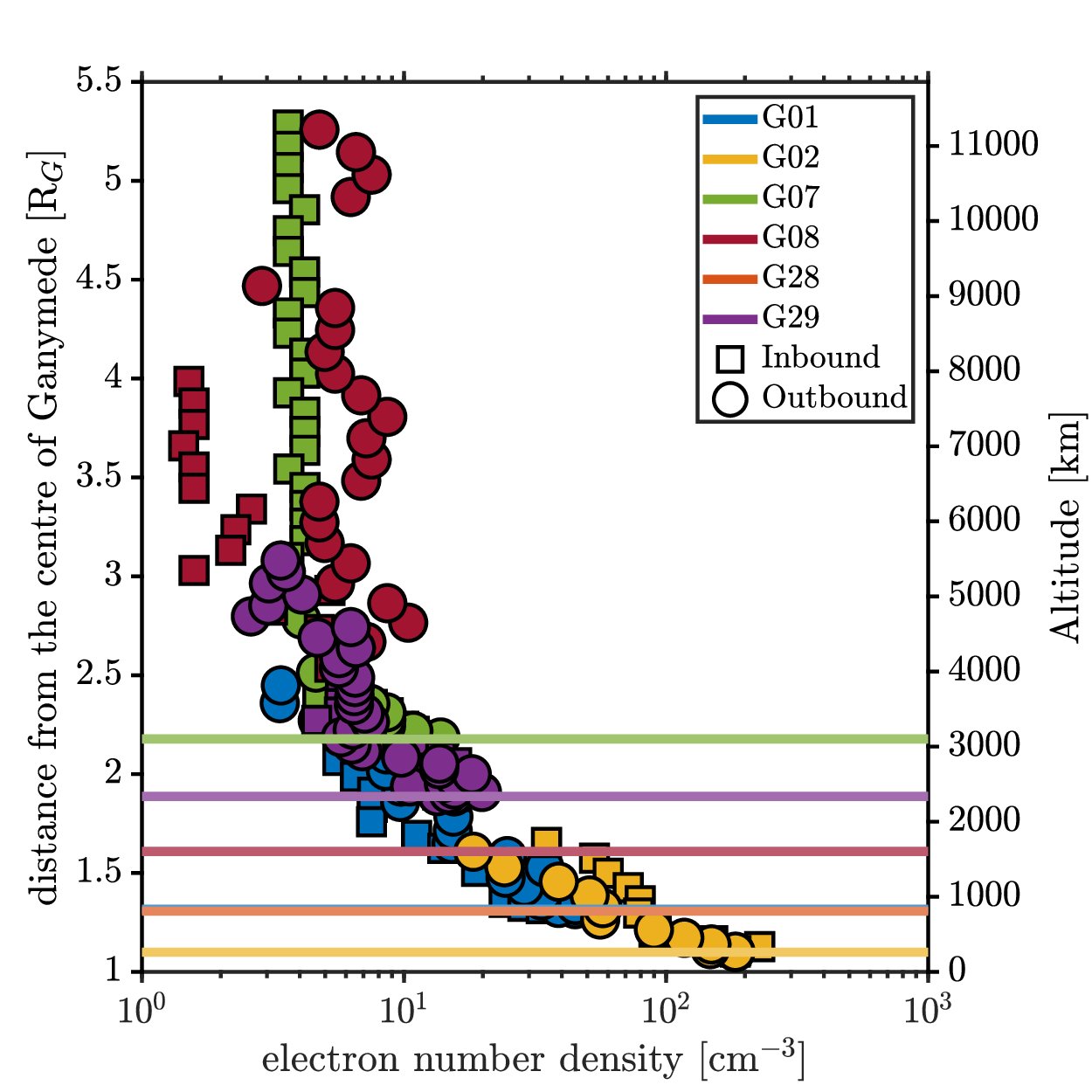}
\caption{Electron number density derived from PWS observations. Electron number density could not be derived from PWS for G28. Colours are associated with a different flyby, horizontal lines mark the distance of closest approach (G01 and G28 overlap), and markers represent either inbound ($\blacksquare$) or outbound ($\bullet$) measurements. Markers' width is of the order of the uncertainty in $n_e$.}\label{Fig3}
\end{figure}

PWS \citep{Gurnett1992} is a a radio and plasma wave sensor that probed the electromagnetic energy spectrum during the flybys. At closest approach, there was a clear resonance that has been attributed to upper hybrid frequency for some flybys. Combined with the magnetic field strength measurements from MAG \citep{Kivelson1992}, one retrieves the plasma frequency and therefore the local electron density from:
\begin{equation*}
\omega^2_\text{uh}=\omega^2_\text{pe}+\omega^2_\text{ce}=\dfrac{q^2n_e}{m_e\varepsilon_0}+\dfrac{qB}{m_e}
\end{equation*}
where $\omega_\text{uh}$, $\omega_\text{pe}$, and $\omega_\text{ce}$ are respectively the upper-hybrid, the electron plasma, and the electron cyclotron pulsations, $q$ is the elementary charge, $n_e$ is the electron number density, $m_e$ is the electron mass, and $\varepsilon_0$ is the vacuum permittivity. The associated error on the number density given by
\begin{equation*}
\dfrac{\Delta n_e}{n_e}=\underbrace{\dfrac{2\Delta \omega}{\omega}}_{\sim 16\%}+\underbrace{\dfrac{2\Delta B}{B}}_\text{negligible}
\end{equation*}
Note that the largest uncertainty is associated with the frequency resolution of PWS \citep{Gurnett1992}. Fig.~\ref{Fig3} shows the electron number densities as a function of the distance from Ganymede's centre, this is similar to Fig.~3  of \citet{Eviatar2001} with all flybys instead. Electron number densities are available at closest approach for all flybys except G28. There is consistency between flybys: below 2.5~$R_G$, the electron number densities increased as Galileo approached the surface and measured a maximum of 230~cm$^{-3}$ at 0.12~$R_G$ above the surface during G02 highlighting the presence of a dense ionosphere around Ganymede.\\
PLS \citep{Frank1992} is the ion energy spectrometer onboard Galileo that provided the ion energy spectrum during each flyby with an effective time resolution of 1~min (due to the main antenna malfunction), with 64 energy bins logarithmically spaced between 0.9~eV and 52704~eV ($\log(52704/0.9)/ 63\approx 0.174$), and an energy resolution of $\Delta E/E\sim0.1$ around each energy bin. As the space between energy bins is larger than the energy resolution, PLS only covers 50-60\% of its energy range. Note that Galileo was travelling at a few km\,s$^{-1}$ (cf. Table~\ref{table1}) which affects the ion' kinetic energy observed and measured in the spacecraft frame. Indeed, for an ion at rest in Ganymede's frame, its kinetic energy would be $E_\text{kin}=0.5\,m_\text{ion} \varv^2=0.52\,m_\text{ion}[\text{u}]$($\varv_\text{SC}$/10~km\,s$^{-1}$)$^2$~eV in Galileo's rest frame where $m_\text{ion}$ is the ion mass expressed in atomic mass unit and the spacecraft speed $\varv_\text{SC}$ in km\,s$^{-1}$. The averaged Galileo speed during flybys varies between 7 and 11~km\,s$^{-1}$ such that protons at rest would be measured at 0.52~eV while O$_2^+$ would be at 16.7~eV.
The processing of the PLS data in this paper is documented in appendix A of \citet{Carnielli2019}. Further information about the PLS data and its calibration is available in \citet{Bagenal2016}.

\section{Results and comparison with Galileo data}\label{section3}

For this section, we have decided to separate simulations with respect to the plasma regions that Galileo flew by. As already summarised in Table~\ref{table1}, Table~\ref{table2}, and Fig.~\ref{Fig2}, G01, G02, G07, and G29 trajectories were mainly within the Alfvén wings and the flux tube over the Northern Hemisphere (see Section~\ref{subsection31}), while G08 and G28 trajectories cross the closed magnetic field lines region upstream (see Section~\ref{subsection32}). The timeline of the trajectory is based on the MAG data archived on the PDS \citep{KivelsonG}. Magnetic field data were archived for each flyby at high time resolution ($\sim0.3$\,s) as well as Galileo's position in GPhiO coordinates with 5 significant digits in Ganymede's radii, much lower than our grid resolution of 0.04~$R_G$. It is critical to not only have a high spatial resolution for our simulations \citep[MHD and test-particle, see][]{Jia2008} but also a very high accuracy regarding the spacecraft position in order to accurately capture sharp transitions and strong gradients in our model.

		
		

\subsection{Within the Alfvén wings: G01, G02, G07, and G29\label{subsection31}} 

\begin{figure*}
\centering
\textcolor{G01}{\textbf{G01}} \textcolor{G02}{\textbf{G02}} \textcolor{G07}{\textbf{G07}} \textcolor{G29}{\textbf{G29}}\\
\stackinset{c}{-40pt}{c}{-40pt}{\Huge $\bigodot$}{%
\includegraphics[width=.65\columnwidth,clip,trim=1cm 1cm 1cm 1cm]{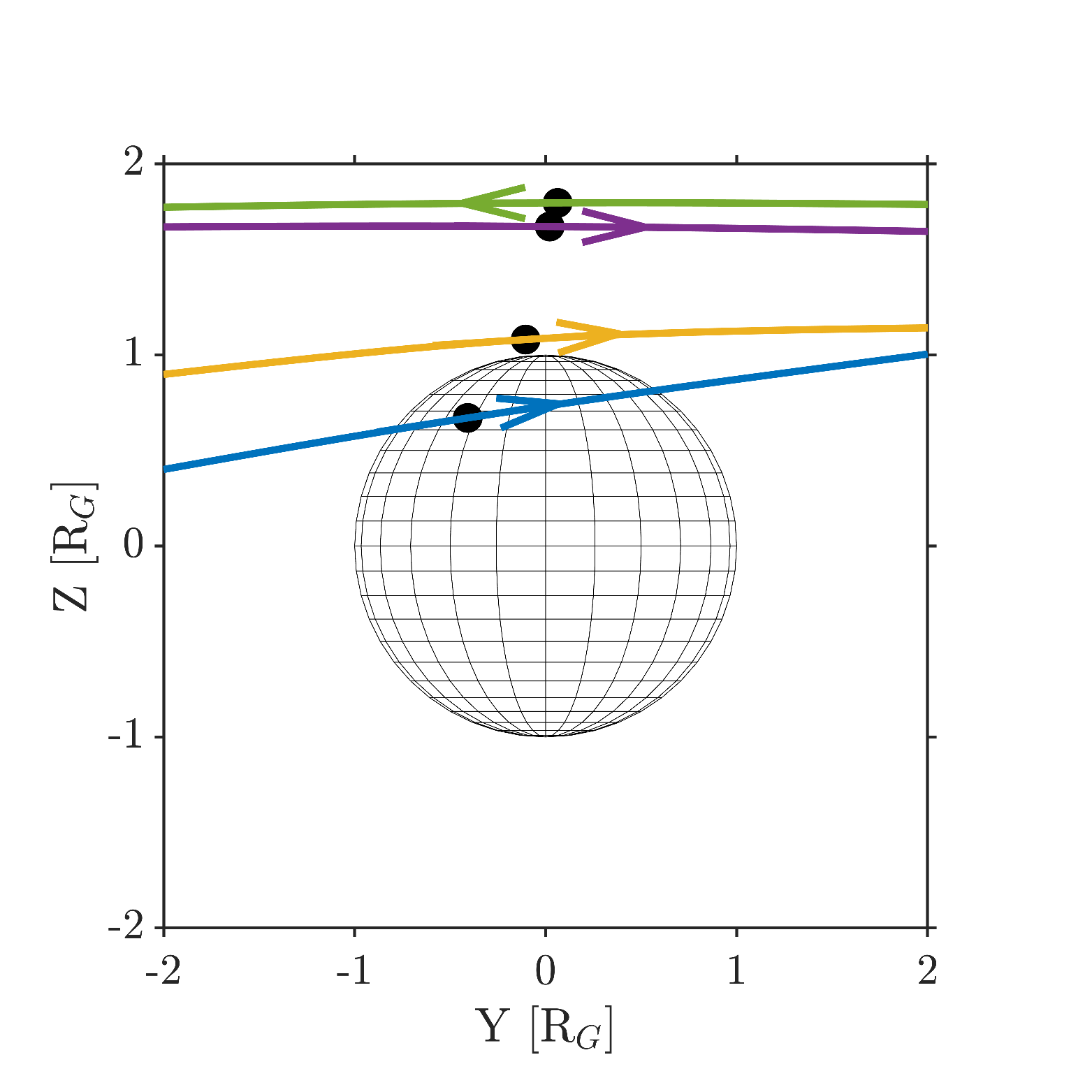}}
\stackinset{c}{-45pt}{c}{0pt}{\Huge $\rightarrow$}{%
\includegraphics[width=.65\columnwidth,clip,trim=1cm 1cm 1cm 1cm]{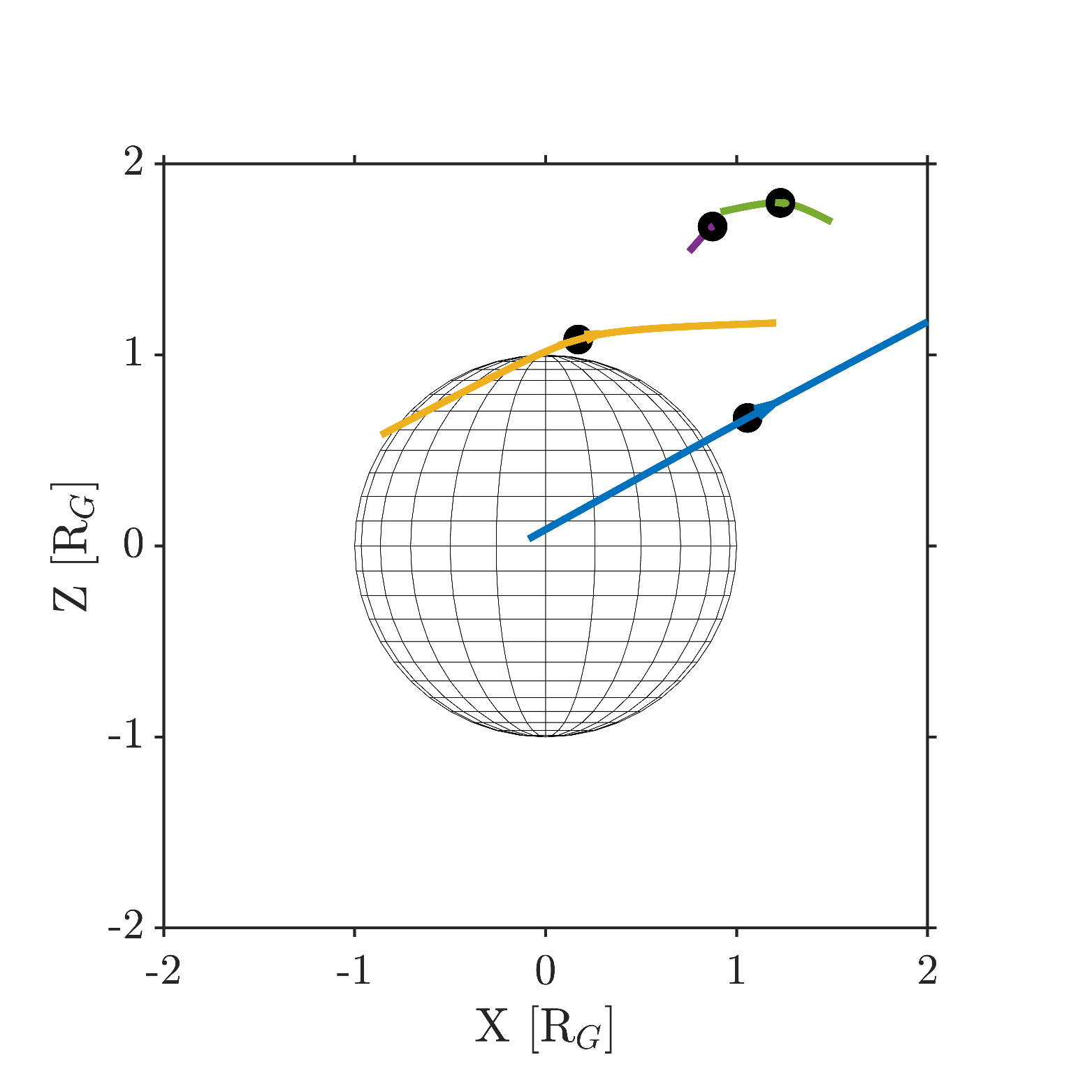}}
\stackinset{c}{0pt}{c}{45pt}{\centering \Huge$\downarrow$ \ \ \ \ \ \ \ \ $\downarrow$}{%
\includegraphics[width=.65\columnwidth,clip,trim=1cm 1cm 1cm 1cm]{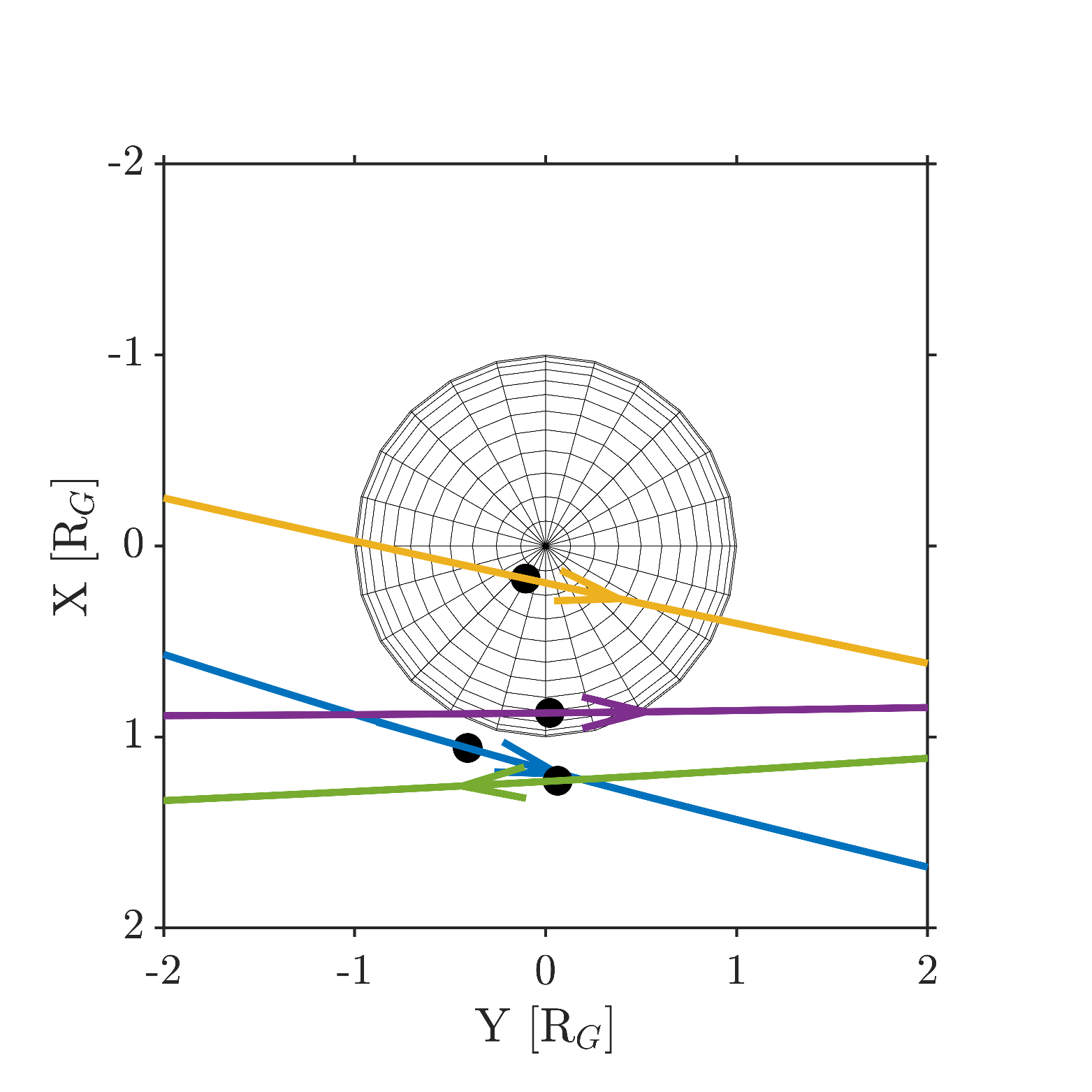}}
\caption{Trajectories of G01, G02, G07, and G29 flybys in GPhiO. The thicker, coloured arrows indicate Galileo's direction during these flybys (the arrow is the projection of the velocity vector and is barely visible on the XZ cut as the motion was mainly along the $y$ axis). The black dot along the trajectory corresponds to the closest approach. The black arrows indicate the direction of the Jovian plasma flow.} \label{Fig4}
\end{figure*}

For a better appreciation of the Galileo flybys, those crossing the Alfvén wings, namely G01, G02, G07, and G29, are shown in Fig.~\ref{Fig4}. The reader should refer to Table~\ref{table1} regarding the configuration and plasma conditions during these flybys. Interestingly, during G29, Ganymede passed through Jupiter's shadow, turning off photo-ionisation and, more critically, water sublimation. The lack of photoionisation is unlikely to have a strong influence as ionisation is dominated by electron impact \citep{Carnielli2019}, except in regions where electron impact is significantly reduced. The lack of sublimation will limit the presence of water and any related neutral and ionised products \citep{Leblanc2023}. These flybys can be typically separated into three sequences: inside the Jovian magnetosphere prior to the inbound MP crossing, inside Ganymede's magnetosphere from the inbound MP crossing and the outbound MP crossing, and outbound in the Jovian magnetosphere. 

\subsubsection{Plasma density and composition}\label{section311}

Fig.~\ref{Fig5} shows the simulated ion plasma densities (upper frame of each panel) and composition (lower frame of each panel) from our test particle simulations for G01, G02, G07, and G29 flybys. Electron number densities derived from PWS \citep{Gurnett1992,KurthPWS} have been added to the top figure as grey squares. Timings of the transition from magnetic field lines only tied to Jupiter to those tied to Jupiter on one end and to Ganymede on the other, referred to as the magnetopause, are indicated by vertical, solid black lines, when crossed, as well as the time of closest approach (cf. Table~\ref{table2} for the timing). Galileo is within Ganymede's magnetosphere between inbound and outbound MP crossings.

Before the inbound MP crossing and after the outbound MP crossing (see Fig.~\ref{Fig5}), Galileo was flying through the Jovian magnetosphere dominated by thermal O$^+$ and H$^+$. The way these Jovian thermal ions are simulated is described at the end of Section~\ref{section21}. The background Jovian plasma flow may vary both in terms of speed and ion number density, not only from flyby to flyby but even during the course of a flyby \citep{Jia2008}. Although \citet{Jia2008} simulated the Jovian plasma flow with a mean mass density of 28~u\,cm$^{-3}$ (corresponding to 2~cm$^{-3}$), we used 4~cm$^{-3}$ instead. This is the value required to have an agreement of the electron density between the simulation and PWS before the G07 inbound MP crossing (cf. Fig~\ref{Fig5}, lower right panel for G07, before 07:00).  Upstream of Ganymede, well beyond the magnetopause at around 1.8~$R_G$, the Jovian ion number density exhibits strong fluctuations (mainly seen before the MP inbound crossing of G01 and G02) along the corotating direction. These are caused by compressional magnetosonic waves propagating upstream of the magnetopause \citep{Jia2008}. As shown in \citet{Jia2008,Jia2009}, the Jovian plasma flow may be already significantly decelerated at $8~R_G$ upstream and lower than the bulk velocity we imposed. It is ignored here.

G01, G02, G07, and G29 exhibit different signatures at the inbound MP crossings. During G07, not only was Ganymede below the plasma sheet (meaning the Jovian magnetic field background had a positive $B_y$ component, the Alfvén wings were tilted differently as they followed $\vec{B}$ direction) but Galileo also travelled mainly along the $-Y$ direction, that is, more along the convective electric field carried by the Jovian plasma (unlike G01, G02, and G29). The sharpest transition amongst the MP inbound crossings is observed for G01 (cf. Fig.~\ref{Fig5}, upper, left panel) with a peak reaching more than 20~cm$^{-3}$. The increase in $n_i$ is seen before the magnetopause crossing at 06:19 and, therefore, outside Ganymede's magnetosphere. However, there is no such peak observed by PWS due to the lack of distinct upper-hybrid waves resonance or the structure being too short in time compared with PWS time resolution. The G01 inbound MP crossing is interesting in terms of ion composition. As we move forward in time from $\sim$06:19, the latter is dominated by ionospheric ions up to 90\%, mainly H$_2^+$, O$_2^+$, and H$_2$O$^+$, followed by an abrupt decrease down to 50\% at the inbound MP crossing. This indicates that the ionospheric ions cannot pass directly through the MP boundary. In contrast, both Jovian O$^+$ and H$^+$ ion number densities do not exhibit such discontinuity. The Jovian ion number density only decreases from 3 to 2~cm$^{-3}$ at the MP, though it decreases more drastically a few minutes later down to 0.3~cm$^{-3}$ around 06:20. The dynamics of ionospheric ions are different from that of Jovian ions: the former primarily leak through the magnetosphere upstream of the magnetopause and drift along the magnetopause, picked up by the Jovian plasma preferentially towards the anti-Jovian flank \citep[e.g. see][]{Stahl2023}. A shy peak in $n_i$ is also seen for G07 but the increase is very minimal. For G02, the MP crossing does not exhibit any local enhancement. 

Once Galileo crossed the inbound MP, all flybys exhibited an enhancement of the ion plasma number density as the spacecraft got closer to Ganymede, consistent with PWS measurements (see also Fig.~\ref{Fig3}). The trend is well captured by our simulations for G01, G07, G29, and G02 to a lesser extent. In addition, within Ganymede's magnetosphere, the ion composition is dominated by O$_2^+$ and H$_2^+$, followed by O$^+$ and H$_2$O$^+$ (except for G29 when Ganymede's was within Jupiter's shadow, suppressing the sublimation of water ices), consistent with the neutral exospheric composition being dominated by O$_2$ near the surface and H$_2$ at higher altitudes. Indeed, the latter has a much larger scale height and extent (see Section~\ref{section22} and Appendix~\ref{AppExo}).

For G01, there is a very satisfactory match between the simulated and observed plasma number densities over the inbound leg between the inbound MP crossing and the closest approach, with $n_i$ going from $\sim4$ cm$^{-3}$ to 60-70~cm$^{-3}$. $n_i$ is overestimated in the simulation over the outbound leg especially as Galileo moved away from the moon. During the outbound leg of G01, PWS electron number density has a slope closer to that of O$_2^+$ than that of H$_2^+$ suggesting we may overestimate the H$_2$ neutral number density: around CA, Galileo crossed the terminator and went on the nightside of Ganymede where there is no observational constraint on the exosphere on this side of the moon due to remote sensing limitations (see also the discussion in Section~\ref{section41}). 

For G02, even though our simulated ion number density is higher than that of \citet{Carnielli2019}, because of our neutral exosphere being ``wetter'' and slightly denser (see Section~\ref{section22}), we still underestimate the plasma number density reported by PWS with a factor of 2 to 5 lower, depending on the location, with a steeper slope in our model for the inbound leg from 18:50 to 19:00. The different slope in PWS data between inbound and outbound can be associated with dayside-nightside asymmetries. While the closest approach happened around 18:59, Galileo also crossed the terminator plane around 19:00. As O$_2^+$ is produced near the surface, the drop in O$_2^+$ could be linked to the photoionisation being turned off. However, as photoionisation is much lower than electron-impact ionisation \citep{Carnielli2019}, the diurnal effect is expected to be minimal. It may be species-dependent as the ratio between electron-impact and photoionisation ionisation frequencies varies from 3 to 10 across the different species. In addition, one expects H$_2$O$^+$ to be more present on the dayside (as seen in the ion composition, Fig.~\ref{Fig5}, upper, right panel, bottom row). Increasing H$_2$O number density on the dayside may help reconcile our modelled ion density with that derived from PWS for G02. Nevertheless, this is still puzzling as we used the same neutral exosphere for G01 and G02 (very similar configuration in terms of phase angle, see Fig.~\ref{Fig2}). A difference between both flybys in our simulations is the MHD electric and magnetic fields (see discussion in Section~\ref{section41}). 

For G07 and G29, it is important to set up an appropriate Jovian plasma background as the simulated ionospheric ion density remains below 10~cm$^{-3}$. Only G07 and G29 have a non-negligible contribution around CA from the Jovian thermal H$^+$ and O$^+$ (5\% to 10\% of the total ion number density) coherent with Galileo's CA being farther away from the moon (cf. Table~\ref{table1}). For instance, during G07, setting up the Jovian thermal plasma density around 4~cm$^{-3}$ provides a very good agreement prior to the inbound MP crossing, as discussed earlier in this section, and within Ganymede's magnetosphere. For G29, while our simulation shows an increase in terms of $n_i$ as Galileo approached the moon, reaching up to 10~cm$^{-3}$, PWS electron number density remains surprisingly flat with two plateaus, one from 08:22 and 08:28 at 15~cm$^{-3}$ and another from 08:30 and 08:33 at 7~cm$^{-3}$. The double plateau is not captured by the test particle simulation and its origin is not clear.

Regarding the outbound MP crossing into the Jovian magnetosphere, for G01 the spacecraft position provided along with the magnetic field data \citep{KivelsonG} unfortunately stops at 06:53 while the outbound MP was reported around 06:59 \citep{Kivelson1998}. However, while $n_i$ was decreasing after CA, $n_i$ increases again around 06:50, probably because Galileo was getting closer to the MP, with the ion composition dominated by O$_2^+$ and H$_2^+$. H$_2$O$^+$ is missing as Galileo went out of Ganymede's magnetosphere on the nightside on the Jovian flank. G02 and G29 do not exhibit this signature. For G07, $n_i$ has a very localised peak right before the outbound MP crossing. Although Galileo was on the nightside at this time, the spacecraft came out of Ganymede's magnetosphere on the anti-jovian side. During G07 flyby, the upstream magnetopause coincided with the subsolar point: H$_2$O$^+$ ions are produced not far from the subsolar point, transported towards the magnetopause upstream, and then drifted along the magnetopause preferentially on the anti-jovian side.

\begin{figure*}
\centering
\fcolorbox{G01}{white}{\parbox{\columnwidth}{
\centering
\includegraphics[width=\linewidth]{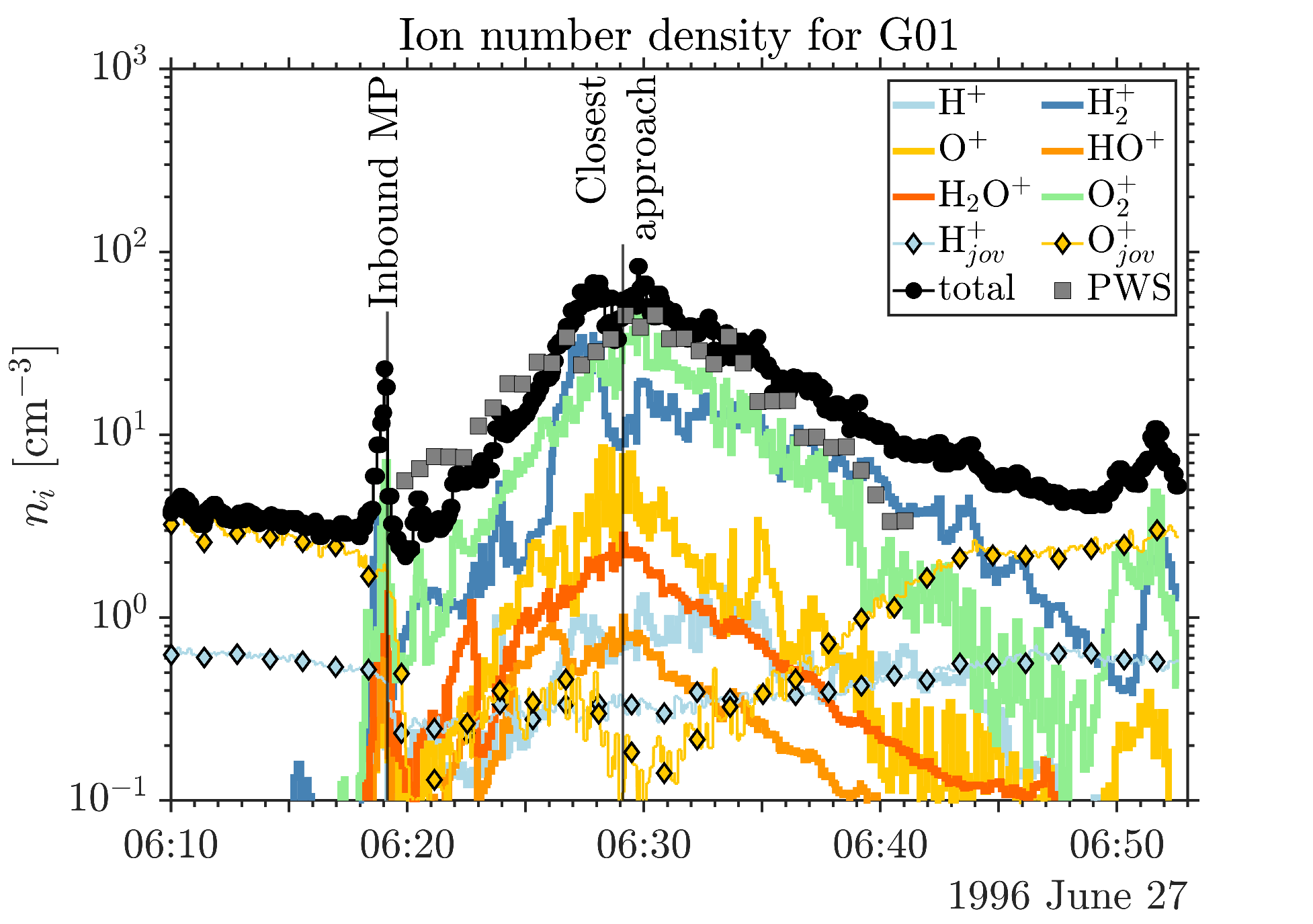}\\
\includegraphics[width=\linewidth]{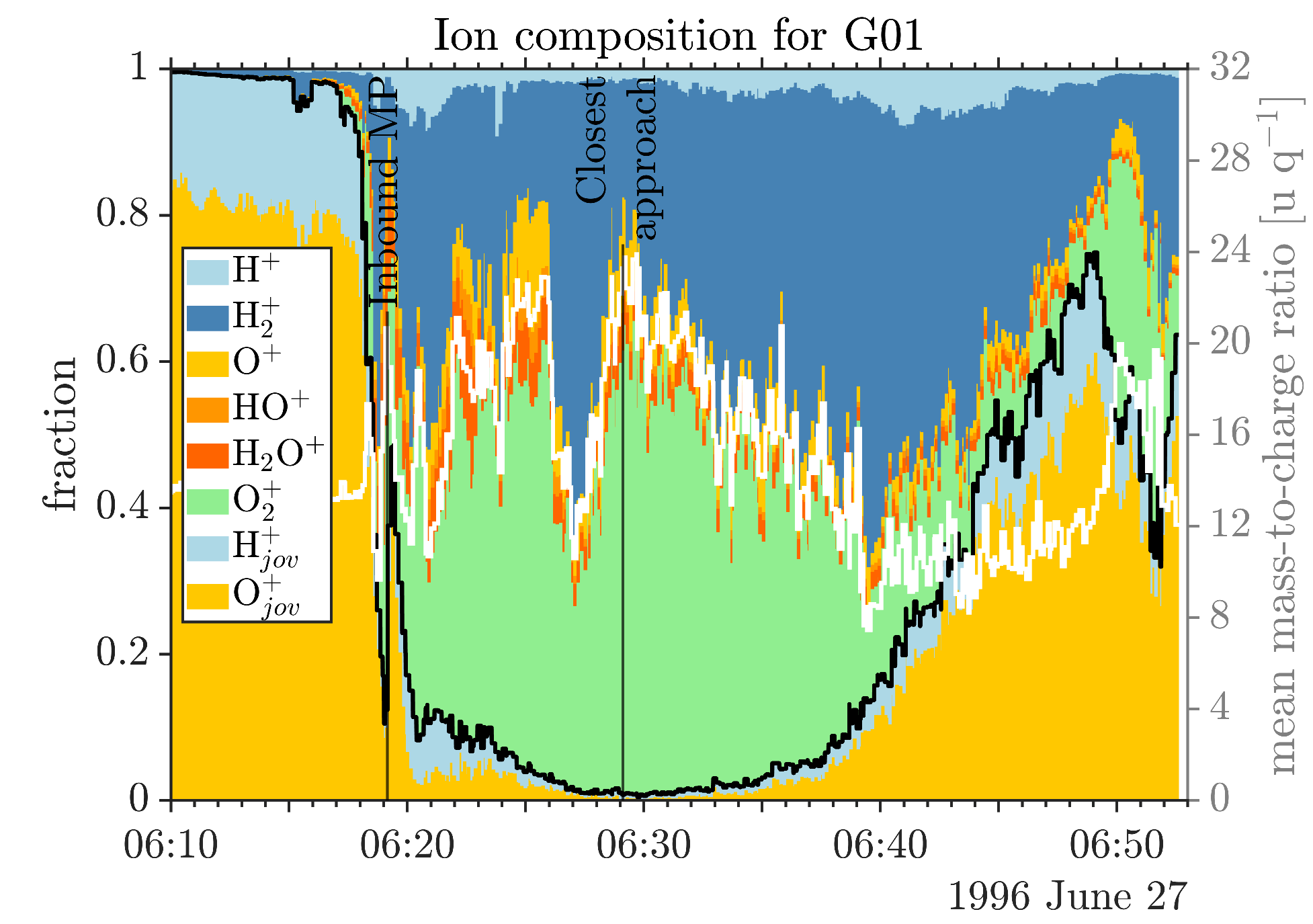}}}\!
\fcolorbox{G02}{white}{\parbox{\columnwidth}{
\centering
\includegraphics[width=\linewidth]{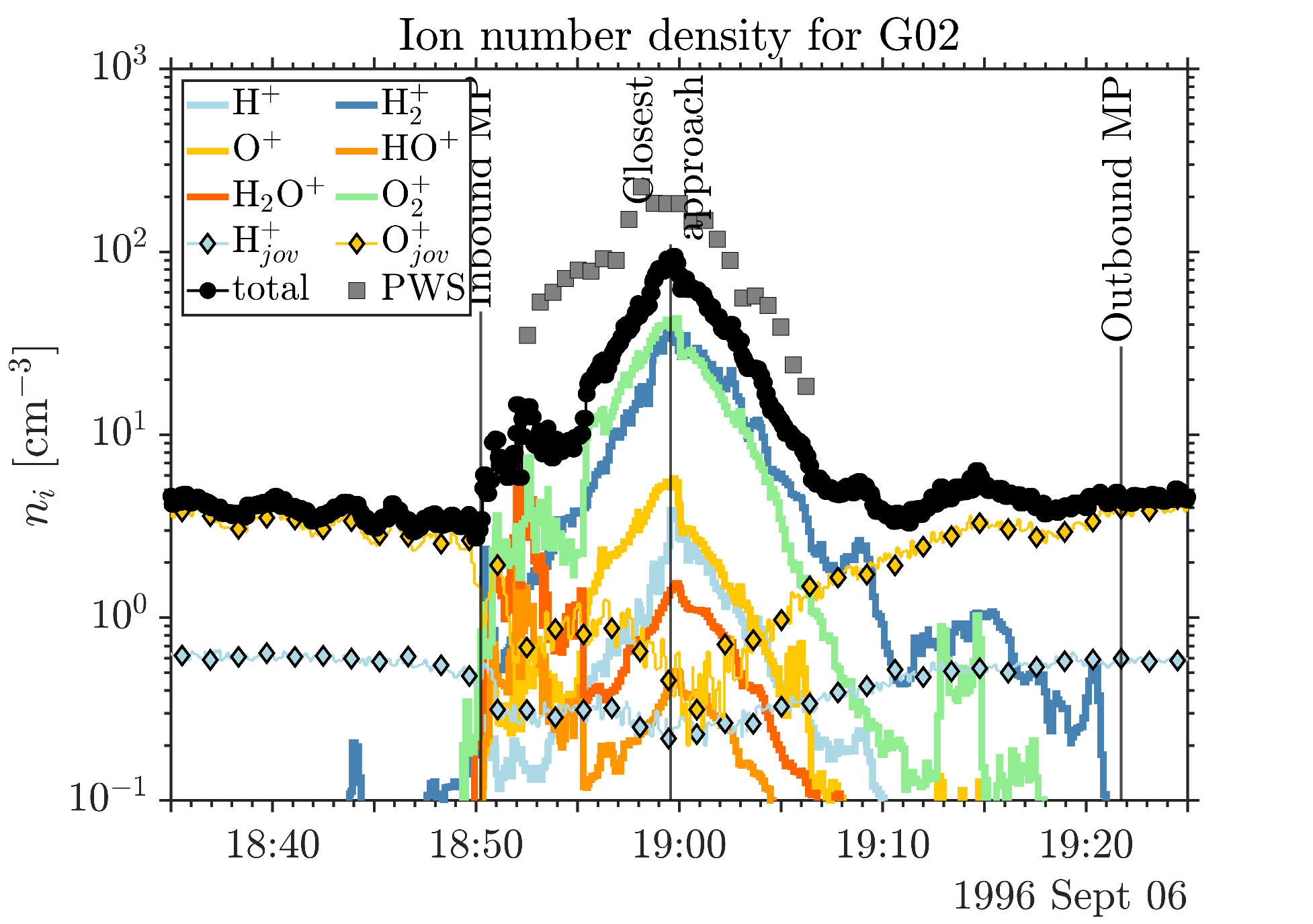}\\
\includegraphics[width=\linewidth]{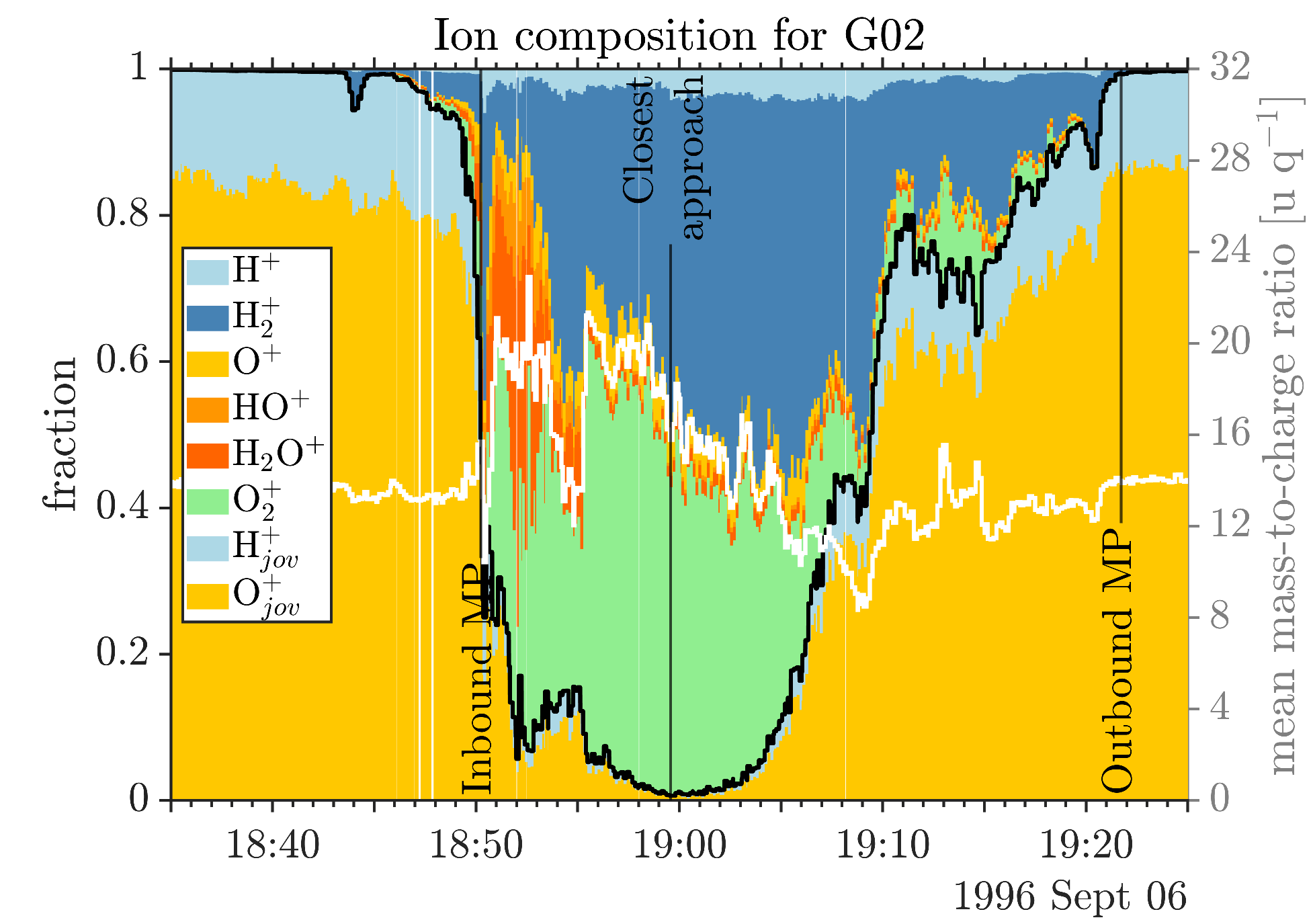}}}
\caption{Within the Alvén wings: simulated ion number densities for Jovian and ionospheric ions compared with in-situ PWS electron number density from \citet{KurthPWS} (top row of each panel) and simulated ion composition (bottom row of each panel) during G01 (top, left panel), G02 (top, right panel), G07 (bottom, left panel), and G29 (bottom, right panel) flybys. Vertical, black lines feature the inbound magnetopause crossing, the closest approach, and the outbound magnetopause crossing (if available from the MAG data), respectively. For the ion composition, the black thick line represents the fraction of Jovian ions. The white thick line must read with the right axis and corresponds to the mean ion mass per charge. Gaps in the magnetic field data and therefore Galileo's location are responsible for the vertical white stripes in some ion composition plots.}\label{Fig5}
\end{figure*}
\begin{figure*}
\ContinuedFloat
\centering
\fcolorbox{G07}{white}{\parbox{\columnwidth}{
\centering
\includegraphics[width=\linewidth]{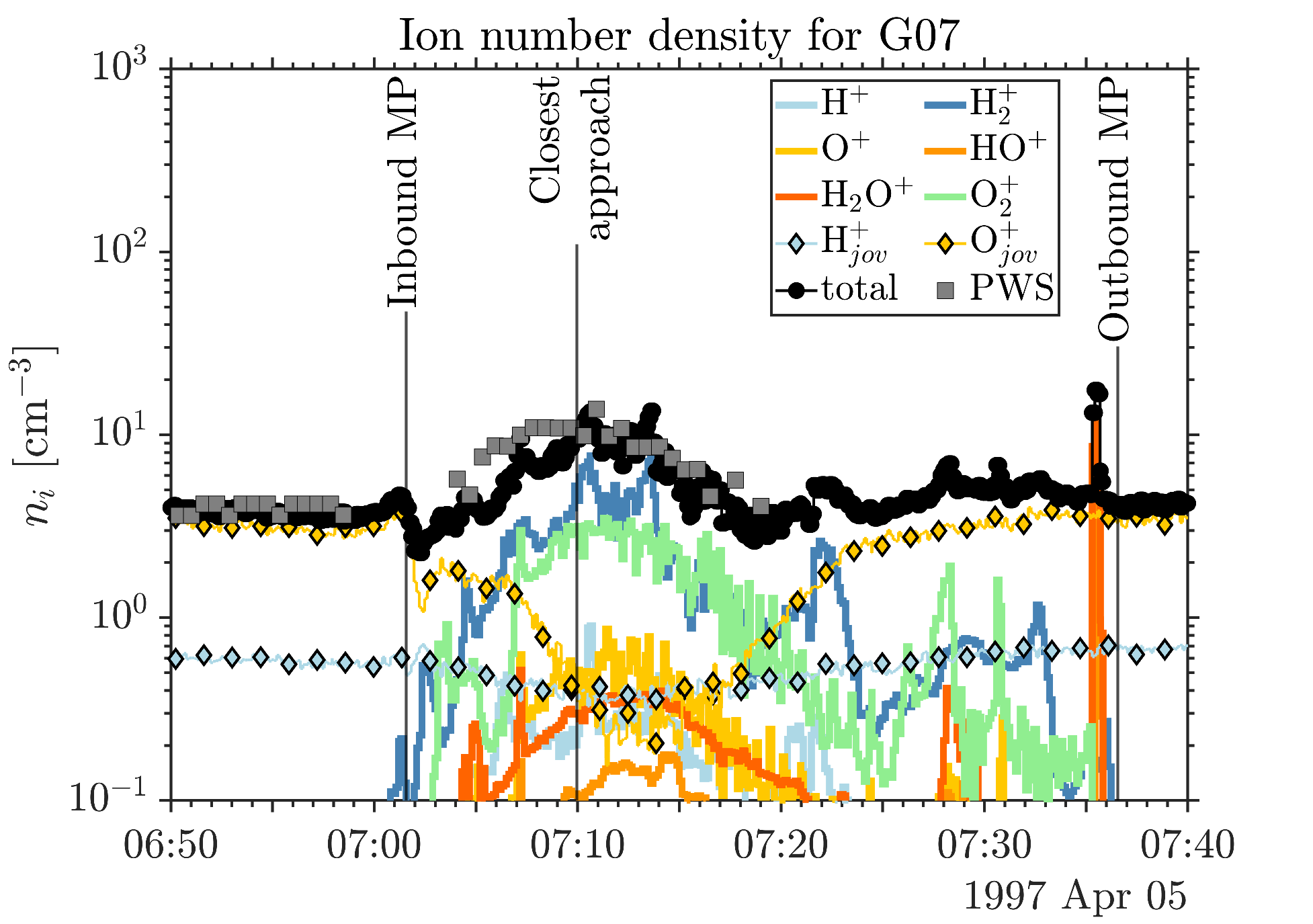}\\
\includegraphics[width=\linewidth]{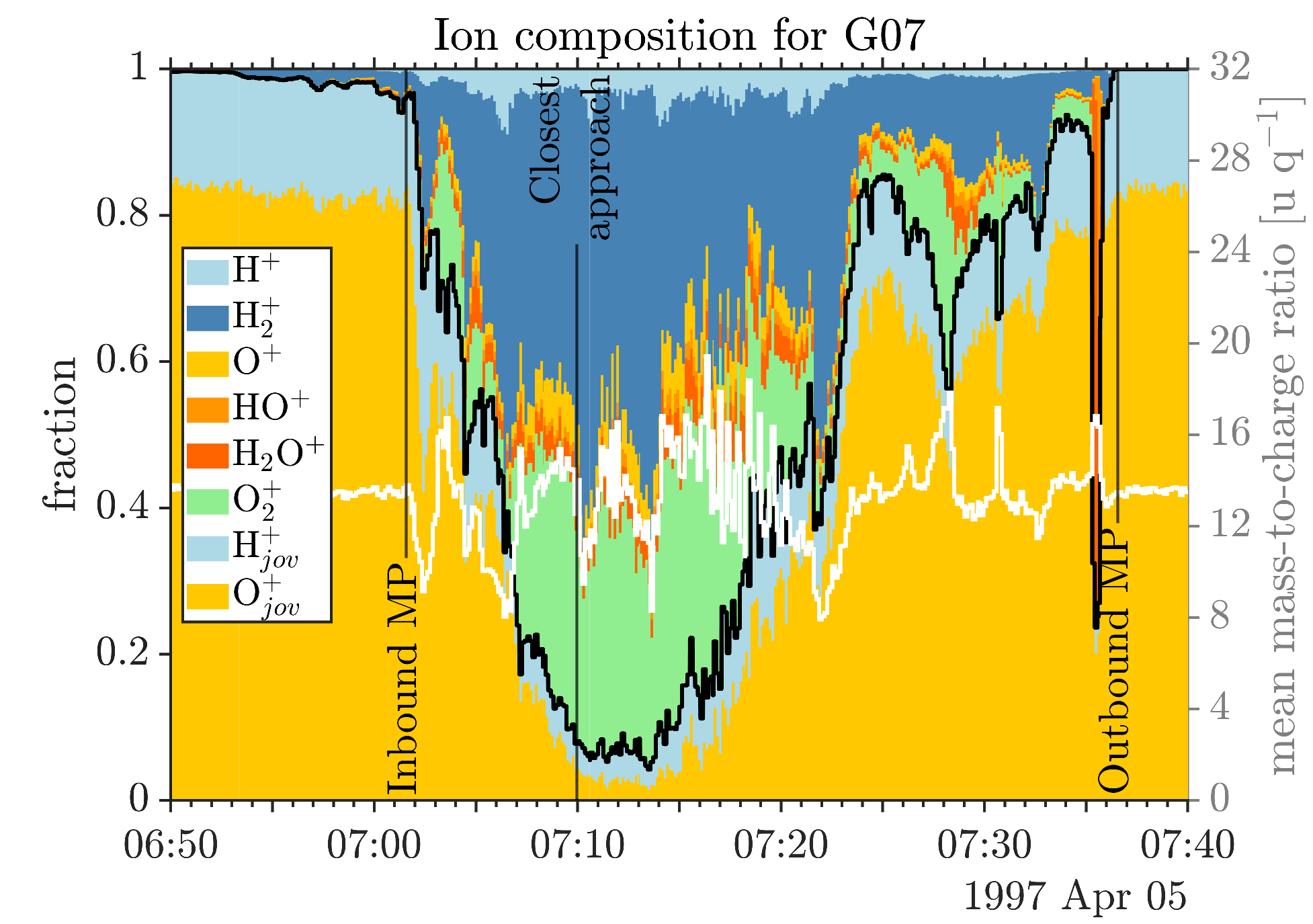}}}\!
\fcolorbox{G29}{white}{\parbox{\columnwidth}{
\centering
\includegraphics[width=\linewidth]{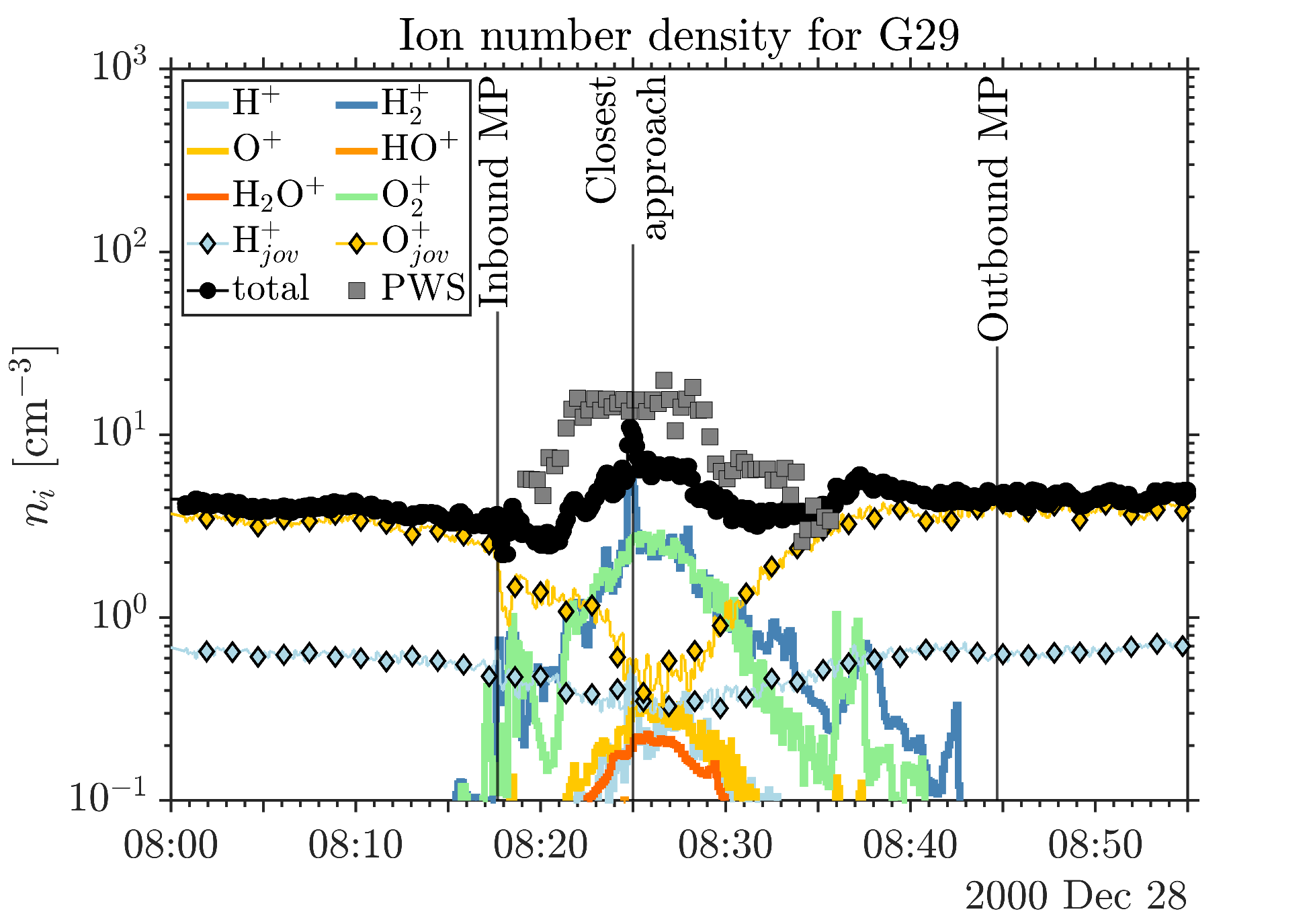}\\
\includegraphics[width=\linewidth]{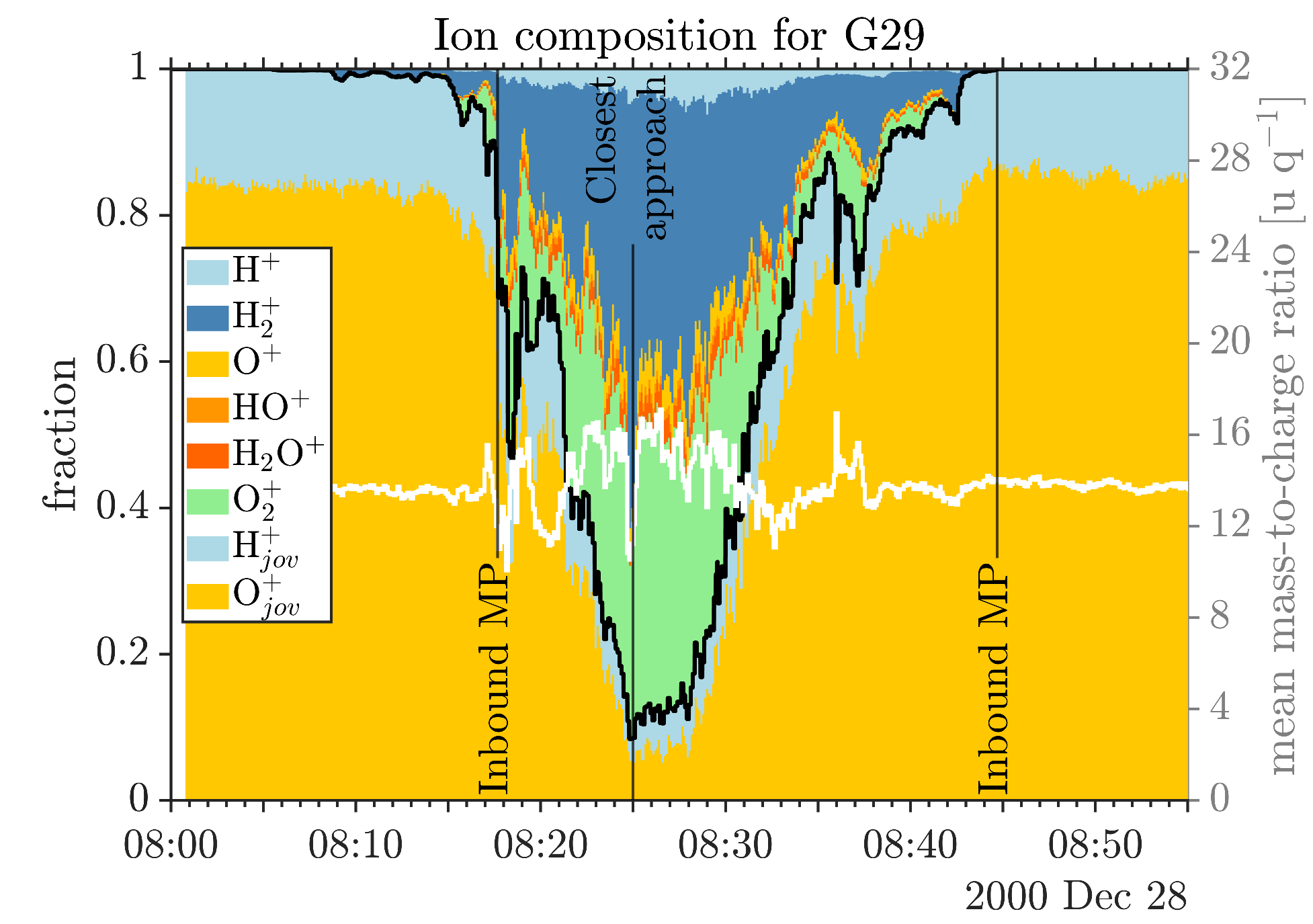}}}\\
\contcaption{\label{Fig5cont}}
\end{figure*}

\subsubsection{Ion velocity}\label{section312}

Fig.~\ref{Fig6} shows the perpendicular (top panel) and parallel (bottom panel) components of the ion bulk velocity to the local magnetic field (resp. $||\vec{\varv}\times \vec{B}||/B$ and $\vec{\varv}\cdot\vec{B}/B$), for each ion species, based on the fields from the MHD simulations. We also plotted in the perpendicular direction the local $\vec{E}\times\vec{B}$ drift,  $||\vec{\varv}_{\vec{E}\times \vec{B}}||$ (top panel of each frame). The latter is critical as it may be used as a sanity check against MHD simulations. Our MHD simulations rely on resistive MHD, that is, $\vec{E}=-\vec{\varv}\times\vec{B}+\eta \vec{J}$ where $\vec{\varv}$ is the plasma bulk velocity is the electric field, $\vec{B}$ is the magnetic field, $\vec{J}$ is the current density, and $\eta$ is the resistivity depending on the radial distance \citep{Jia2009}. Therefore, the perpendicular ion bulk velocity is expected to follow $\vec{\varv}_{\vec{E}\times \vec{B}}\approx (\vec{E}-\eta \vec{J})\times\vec{B}/B^2=\vec{E}\times\vec{B}/B^2$ except near the surface and at the magnetopause due to the addition of an anomalous resistivity \citep{Jia2009}. $\vec{\varv}_{\vec{E}\times \vec{B}}$ also gives insights on the Galileo position with respect to the Alfvén wings: the lower $||\vec{\varv}_{\vec{E}\times \vec{B}}||$ is, the closer Galileo was from the centre of the wings.

 Every flyby crossing the Alfvén wings exhibits very similar trends. In general, Galileo was first within the undisturbed Jovian magnetosphere where the ion plasma flow, dominated by Jovian H$^+$ and O$^{+}$ (not shown), drifts around 140~km\,s$^{-1}$ \citep[130~km\,s$^{-1}$ for G07, cf. ][]{Jia2008}, perpendicular to the ambient magnetic field, in the positive $x$ direction in GPhiO coordinates. As the Jovian plasma flow approaches the magnetopause, it is strongly decelerated without forming a shock. The deceleration is more significant for G01 and G02 than for G07 and G29 as the former flybys penetrated Ganymede's magnetosphere much closer to the moon. $||\vec{\varv}_{\vec{E}\times \vec{B}}||$ exhibits strong gradients right after the magnetopause crossings (indicated by a vertical, solid line). As shown in Fig.~\ref{Fig1}, these gradients are particularly strong in the open magnetic field line region (cf. the topology of magnetic field lines connected to Ganymede or not) and would correspond to a cusp-like region. G01 and G02 crossed this region, while G07 and G29 do not, only flying above it according to the MHD model. As Galileo approached the moon, $||\vec{\varv}_{\vec{E}\times \vec{B}}||$ continued to decrease down to a few km\,s$^{-1}$, that is, of the same order of magnitude as Galileo's speed (7-10~km\,s$^{-1}$); this means that ion observation may be strongly biased and great care must be taken interpreting ion energy data in this region (see Section~\ref{section313}). The minimum perpendicular speed is not always reached near the closest approach and is sometimes significantly delayed (e.g. G07 around 07:15). As Galileo flew away from the moon, $||\vec{\varv}_{\vec{E}\times \vec{B}}||$ increases again, sometimes with a weak drop before the magnetopause outbound crossing, such as for G02 and G29. 

Ion speed $||\vec{\varv}_\perp||$ perpendicular to the magnetic field  (see Fig.~\ref{Fig6}, top panel of each frame) follows extremely well, even overlaping, $\vec{\varv}_{\vec{E}\times \vec{B}}$ within the Alfvén wings. As Galileo flew away from Ganymede and got closer to the magnetopause, $||\vec{\varv}_\perp||$ tends to separate from $||\vec{\varv}_{\vec{E}\times \vec{B}}||$. This separation is larger where $||\vec{\varv}_{\vec{E}\times \vec{B}}||$ has strong gradients near the magnetopause. This effect is very likely associated with finite Larmor radius effect. Indeed, as the ions approach the magnetopause, the magnetic field strength decreases such that the ion gyroradii increase and ionospheric ions become less and less magnetised as they move away from Ganymede and from their production region. The effect depends on the ion gyroradius and mass. We observe different separation for different ion species near the magnetopause. The higher the ion mass is, the more $\vec{\varv}_\perp$ departs from $\vec{\varv}_{\vec{E}\times \vec{B}}$ as seen with O$_2^+$. Once outside Ganymede's magnetosphere and Alfvén wings, ions are accelerated and picked up by the thermal Jovian plasma flow. Our simulation shows strong variations resulting from the low density and statistics. Ionospheric ions leak with difficulty through the magnetopause in the Jovian magnetosphere.

Fig.~\ref{Fig6}, bottom panel of each frame, shows the ion speed in the direction of the local magnetic field. For G01, ${\varv}_\parallel$ (or more appropriately $\vec{\varv} \cdot \vec{B}/||\vec{B}||$) is rather constant and flat within Ganymede's magnetosphere compared to G02, G07, and G29. For all cases, ${\varv}_{\parallel}$  is more and more negative as the spacecraft position gets farther away from Ganymede and closer to the magnetopause, meaning that all ionospheric ions move away from Ganymede as the magnetic field points southwards while Galileo crosses Ganymede's magnetosphere over the Northern Hemisphere. The fact that ${\varv}_\parallel$ and $\vec{\varv}_\perp$ are of the same order of magnitude is coherent for ions drifting along the Alfvén wings, as the wings are tilted with respect to the magnetic field lines \citep{Kivelson2004}. Therefore, one would expect that $\arctan(\varv_{\parallel}/\varv_\perp)$ represents the angle between the magnetic field lines and the direction of the Alfvén wings. Once outside of the Alfvén wings, ${\varv}_\parallel$ returns close to 0. The inbound MP crossings mark the largest discontinuities in $\varv_\parallel$ for G01 and G02 as it reverses its sign with a strong spatial gradient: either side of the magnetopause, ions flow with an opposite component, that said $\varv_\parallel=\vec{\varv} \cdot\vec{B}$, along $\vec{B}$. That indicates that ionospheric ions are not leaking through the magnetopause at this location and drift in opposite directions on either side of the magnetopause. The ionospheric ions seen outside of the Alfvén wings, before crossing the inbound MP, are mainly leaking from upstream of the magnetopause, picked up, and then drifting preferentially on the anti-Jovian flank \citep[e.g. see][]{Stahl2023}

\begin{figure*}
\centering
\fcolorbox{G01}{white}{\parbox{\columnwidth}{
\centering
\includegraphics[width=\linewidth]{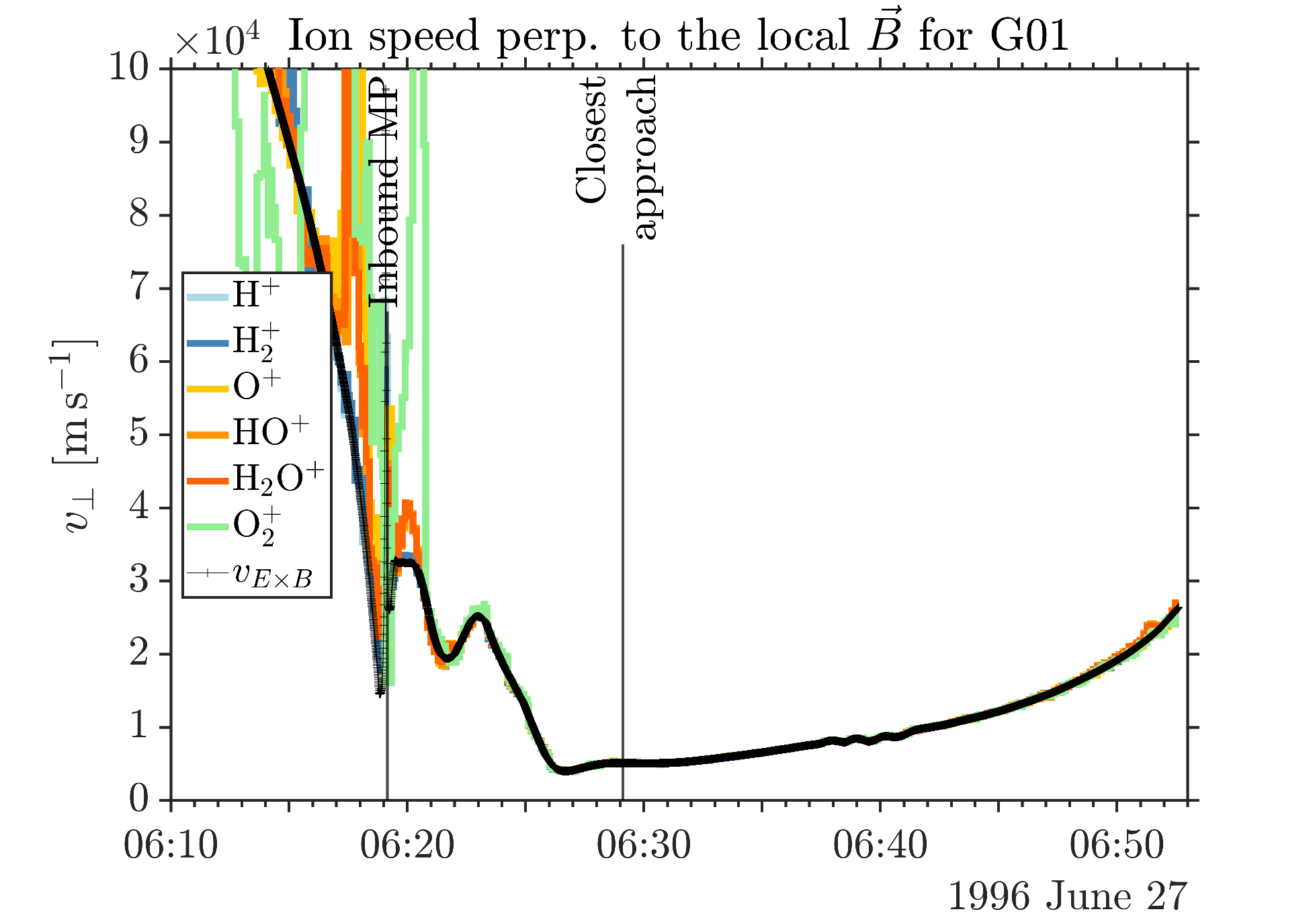}\\
\includegraphics[width=\linewidth]{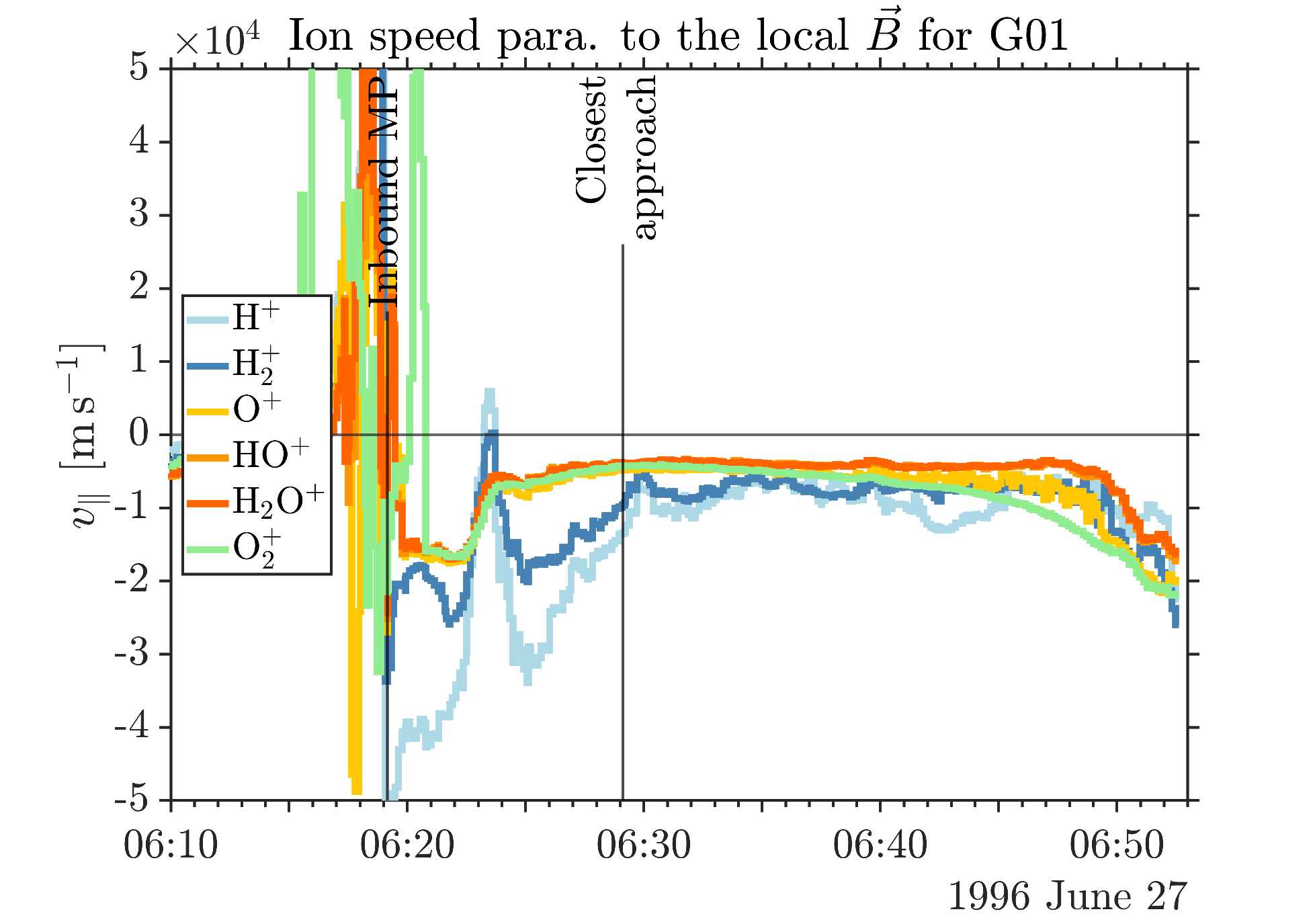}}}\!
\fcolorbox{G02}{white}{\parbox{\columnwidth}{
\centering
\includegraphics[width=\linewidth]{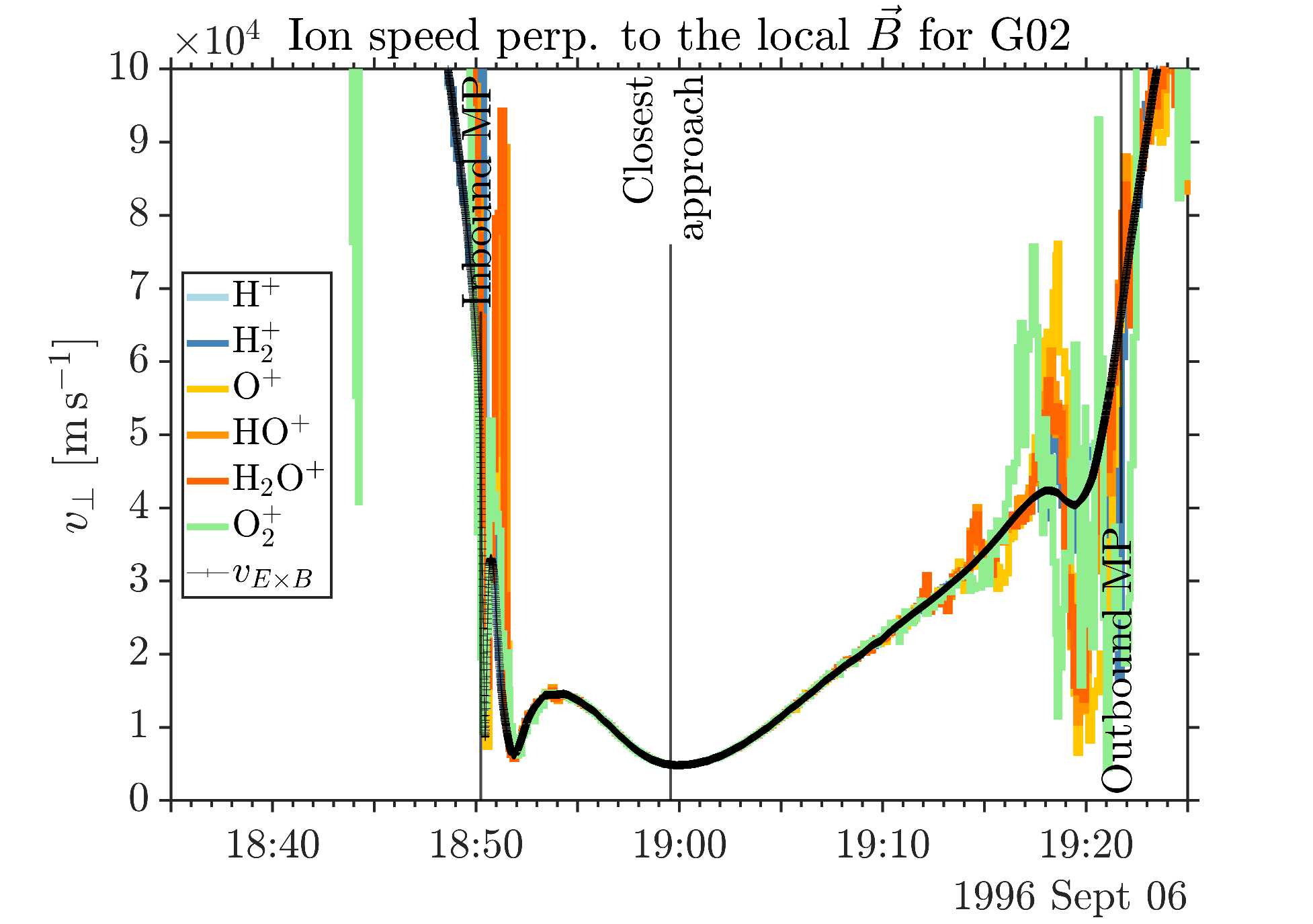}\\
\includegraphics[width=\linewidth]{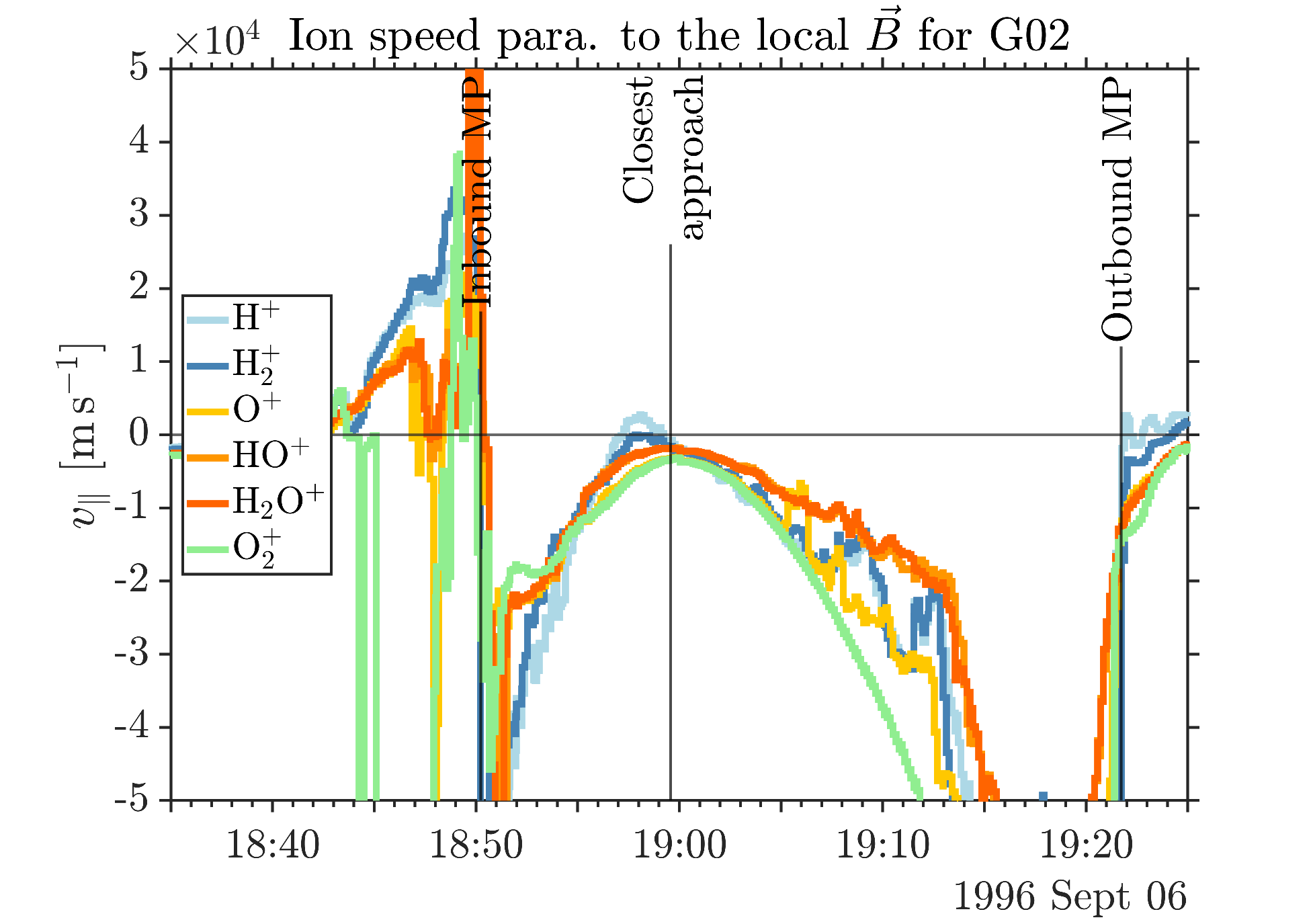}}}\\
\caption{Within the Alvén wings: same as Fig.~\ref{Fig5}, but for ion speed perpendicular (top panel in each frame) and parallel (bottom panel of each frame) to the local magnetic field. Only ionospheric ions are shown. \label{Fig6}}
\end{figure*}
\begin{figure*}
\ContinuedFloat
\centering
\fcolorbox{G07}{white}{\parbox{\columnwidth}{
\centering
\includegraphics[width=\linewidth]{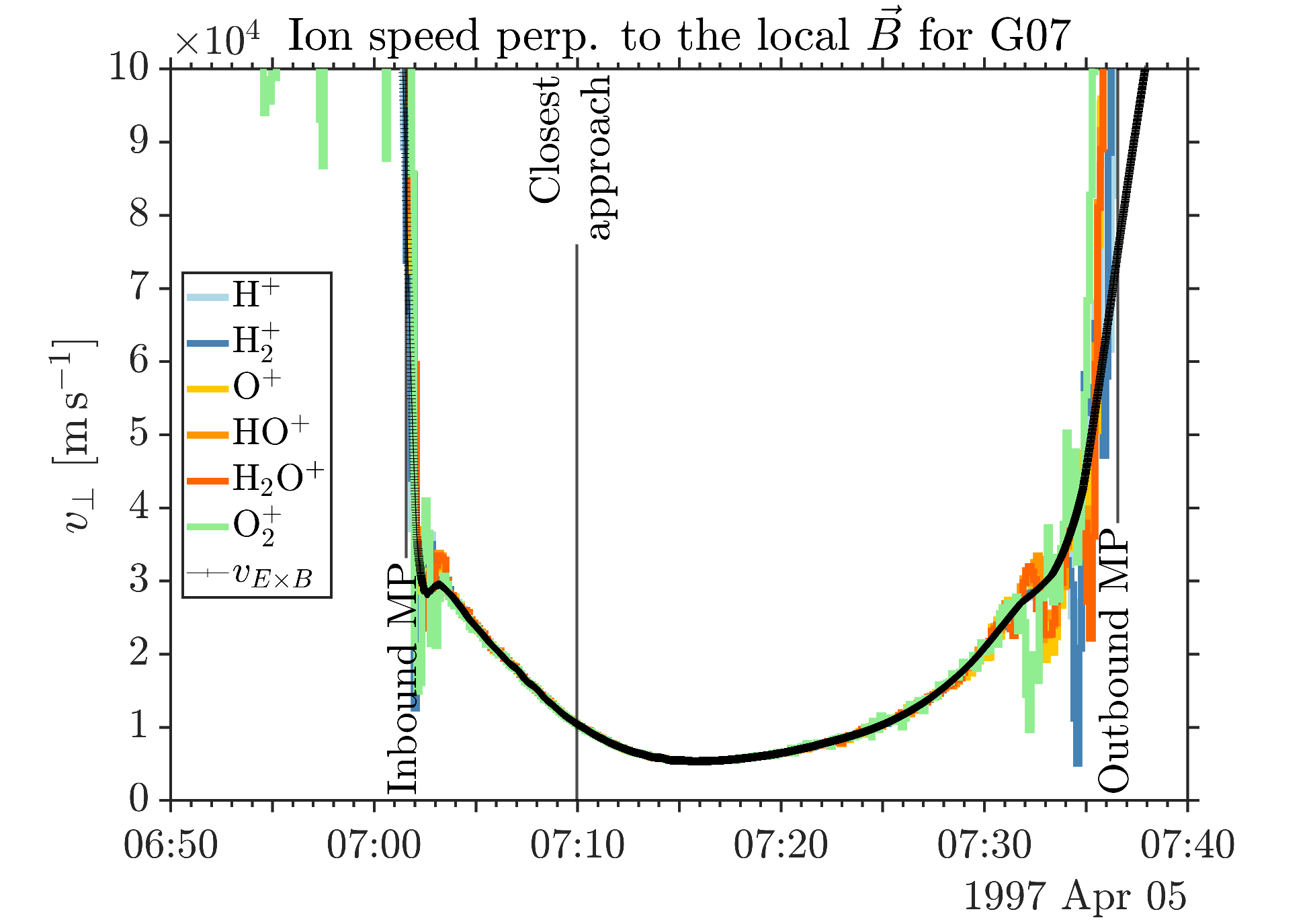}\\
\includegraphics[width=\linewidth]{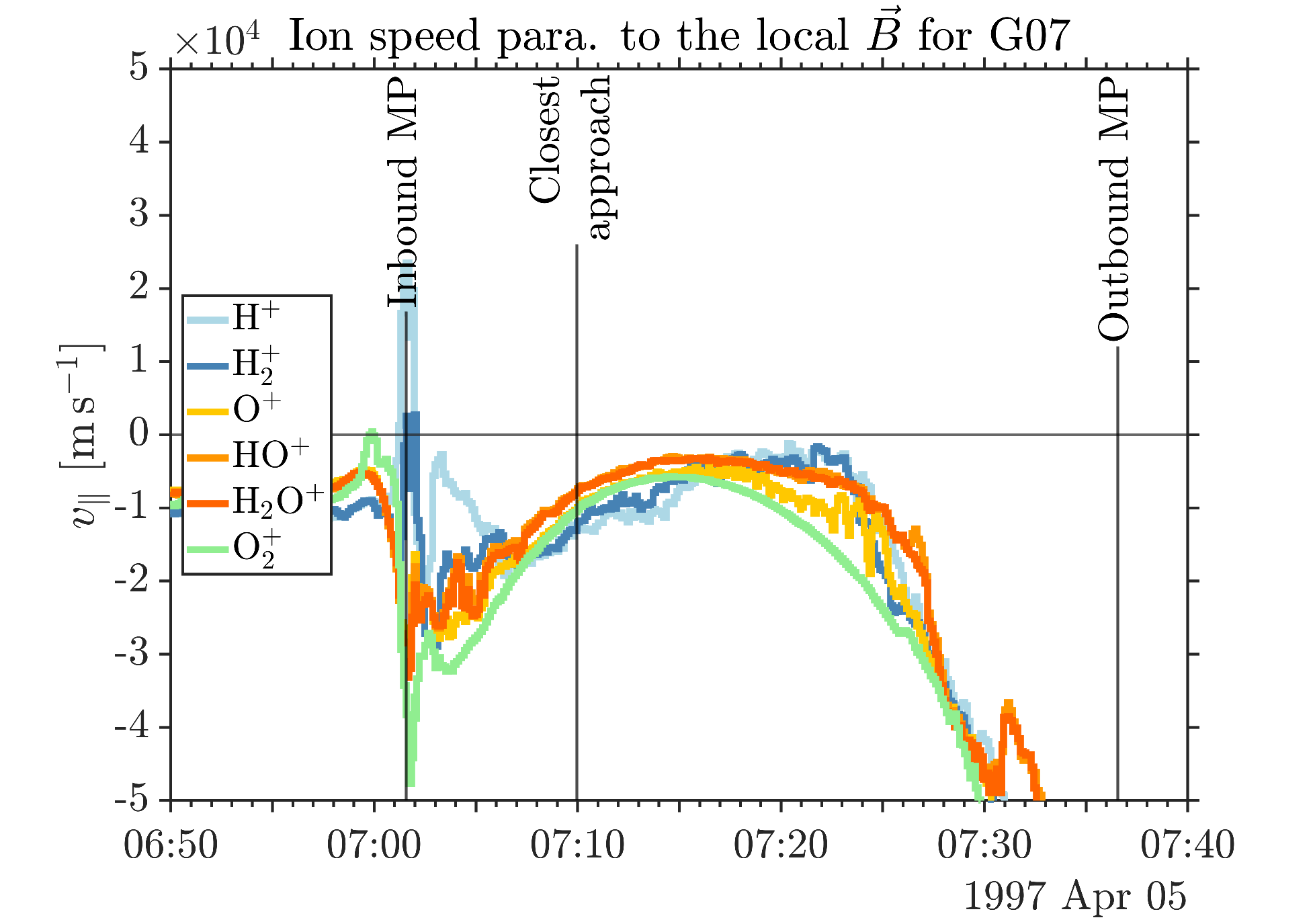}}}\!
\fcolorbox{G29}{white}{\parbox{\columnwidth}{
\centering
\includegraphics[width=\linewidth]{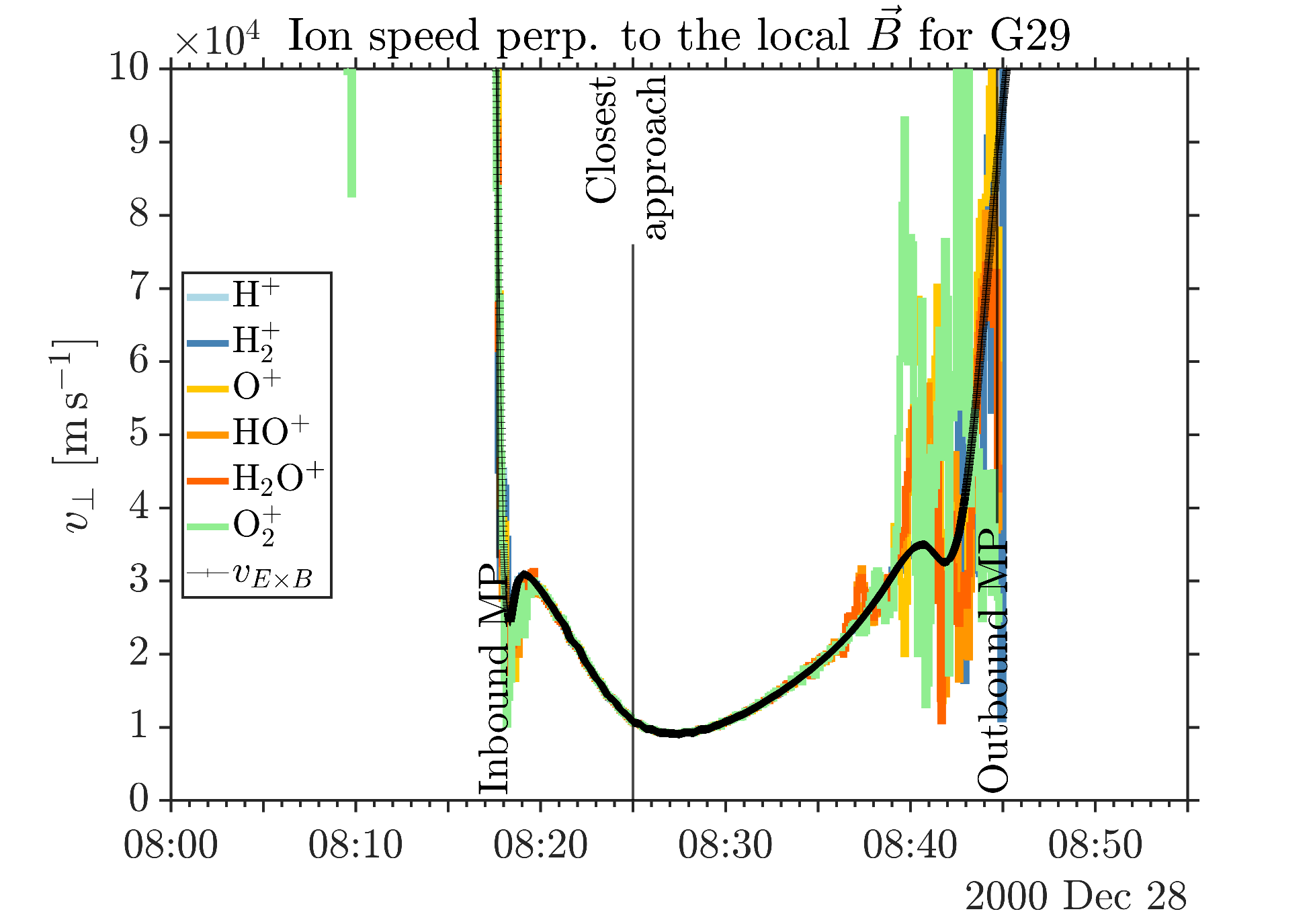}\\
\includegraphics[width=\linewidth]{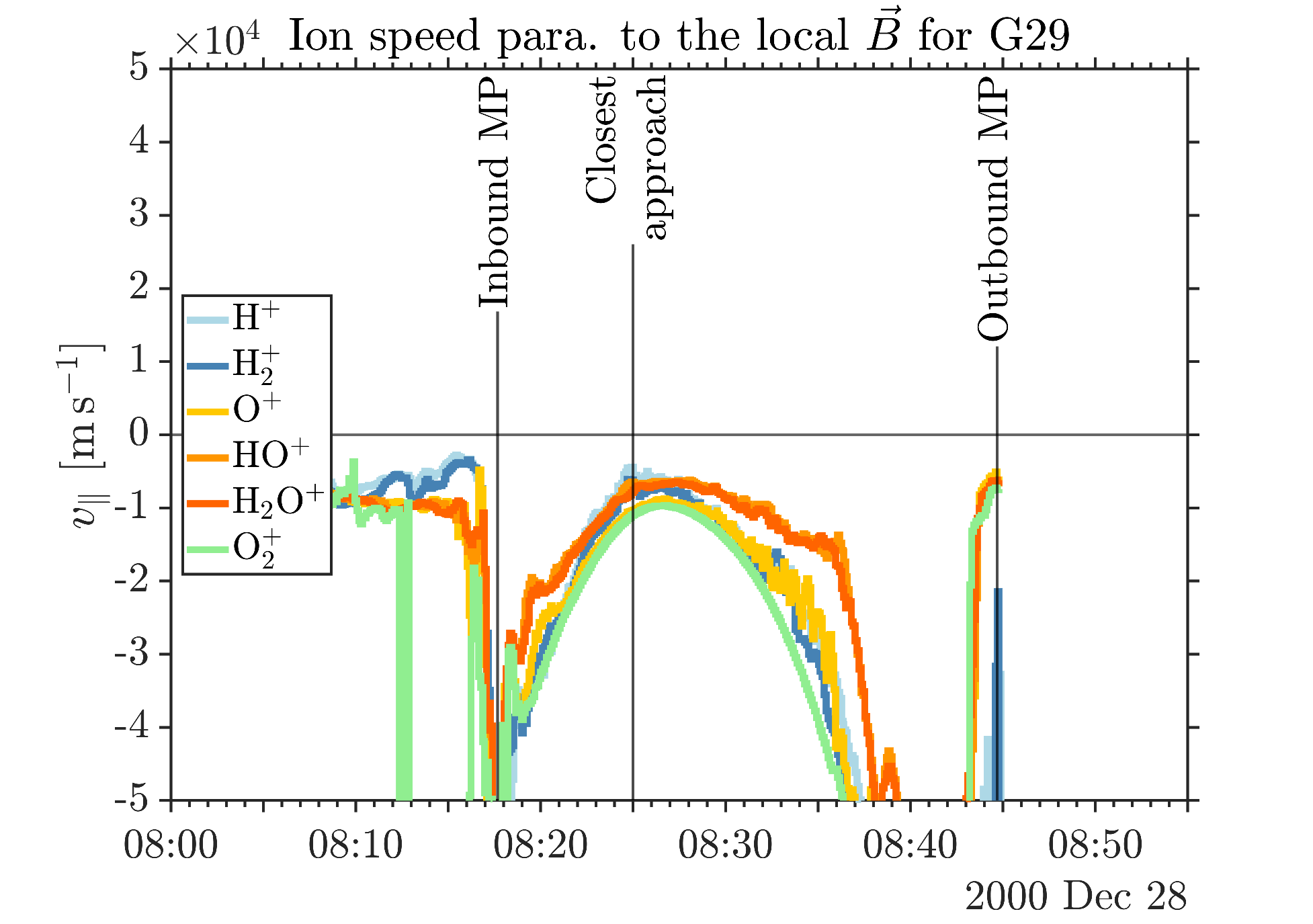}}}
\contcaption{\label{Fig6cont}}
\end{figure*}

\subsubsection{Ion energy spectra}\label{section313}

Fig.~\ref{Fig7} shows the simulated total ion (thermal Jovian ions and those from Ganymede) energy spectra (top row of each frame) and PLS (bottom row) ion energy spectra for G01, G02, G07, and G29. In order to help and support our interpretation, we superimposed the kinetic energy of O$_2^+$ ion drifting at $\vec{\varv}_{\vec{E}\times \vec{B}}$ in the spacecraft frame, that is:
\begin{equation}
    E_{\perp,{\text{O}_2^+}}=\dfrac{1}{2}m_{\text{O}_2^+}(\vec{\varv}_{\vec{E}\times \vec{B}}-\vec{\varv}_{SC})^2
    \label{Ek}
\end{equation}
where $\vec{\varv}_{SC}$ is the spacecraft velocity. For our simulated spectra, $\vec{\varv}_{SC}$ is assumed to be constant and corresponds to the averaged Galileo flyby speed (see Table\,\ref{table1}). In contrast, for PLS ion energy spectra, $\vec{\varv}_{SC}$ is based on the spacecraft's position provided with MAG data on the PDS \citep{KivelsonG}. As ion energy spectra are specific to each flyby, they are treated separately in the following paragraphs. 

Fig.~\ref{Fig7}, top left panel, shows the simulated and PLS ion energy spectra during G01. The ion flux within the Jovian magnetosphere before 06:15 is dominated by thermal Jovian O$^+$ ($10^3-10^4$~eV). Closer to the magnetopause, the modelled ion energy flux decreases and peaks at lower energy, around a few hundred eVs. Such a trend is not seen in the corresponding PLS data (lower panel), though there may be too much noise to do so. At the inbound MP crossing, there is a thin but abrupt transition where we observe a clear energisation of ionospheric ions with ion fluxes over the range 10~eV to 1~keV. The higher the ion mass is, the higher their energy goes. Our simulations show that the signature in the data at 10~keV at the inbound magnetopause crossing (06:19) and up to 50~keV is associated with ionospheric ions, such as O$_2^+$. It is not driven by reconnection, in contrast with \citet{Collinson2018} and \citet{Fatemi2022}, as the fields are from a snapshot of the MHD simulations. This is further discussed in Section~\ref{section4}.

Even though O$_2^+$ ions are expected to be at $E_{\perp,{\text{O}_2^+}}\sim 3.6~\text{keV}$, that is true only if their gyroradius $r_g$ is negligible with respect to the magnetic field variation (i.e. $r_g||\nabla B / B||\ll 1$). However, supposing that O$_2^+$ is close to rest in Ganymede's frame once it enters the Jovian magnetosphere, its gyroradius should be calculated in the local plasma frame and is given by:
\begin{equation}   
r_{g,\text{O}_2^+}=\dfrac{m_{\text{O}_2^+}||\vec{\varv}_{\vec{E}\times \vec{B}}||}{qB}\approx \dfrac{m_{\text{O}_2^+}{\varv}_\text{jov, flow}}{qB_\text{jov}} \sim 0.17 R_G
\end{equation}
where ${\varv}_\text{jov, flow}\approx\,140$\,km\,s$^{-1}$ is the speed of the thermal Jovian plasma and $B_\text{jov}\approx\,110\,\text{nT}$ is the Jovian magnetic field strength at Ganymede's location \citep{Jia2008}. Its speed varies between 0 and 2$||\vec{\varv}_{\vec{E}\times \vec{B}}||$. Therefore, O$_2^+$ is varying between 0 and $4E_{\perp,{\text{O}_2^+}}$ with an average energy of $2E_{\perp,{\text{O}_2^+}}$ over a gyroperiod. After the inbound MP crossing (06:19), the flux is dominated by ionospheric ions, according to our simulation (Fig.~\ref{Fig7}). In the simulated spectrum, between 06:19 and 06:25, we can distinguish 3 bands parallel to each other (with 2 bands very close near the curved dashed line). At low energy below 30 eV, the ion flux is associated with H$^+$ and H$_2^{+}$. At higher energy above 30~eV, we observe 2 bands. The lower one is associated with the ion group around $m/z\approx16$ u\,q$^{-1}$, that is O$^{+}$, HO$^{+}$, and H$_2$O$^{+}$, while the upper one is due to O$_2^+$. Towards CA, the modelled ion flux peaks at lower and lower energy following the decrease in $\vec{\varv}_{\vec{E}\times\vec{B}}$ (cf. Fig.~\ref{Fig6}). Such a trend is not observed in the PLS data. There is only a weak ion flux around 30-40~eV at 06:21-23. Near the closest approach, only modelled O$_2^+$ remain above 15~eV. However, PLS data shows a large ion flux between 10~eV and 100~eV, from 06:25 to 06:35. These energies are much larger than those simulated. The possible causes of this discrepancy are numerous: exospheric neutral density and composition, spacecraft potential, field-aligned acceleration, and/or lack of terms in Ohm's law of the MHD simulations. They are discussed in more detail in Section~\ref{section41}. From 06:35 onwards, the simulated energy spectrum remains flat as reflected by the velocity in Fig.~\ref{Fig5}. PLS data shows a weak flux around 100~eV starting from 06:40 and a band at lower energy starts to appear at the same time onwards, going from 10~eV to 20-30~eV at 06:50. The former is likely associated with O$_2^+$, as O$_2^+$ has the largest mass and kinetic energy, and the latter to H$_2^+$, though we cannot reconcile the energies with our test-particle simulation as they are observed at much higher energies, above the dashed curve (cf. caption).

Fig.~\ref{Fig7}, top right panel, shows the simulated and PLS ion energy spectra during G02. From 18:35 to 18:50, the modelled ion energy spectrum is very similar to that of G01 with a band associated with Jovian O$^+$ ions around 2~keV consistent with $0.5E_{\perp,\text{O}_2^+}$. In contrast, PLS ion energy flux peaks at lower energy, around 200-300~eV. Modelled and PLS ion energy fluxes show a consistent signature at the crossing, similar to G01, with an increased flux at all energies for a short period of time around 18:51. Between 18:51 and 18:57, the PLS ion energy flux peaks at 100~eV at 08:53 and then decreases down to 20~eV. Our simulated spectrum for O$_2^+$, not low mass ions, agrees with this. This is consistent with the change in $E_{\perp,\text{O}_2^+}$ though there is still a significant contribution to the kinetic energy in the parallel direction to the magnetic field. Around CA, from 18:58 to 19:03, PLS shows enhanced flux up to 50~eV whereas our simulations remain limited to 30~eV with much lower fluxes. That might be again associated with the same causes as for G01 (see discussion in Section~\ref{section41}). At 19:07 onwards, PLS exhibits one faint band above $E_{\perp,\text{O}_2^+}$ and another one starting at 10~eV due to O$_2^+$ and H$_2^+$ respectively, as previously demonstrated by \citet{Carnielli2019}.

Fig.~\ref{Fig7}, bottom left panel, shows the simulated and PLS ion energy spectra during G07. The sharp transition, as observed for G01, is not apparent at the inbound MP crossing for G07 in both the simulated and PLS spectra. There is, however, a clear drop in energy. From 07:03 to 07:15, the PLS ion energy spectrum exhibits only one band between 30 and 100~eV, broader before CA. Our simulation predicts that the H$_2^+$ ion energy flux peaks at energies too low to be reliable within PLS data from 07:07 to 07:25 (horizontal dashed white line). While the main signature up to 07:15 is associated with O$_2^+$ above 15~eV, our simulations show that the flux should peak at lower and lower energy, from 200~eV down to 20~eV, while PLS ion flux does not go so low and stops around 30~eV at 07:08. $||\vec{\varv}_{\vec{E}\times\vec{B}}||$ is not minimal at closest approach, only later on at 07:15 (see Fig.~\ref{Fig6}). The peak in the PLS ion energy flux seems to drift towards higher energies from 07:10 to 07:15. This might be again evidence for field-aligned acceleration. From 07:15 to 07:26, there is no ion signature in the PLS data, which is not understood. From 07:27, ion flux is detected by PLS around 200~eV and is likely associated with O$_2^{+}$, O$^{+}$, HO$^{+}$, and H$_2$O$^{+}$ ion species. Another band associated with H$_2^+$ appears later on at 07:30. Both signatures and their energy range are consistent with our simulations.

Fig.~\ref{Fig7}, bottom right panel, shows the simulated and PLS ion energy spectra during G29. Before the inbound magnetopause crossing around 08:17-08:18, PLS detected a larger ion flux around 2~keV than during other flybys. It would be consistent with O$^+$ moving at $\sim$140~km\,s$^{-1}$. At the magnetopause, neither our modelled ion energy flux nor PLS data are marked by a strong transition like for G01 and G02. Between the magnetopause (crossed at 08:17) and the closest approach (08:25), one band is visible in the PLS data going from 300~eV down to 30~eV while our simulation shows two distinct ones, one from H$_2^+$ between 10 and 50~eV and another, fainter, from 500~eV down to 70~eV at closest approach due to O$_2^+$. Around 08:24, there is a clear split into two energy bands in the PLS data, with one at 20~eV and the other at 80~eV. While the model captures this separation, associating the lowest energy to H$_2^+$, there is a difference between observed and simulated energies, the difference being more significant for H$_2^+$ (see Section~\ref{section41}). It may affect O$_2^+$ as well, however less pronounced owing to the logscale and O$_2^+$ having already a much larger perpendicular kinetic energy. From 08:30 to 08:40, both energy bands in PLS data move towards higher energy and become more and more consistent with those simulated. From 08:40 to the inbound magnetopause crossing, they become weaker, as in our simulation. After the outbound magnetopause crossing, a large ion flux between 1~keV and 2~keV is observed like before the inbound crossing, corresponding to thermal Jovian ions (cf. Fig.~\ref{Fig5}).
\begin{figure*}
\fcolorbox{G01}{white}{\parbox{\columnwidth}{
\centering
\includegraphics[width=\linewidth]{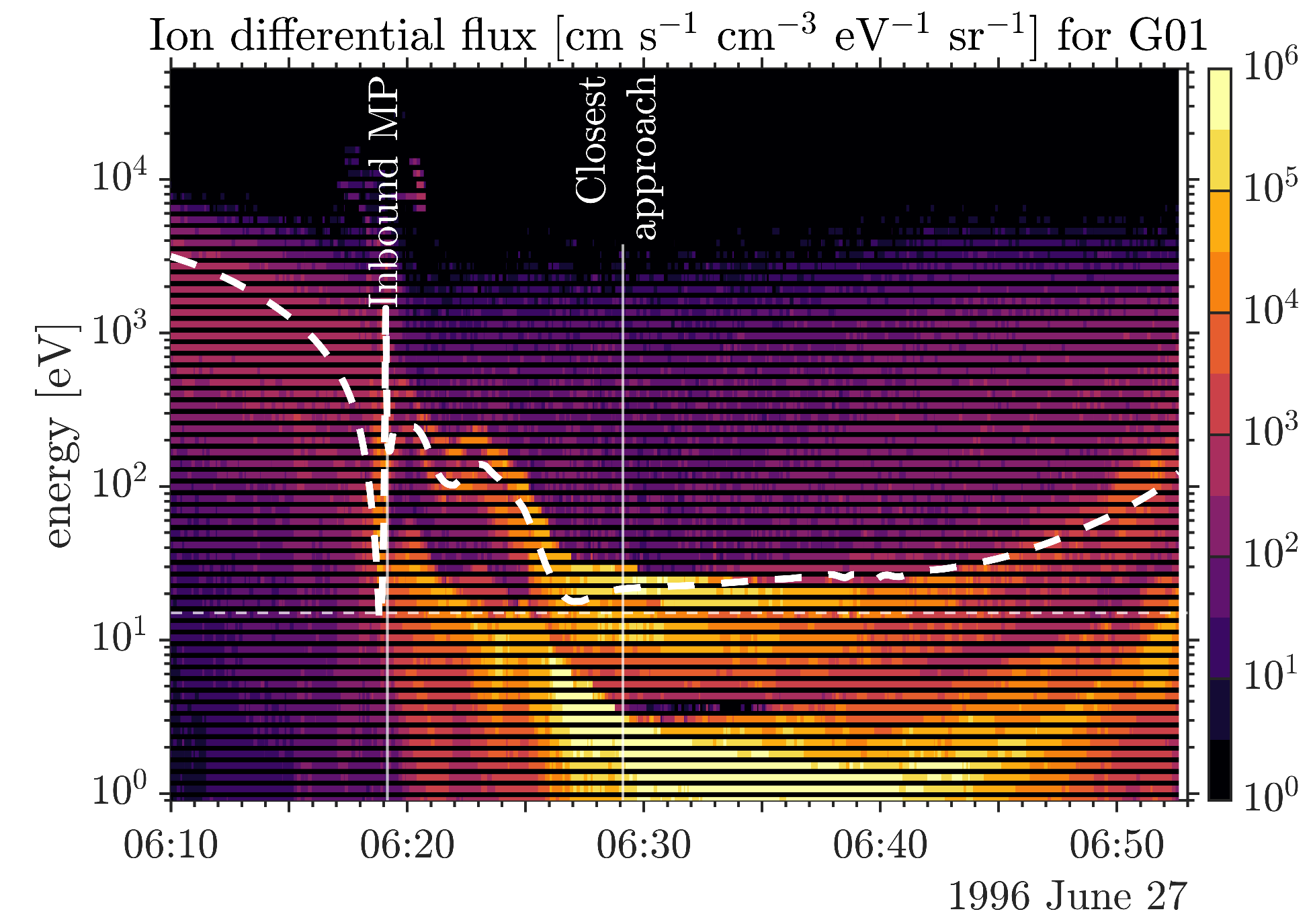}\\
\includegraphics[width=\linewidth]{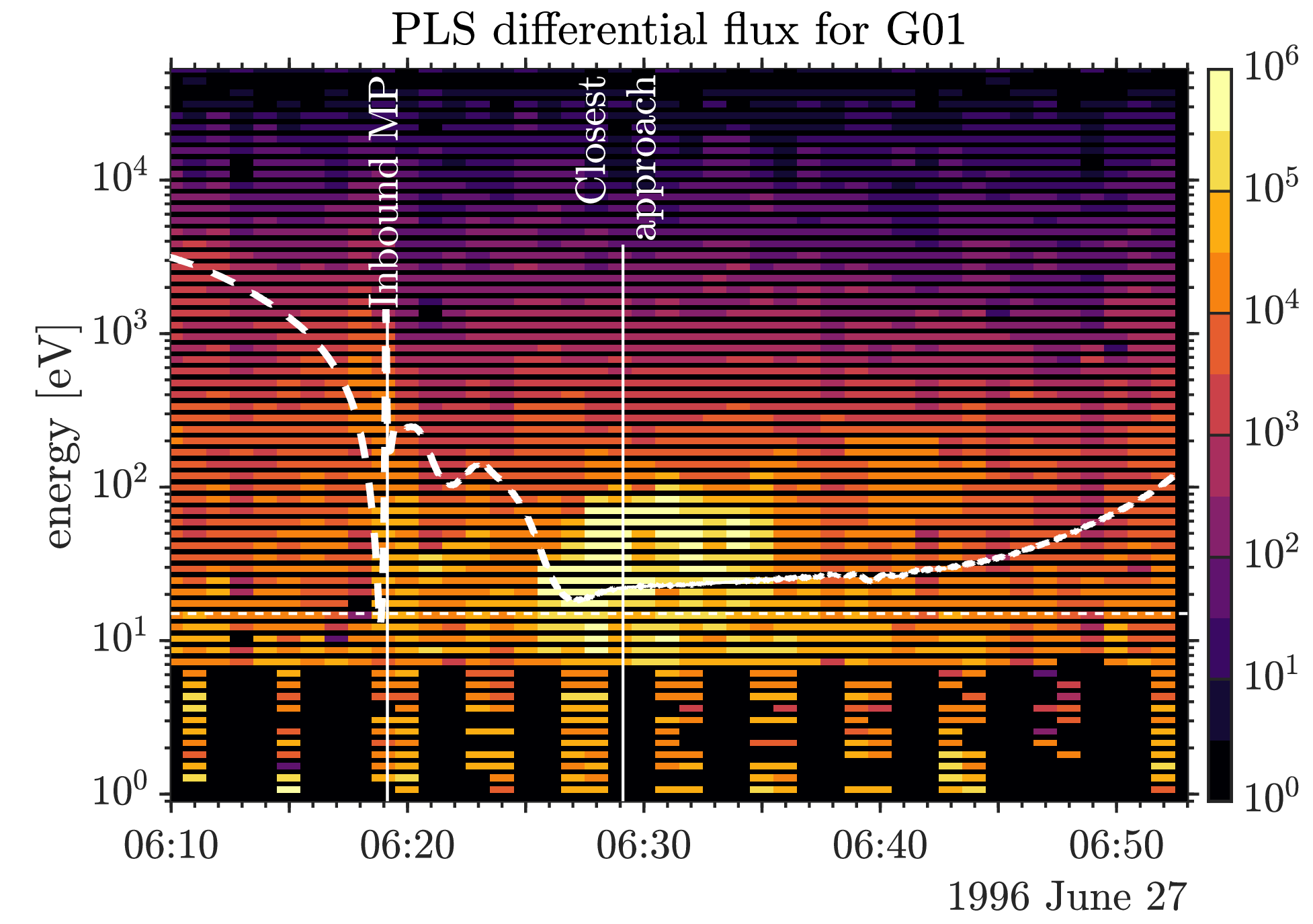}}}\!
\fcolorbox{G02}{white}{\parbox{\columnwidth}{
\centering
\includegraphics[width=\linewidth]{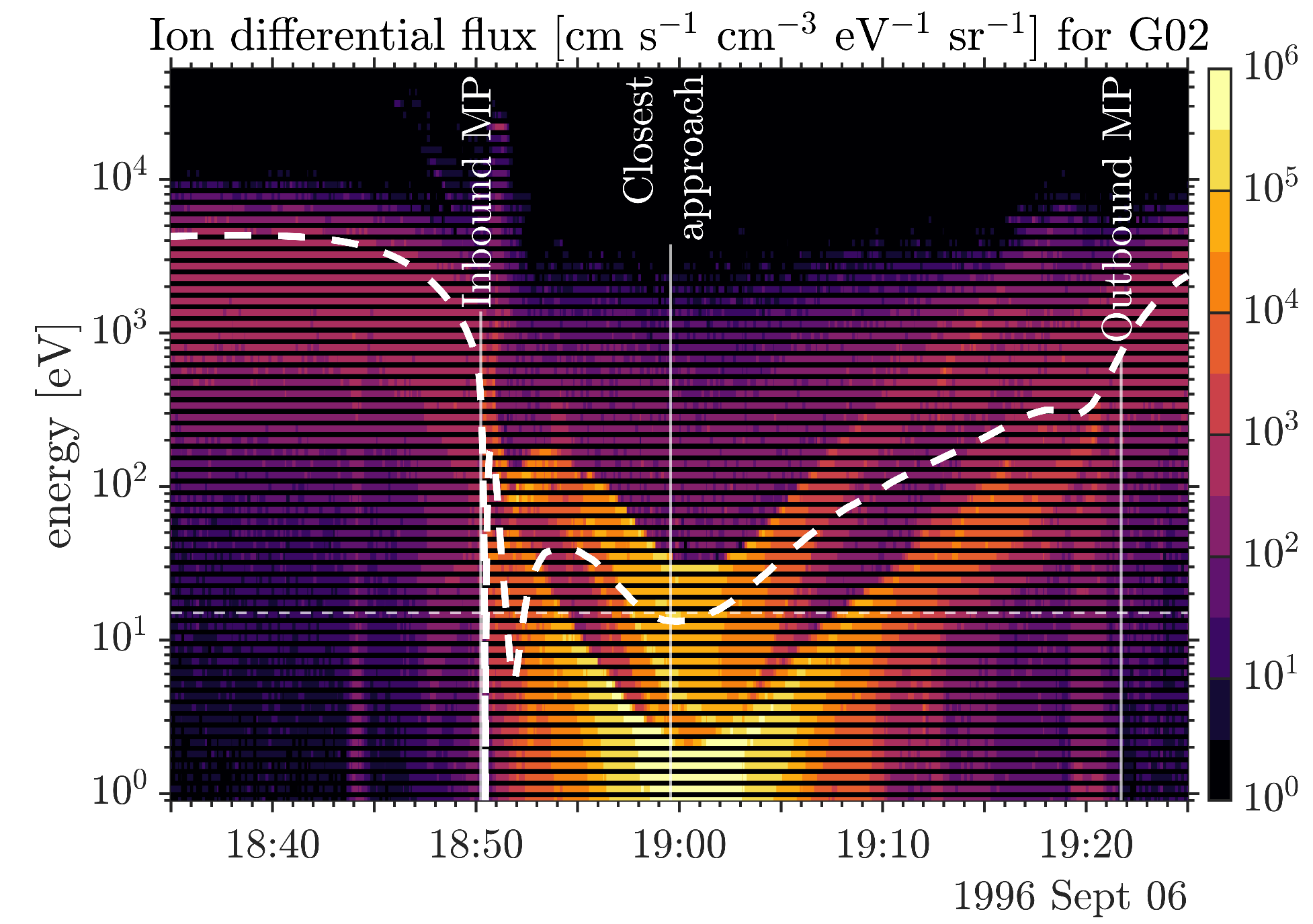}\\
\includegraphics[width=\linewidth]{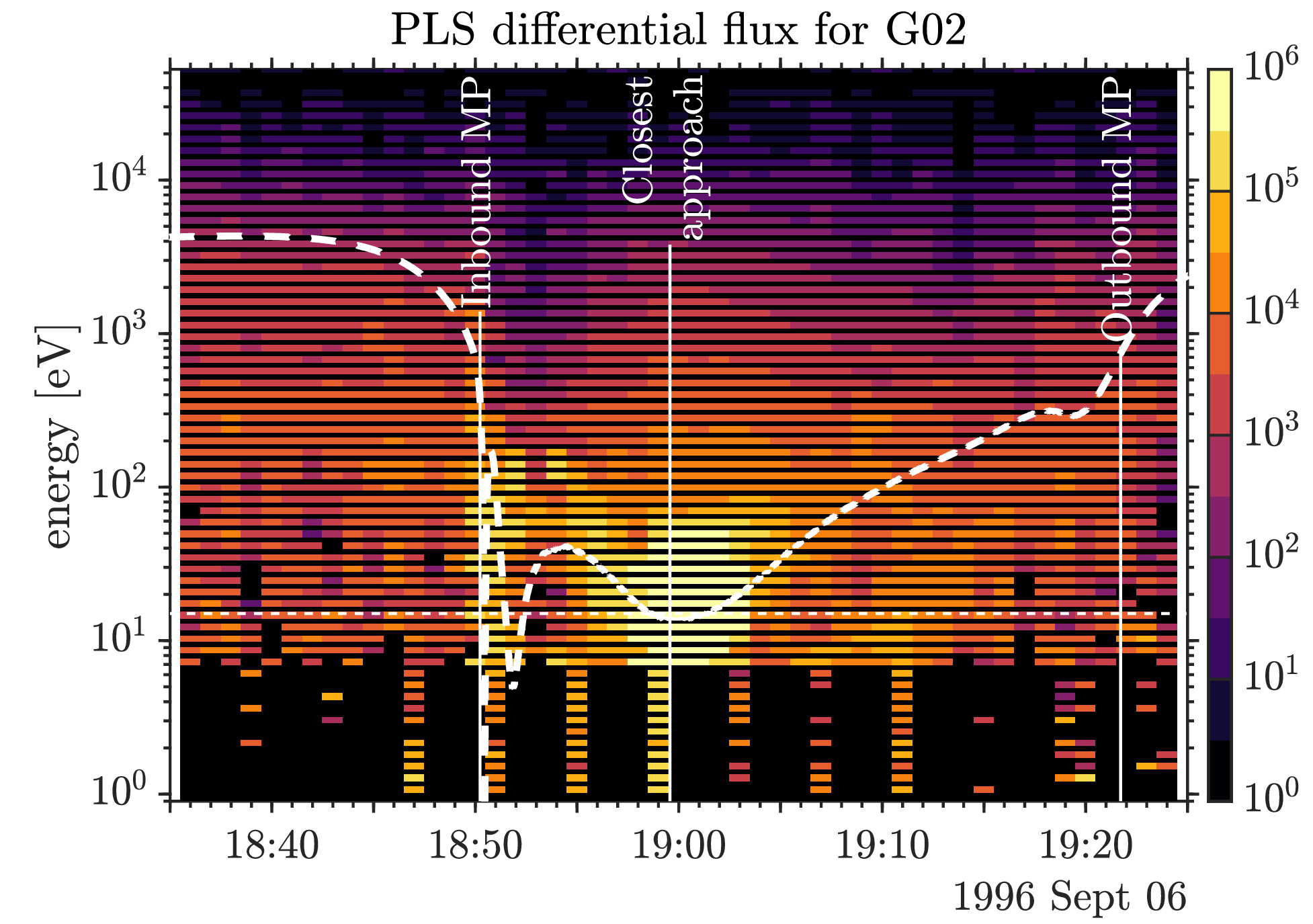}}}\\
\caption{Within the Alvén wings: same as Fig\,\ref{Fig5} but for the simulated ion energy spectra (top panel of each frame) vs PLS ion energy spectra (bottom panel of each panel). The curved dashed line represents the kinetic energy of O$_2^+$ ion drifting at $\vec{\varv}_{\vec{E}\times \vec{B}}$ in the spacecraft frame (cf. Eq.~\ref{Ek}). \label{Fig7}}
\end{figure*}
\begin{figure*}
\centering
\ContinuedFloat
\fcolorbox{G07}{white}{\parbox{\columnwidth}{
\centering
\includegraphics[width=\linewidth]{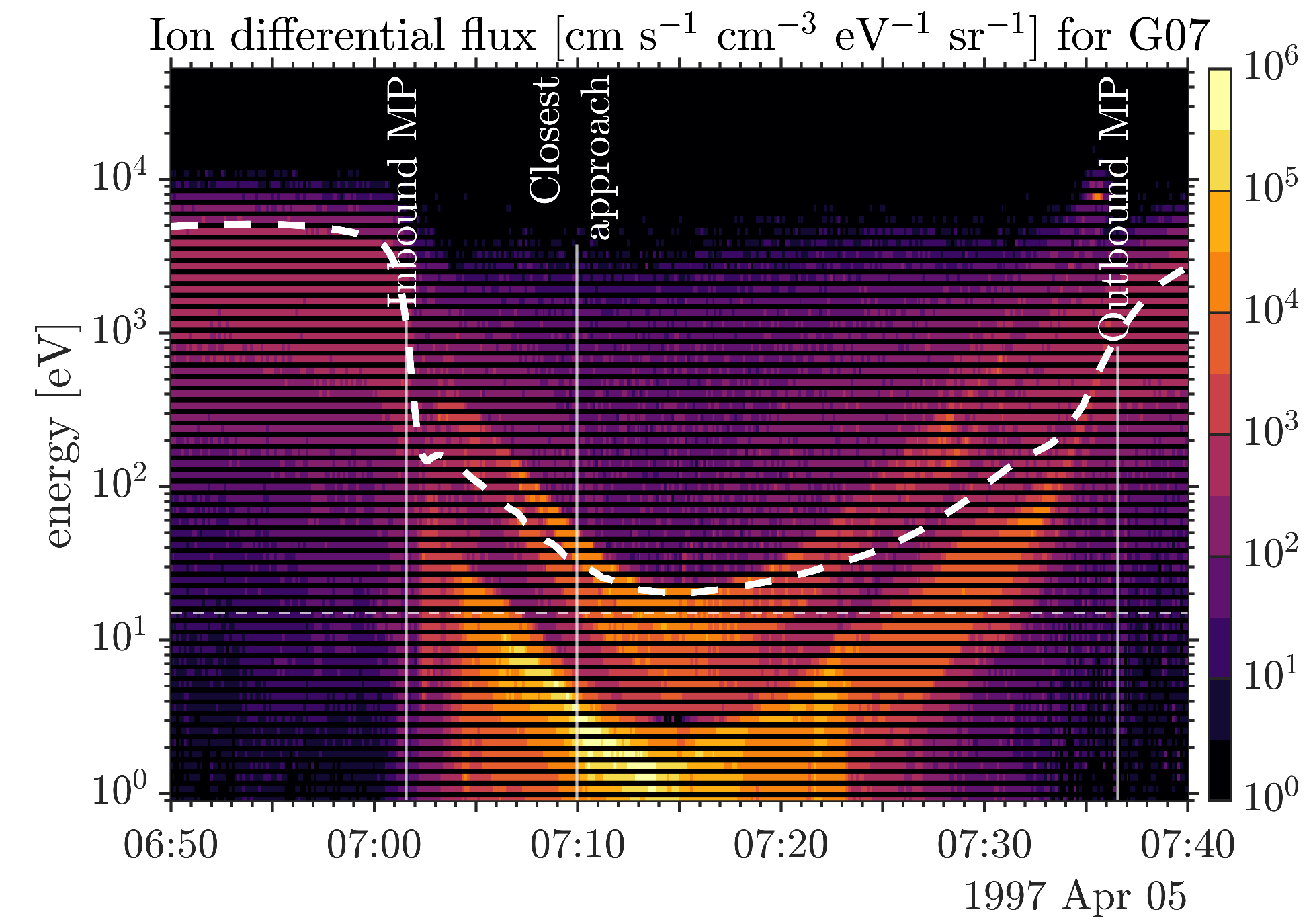}\\
\includegraphics[width=\linewidth]{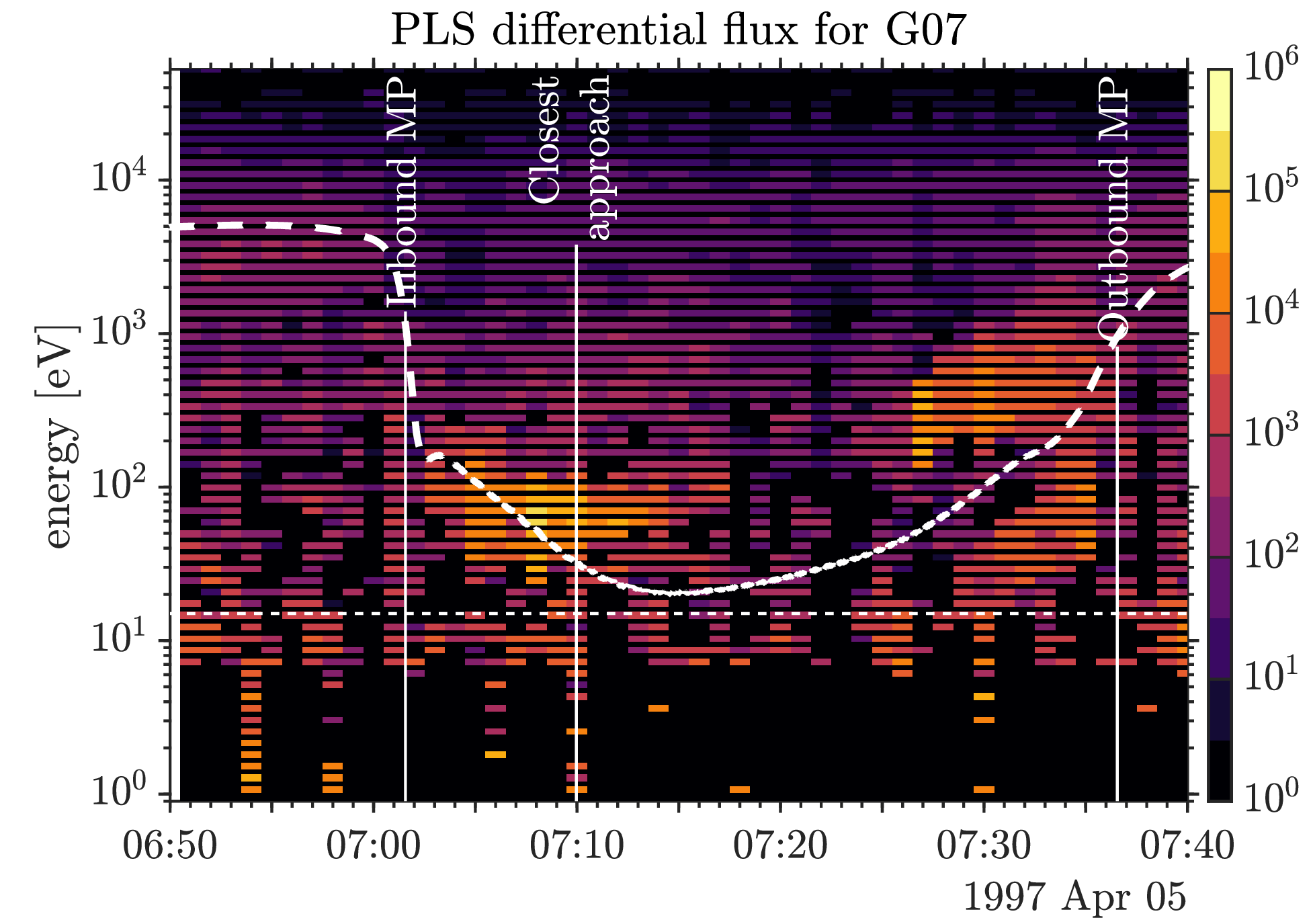}}}\!
\fcolorbox{G29}{white}{\parbox{\columnwidth}{
\includegraphics[width=\linewidth]{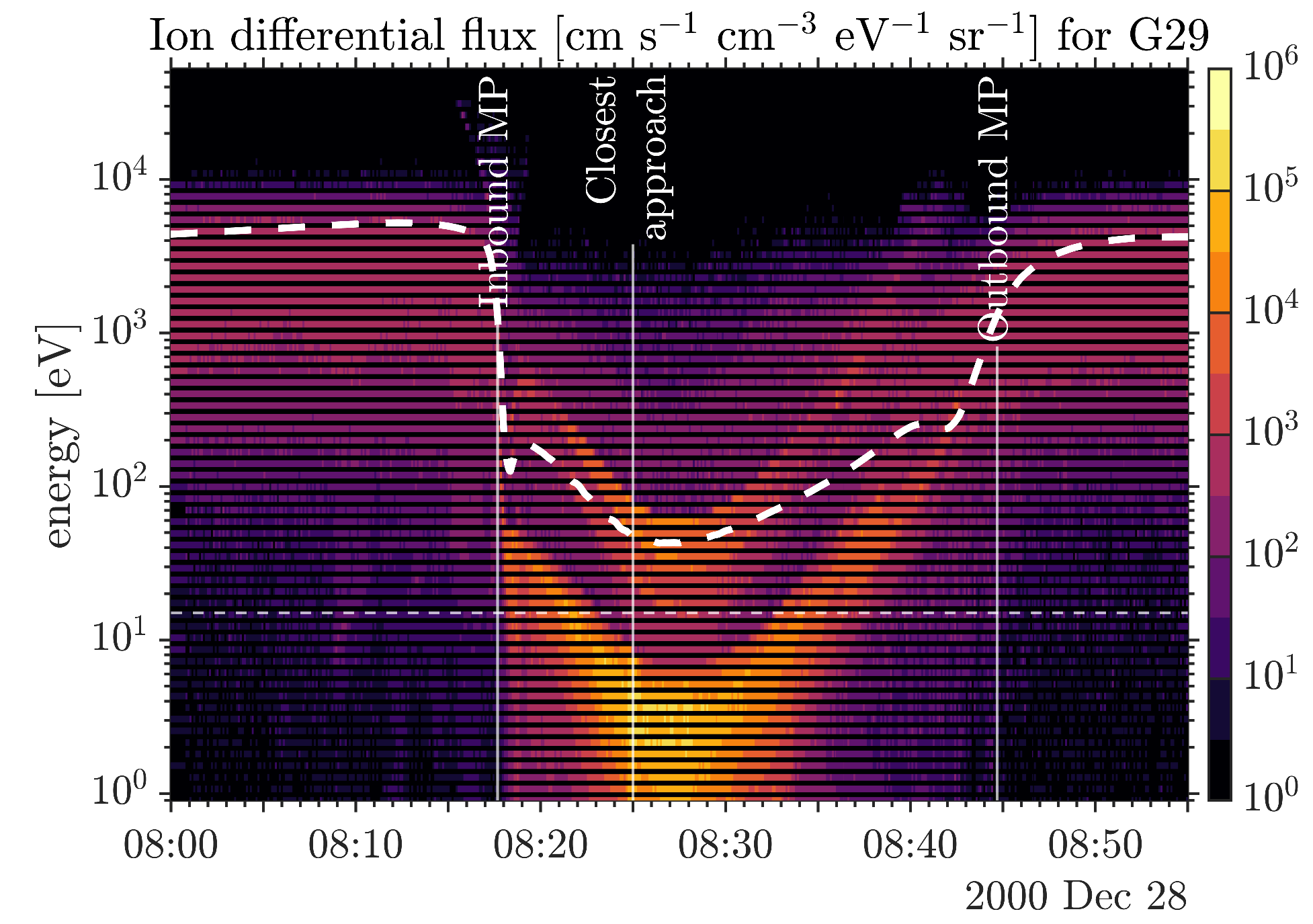}\\
\includegraphics[width=\linewidth]{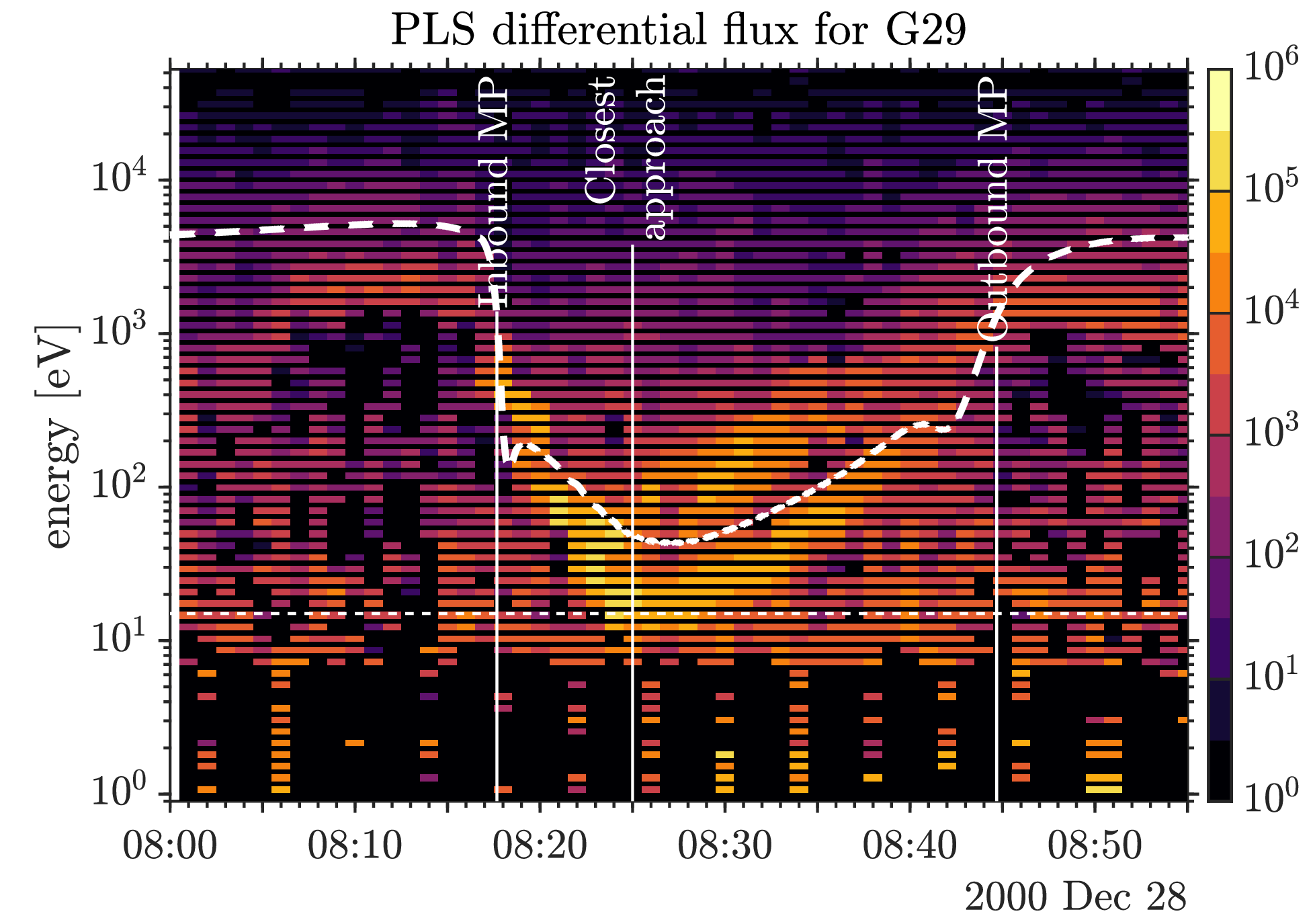}}}\\
\contcaption{\label{Fig7cont}}
\end{figure*}

\subsection{Within the closed field line regions: G08, G28\label{subsection32}} 

\begin{figure*}
\centering
\textcolor{G08}{\textbf{G08}} \textcolor{G28}{\textbf{G28}}\\
\stackinset{c}{-40pt}{c}{-40pt}{\Huge $\bigotimes$}{%
    \includegraphics[width=.65\columnwidth,clip,trim=1cm 1cm 1cm 1cm]{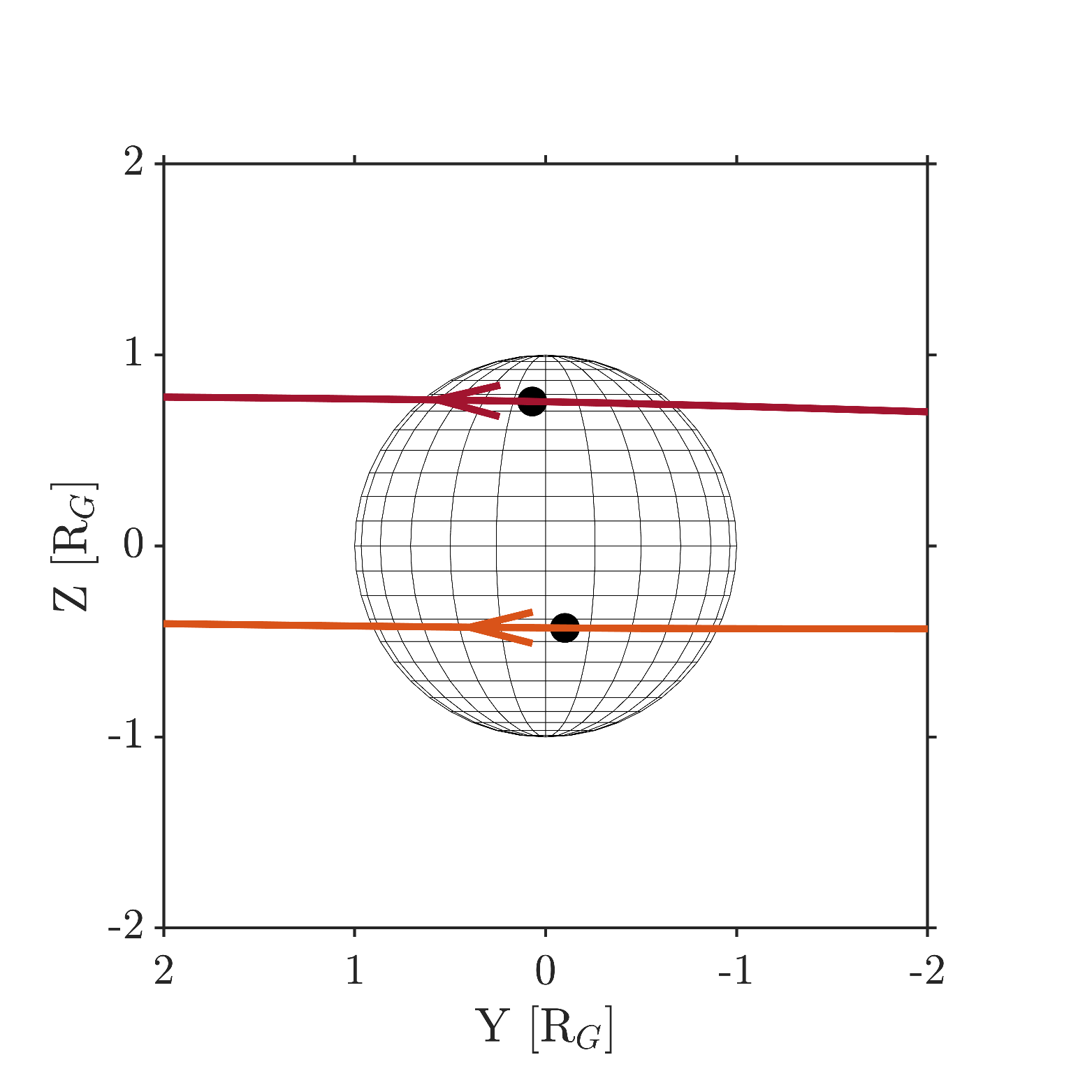}}
    \stackinset{c}{-45pt}{c}{0pt}{\Huge $\rightarrow$}{%
    \includegraphics[width=.65\columnwidth,clip,trim=1cm 1cm 1cm 1cm]{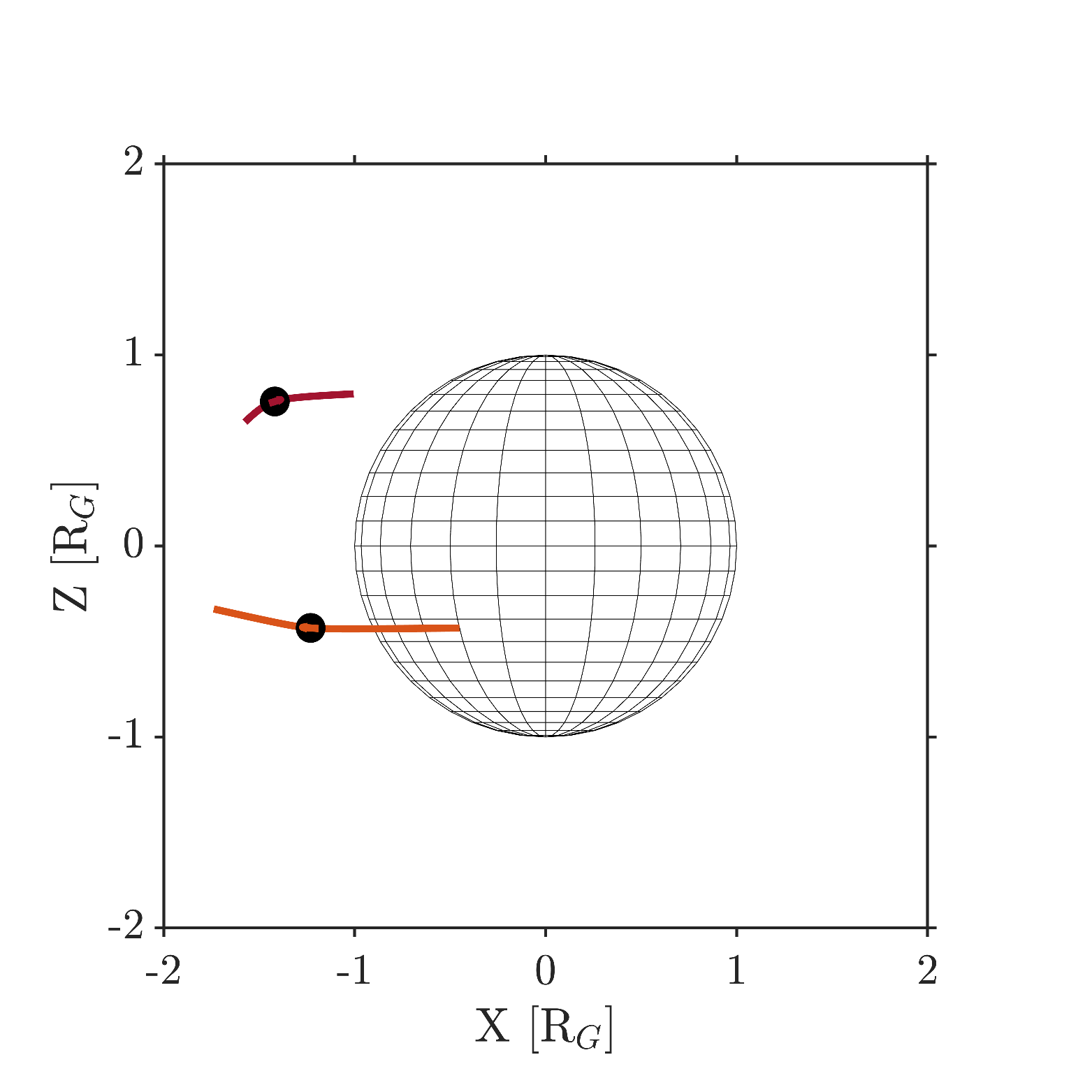}}
    \stackinset{c}{0pt}{c}{45pt}{\centering \Huge$\downarrow$ \ \ \ \ \ \ \ \ $\downarrow$}{%
    \includegraphics[width=.65\columnwidth,clip,trim=1cm 1cm 1cm 1cm]{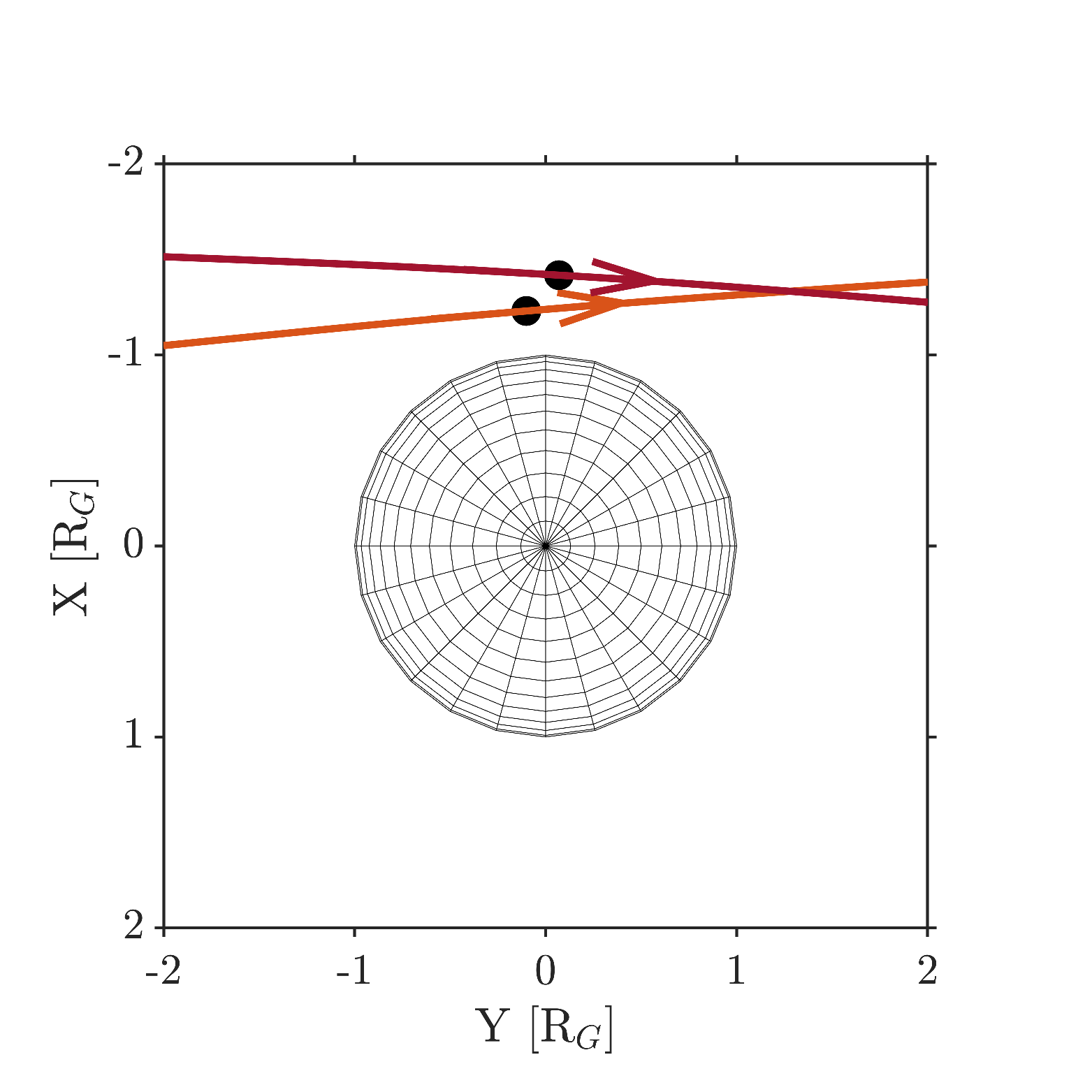}}
\caption{Trajectories of G08 and G28 flybys in GPhiO. The thicker coloured arrows indicate Galileo's direction during these flybys (the arrow is the projection of the velocity vector and is barely visible on the XZ cut as the motion was mainly along the $y$ axis). The black dot along the trajectory corresponds to the closest approach. The black arrows indicate the direction of the Jovian plasma flow.} \label{Fig8}
\end{figure*}

In this section, we present G08 and G28. Unlike the other flybys analysed, they crossed through the closed field region. G08 occurred when Ganymede was within the plasma sheet while G28 happened when the moon was above it. A visualisation of both trajectories is shown in Fig.~\ref{Fig8}.

\subsubsection{Plasma density and composition}\label{section321}

\begin{figure*}
\fcolorbox{G08}{white}{\parbox{\columnwidth}{
\centering
\includegraphics[width=\linewidth]{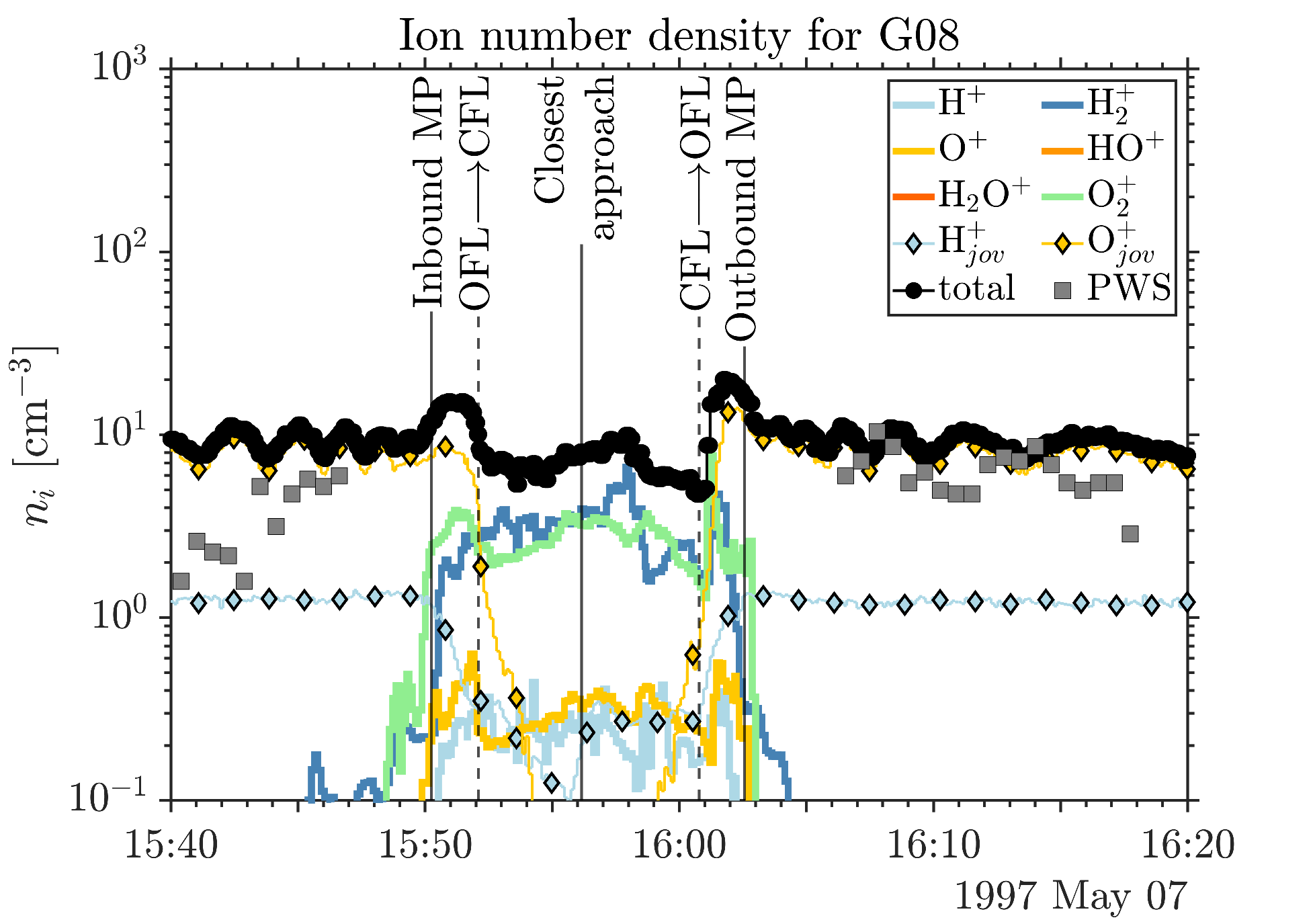}\\
\includegraphics[width=\linewidth]{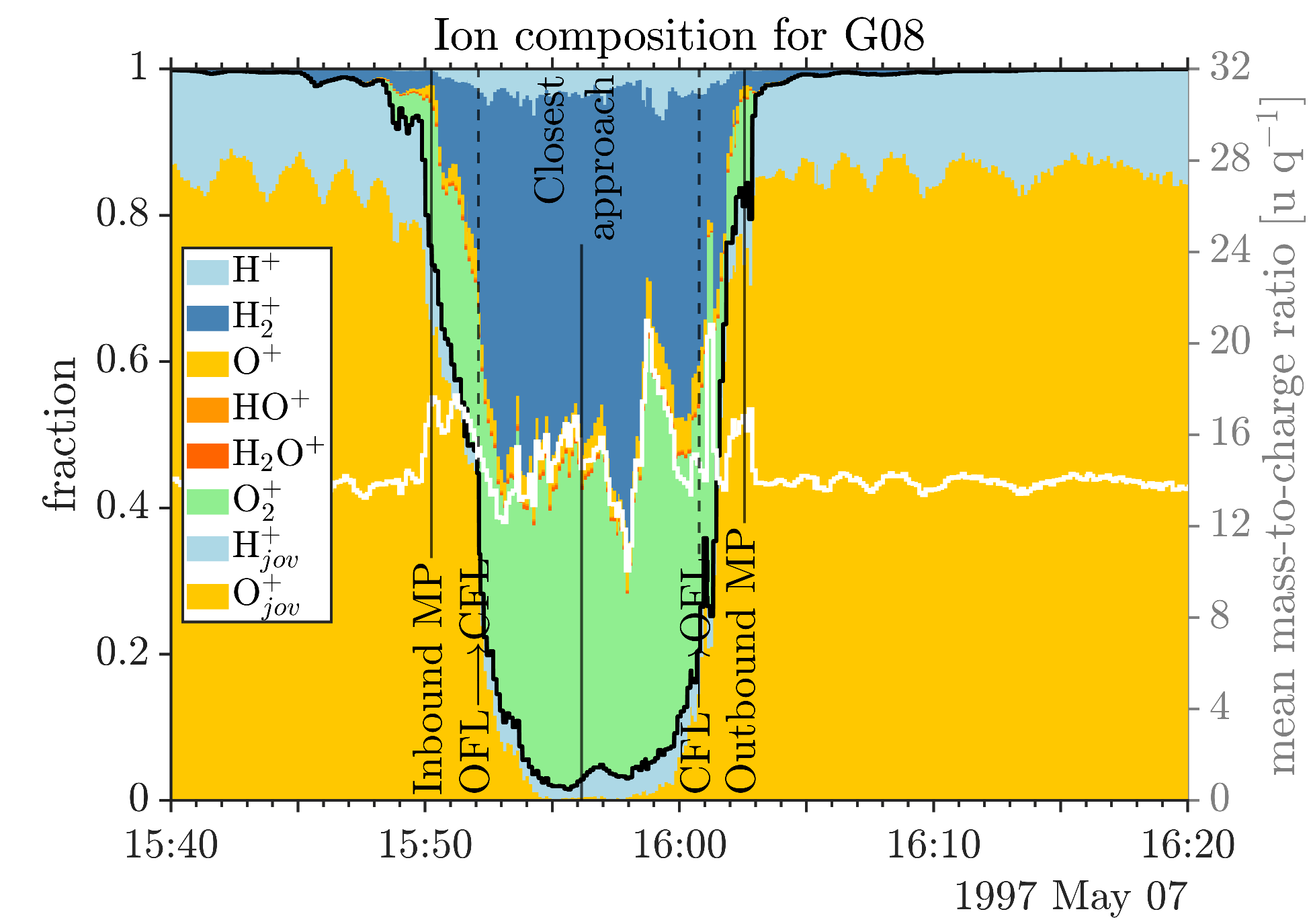}}}\!
\fcolorbox{G28}{white}{\parbox{\columnwidth}{
\centering
\includegraphics[width=\linewidth]{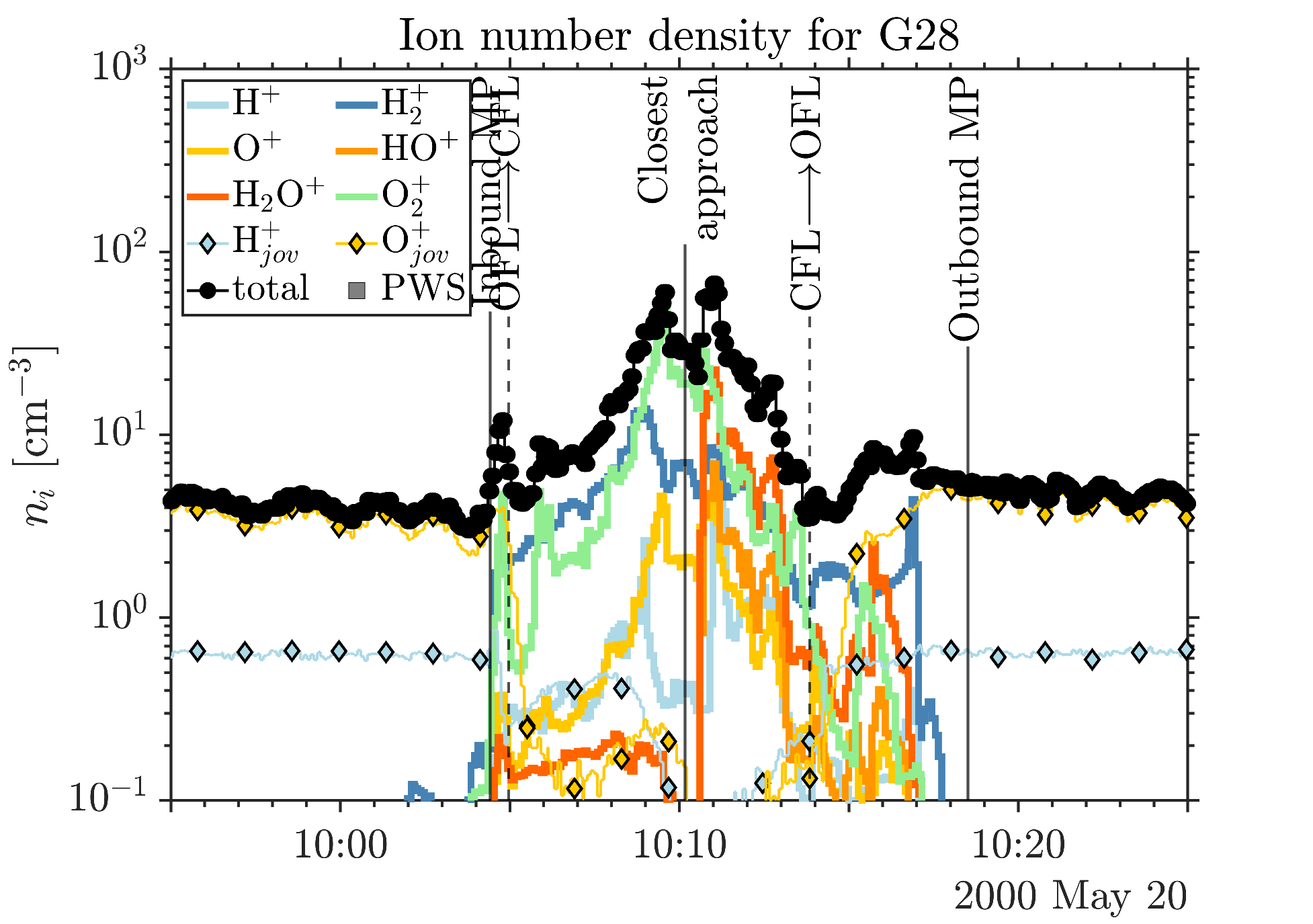}\\
\includegraphics[width=\linewidth]{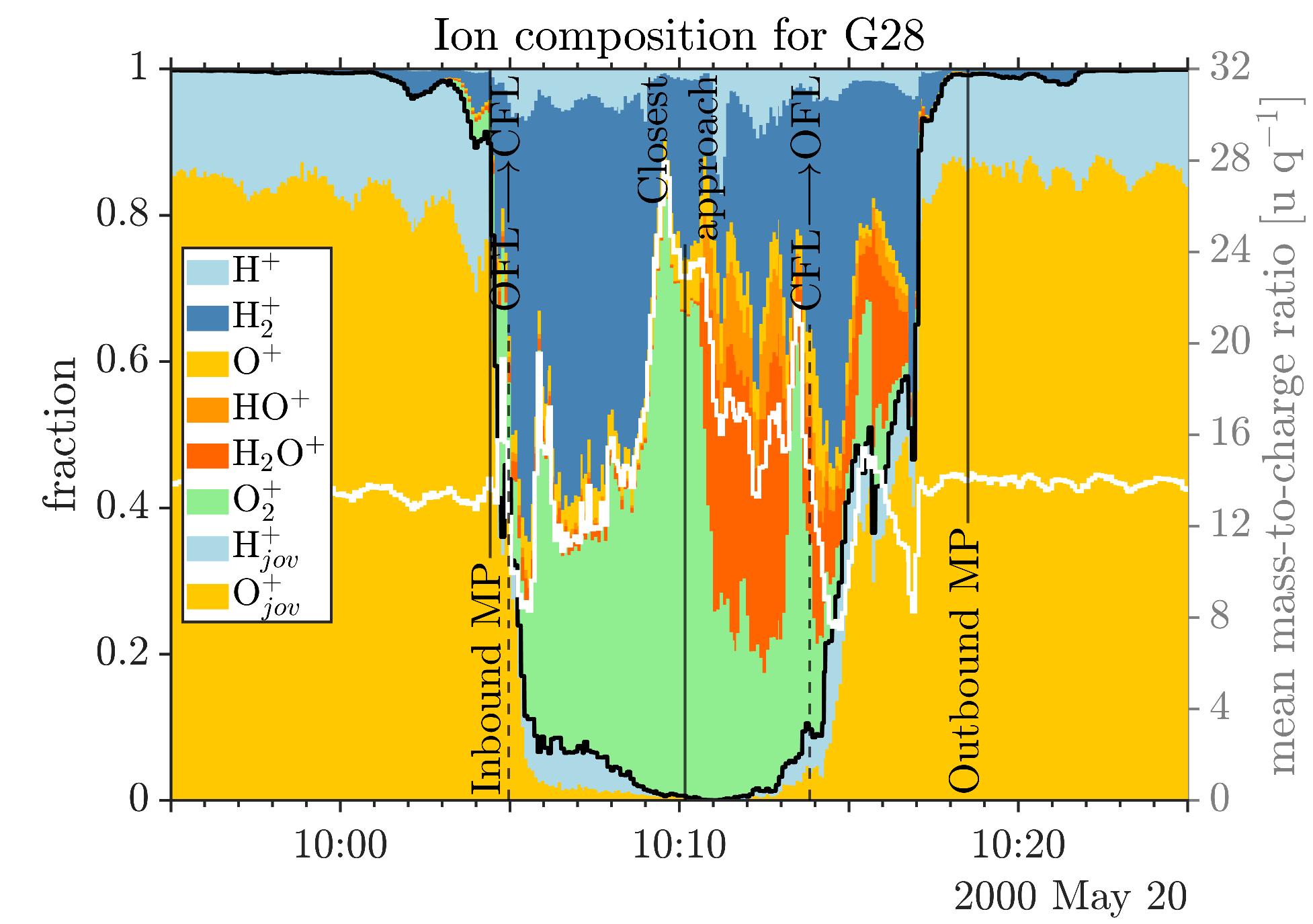}}}
\caption{Within the CFL region: same as Fig.~\ref{Fig5} but for G08 (left panel) and G28 (right panel). The vertical dashed lines correspond to the transition between the open and closed field line regions. \label{Fig9}}
\end{figure*}

Fig.~\ref{Fig9} shows the ion plasma densities and composition from our test particle simulations for G08 (left panel) and G28 (right panel) flybys. PWS electron number density could not be derived from spectra during the G28 flyby and is only available within the Jovian magnetosphere for G08. During G08, the PWS electron number density is ranging from 2~cm$^{-3}$ to 6~cm$^{-3}$ before the inbound MP crossing and from 3~cm$^{-3}$ to 10~cm$^{-3}$ after the outbound MP crossing. Therefore, the plasma conditions, within the plasma sheet, which are different during each crossing, varied during the flyby, unlike what was assumed for the magnetospheric and ionospheric simulations(see Table~\ref{table1}).

For G08 (Fig.~\ref{Fig9}, left panels), the simulated inbound MP crossing is characterised by a net increase in ionospheric ions, dominated by O$_2^+$ and H$_2^+$. The absence of H$_2$O$^+$ results from Galileo flying through Ganymede's magnetosphere while within Ganymede's shadow (Fig.~\ref{Fig2}): the dayside coincides more with the leading hemisphere whereas Galileo flew by the trailing hemisphere. There is a drop in Jovian ions number density, more marked for O$^+$ (factor 4 lower, slightly delayed) than for H$^+$ (factor 2-3). As the ionospheric O$_2^+$ and H$_2^+$ number densities (both around 3-4~cm$^{-3}$) are of the same order in Ganymede's magnetosphere as that of Jovian ions outside, it may be again important to properly adjust/constrain the thermal Jovian ion number density background flow. While the simulated total ion number density remains relatively flat within Ganymede's magnetosphere, it is discontinuous around 16:01 when Galileo went from the closed magnetic field line region to the open field line region (second dashed vertical line, CFL$\longrightarrow$OFL). 
Although it is not perfectly synchronised, the ion number density of each ion species exhibits a discontinuity near the CFL$\longrightarrow$OFL crossing ($\sim$30~s before for the Jovian ions,  $\sim$1~min after for the ionospheric ions).
The feature looks purely numerical: it is very weakly present in the Jovian ion number density (mainly O$^+$) and strongly in ionospheric ion number densities (mainly H$_2^+$, O$_2^+$, and O$^+$). Therefore, this feature is unlikely associated with the MHD electromagnetic fields, as it does not appear symmetrically for the OFL$\longrightarrow$CFL crossing (only for ionospheric O$^+$ though), or perhaps only within Ganymede's magnetosphere at this particular boundary crossing where the MHD simulation spatial resolution may be too limited. 

G28 is more challenging to interpret as we lack observations. At the inbound MP crossing, there is a localised peak at 10~cm$^{-3}$ (around 10:05): it is pretty thin and short, as Galileo barely spent time within the open field line region. Once inside the closed field region, the ion number density increases as Galileo approaches the moon, reaching $\sim$60~cm$^{-3}$, dominated by O$_2^+$, H$_2^+$, and Jovian O$^+$. After CA, there is a clear addition of H$_2$O$^+$ to the ion composition: It does not drastically affect the ion number density but rather its slope, which is less steep than inbound. During G28, Galileo went from the nightside to the dayside of Ganymede (see Fig.~\ref{Fig2}), which yields the H$_2$O$^+$ asymmetry seen in the ion composition. Around 10:15, $n_i$ weakly increases, only seen in H$_2$O$^+$ and O$_2^+$ number densities. It could be caused by either the exosphere (and the noise from the DSMC model) or the MHD fields. No particular signature is associated with the outbound MP crossing seen in n$_i$ besides the change in ion composition.

\subsubsection{Ion velocity}\label{section322}

\begin{figure*}
\fcolorbox{G08}{white}{\parbox{\columnwidth}{
\centering
\includegraphics[width=\linewidth]{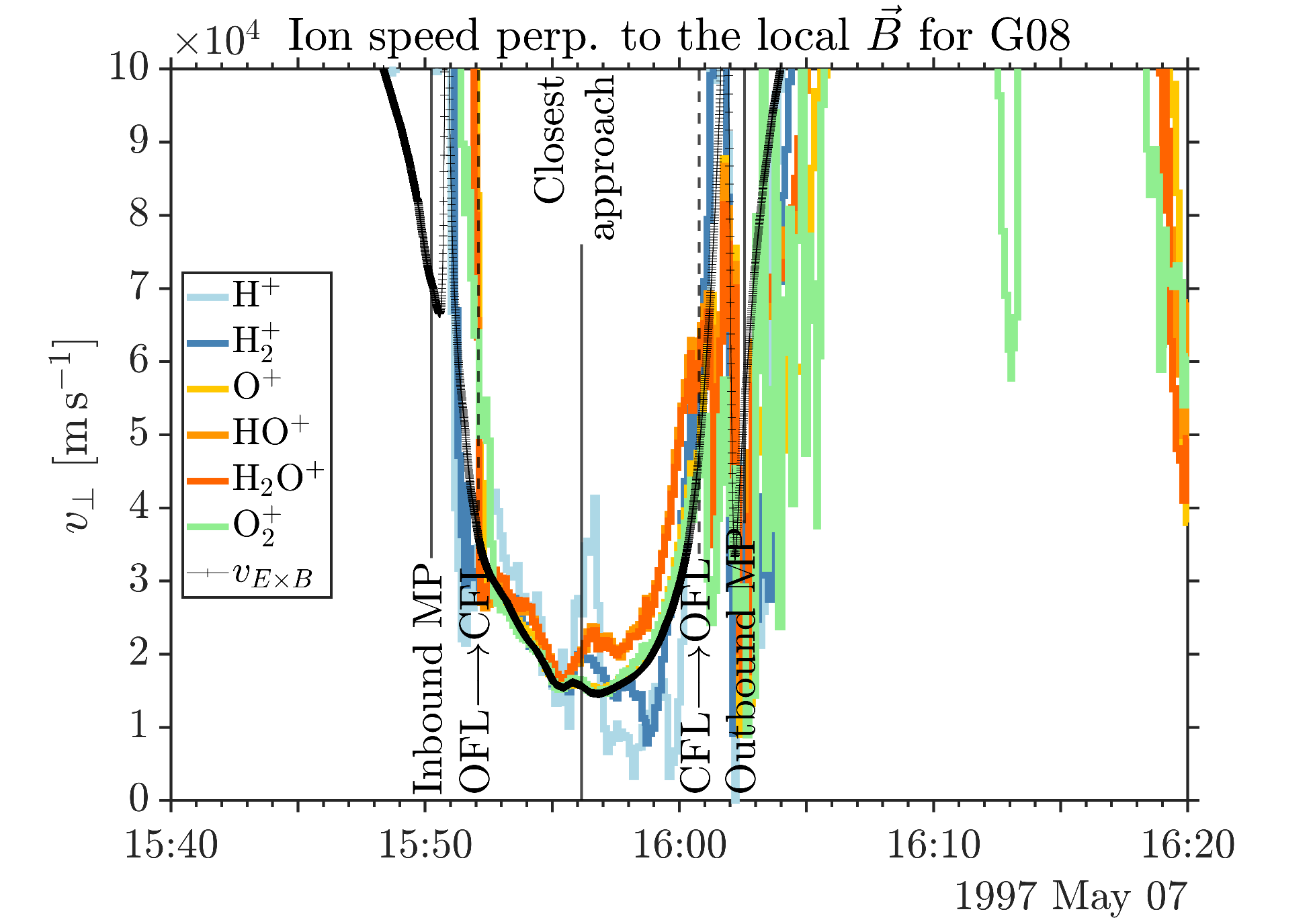}\\
\includegraphics[width=\linewidth]{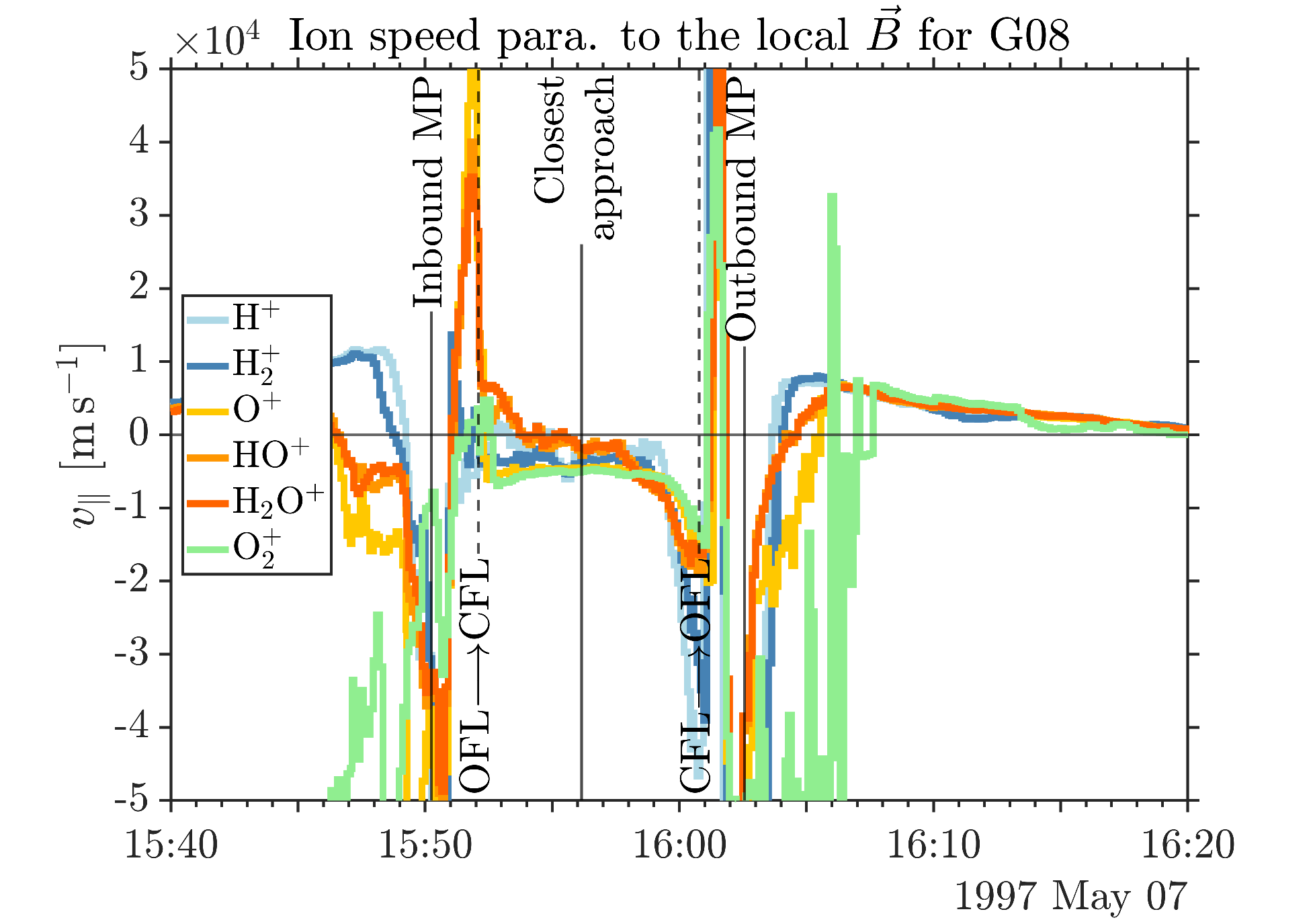}}}\!
\fcolorbox{G28}{white}{\parbox{\columnwidth}{\centering
\includegraphics[width=\linewidth]{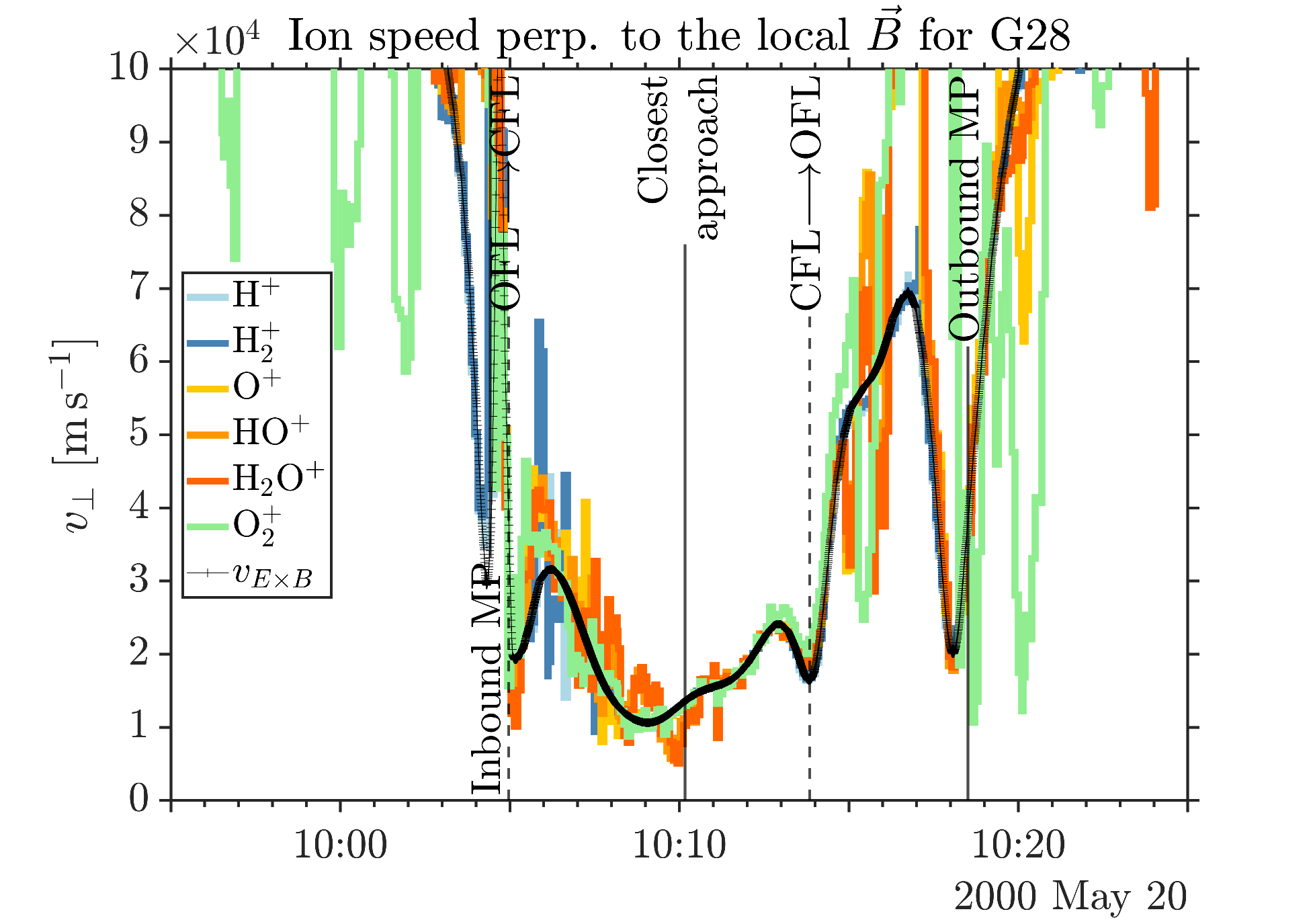}\\
\includegraphics[width=\linewidth]{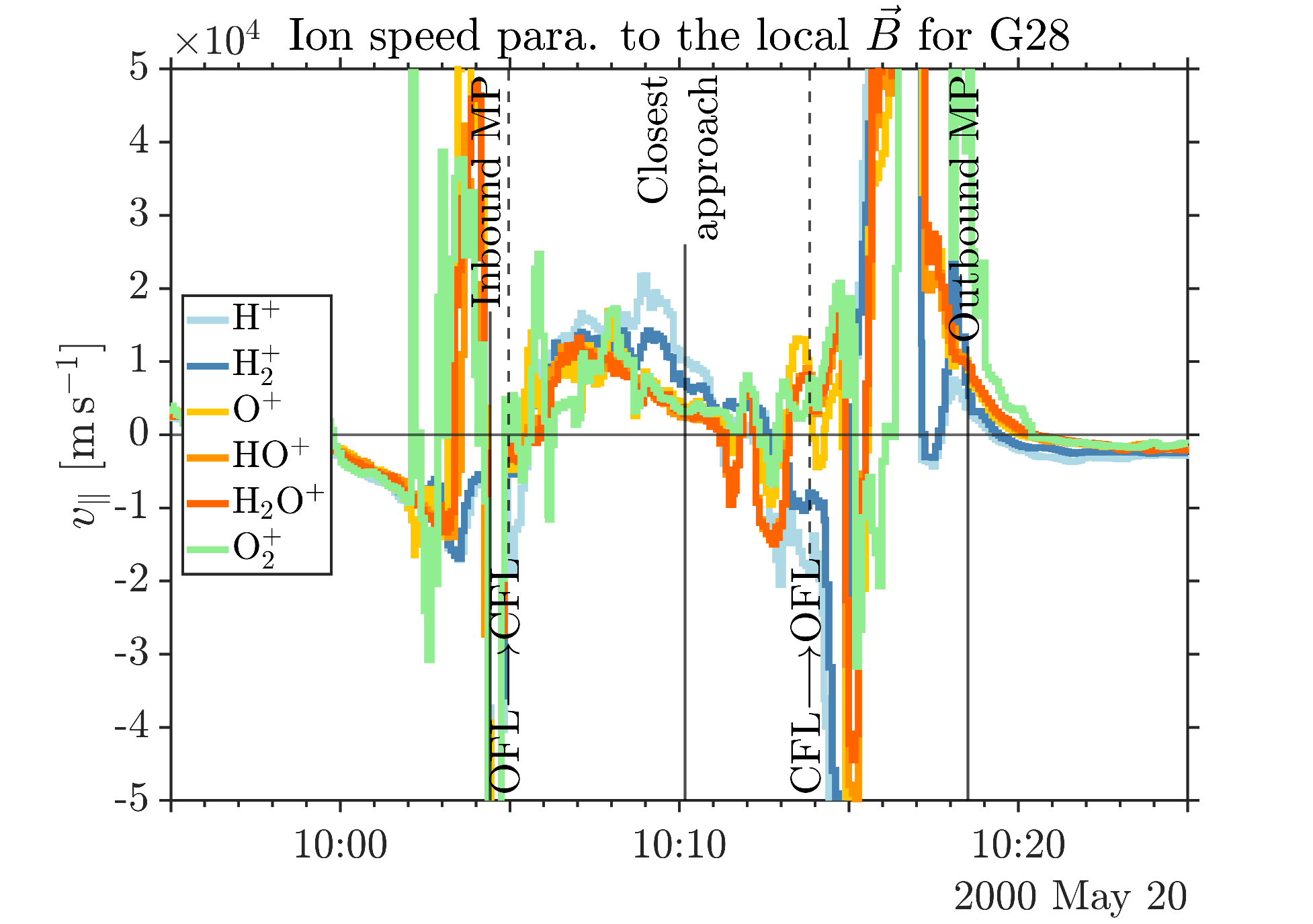}}}
\caption{Within the CFL region: same as Fig.~\ref{Fig6} but for G08 (left) and G28 (right panel) \label{Fig10}}
\end{figure*}

Fig.~\ref{Fig10} shows the modelled ion speeds perpendicular (top panel) and parallel (bottom panel) to the local magnetic field for G08 (left) and G28 (right). Unlike what was found within the Alfvén wings (cf. Fig.~\ref{Fig6}), the modelled speeds for each species appear here extremely variable and numerically noisy. Regarding the perpendicular speed, ions roughly follow $\vec{\varv}_{\vec{E}\times\vec{B}}$ between inbound and outbound MP crossings, but not as well as for G01, G02, G07, and G29 (Section~\ref{section312}). For the parallel speed, it seems to remain near $0$. Within the closed field line region, ion plasma flow is expected to remain in the $xy$ plane, moving perpendicular to the magnetic field, which points towards the $+z$ direction as it is dominated by Ganymede's dipole field. In this region, the curvature of the magnetic field is important and may play a significant role in the response of the different species to the local electromagnetic fields.

\subsubsection{Ion energy spectra}\label{section323}

\begin{figure*}
\fcolorbox{G08}{white}{\parbox{\columnwidth}{
\centering
\includegraphics[width=\linewidth]{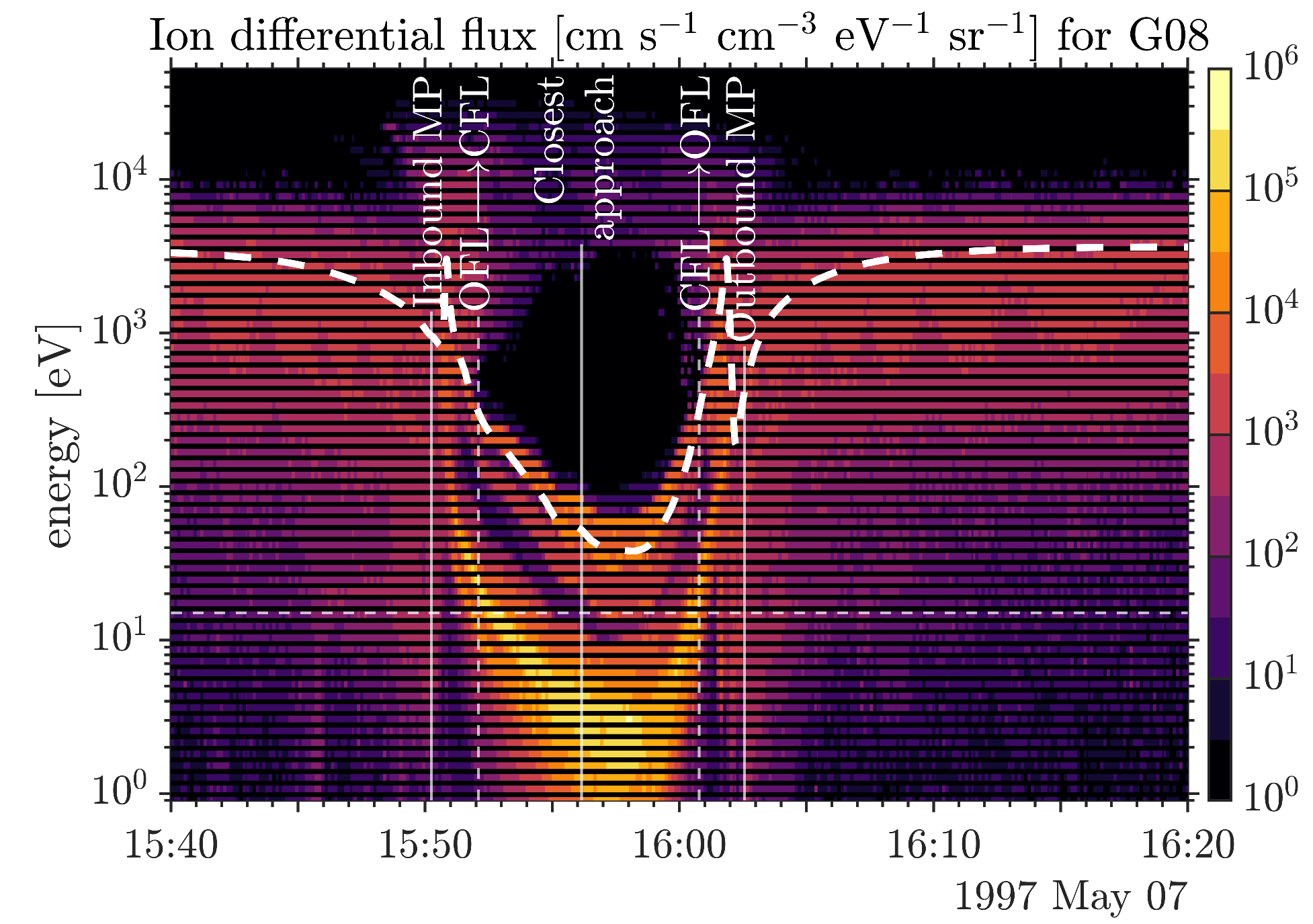}\\
\includegraphics[width=\linewidth]{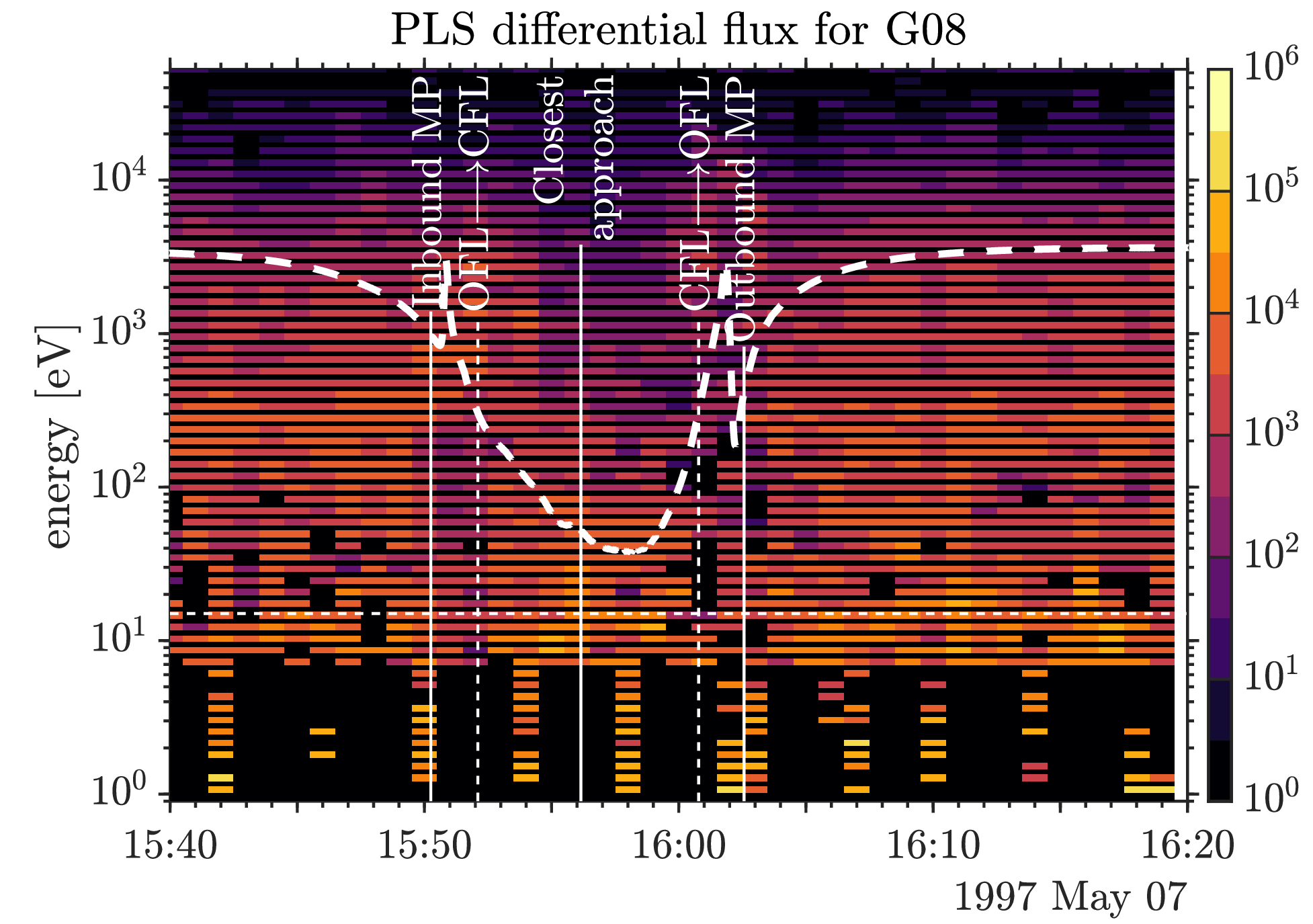}}}\!
\fcolorbox{G28}{white}{\parbox{\columnwidth}{
\centering
\includegraphics[width=\linewidth]{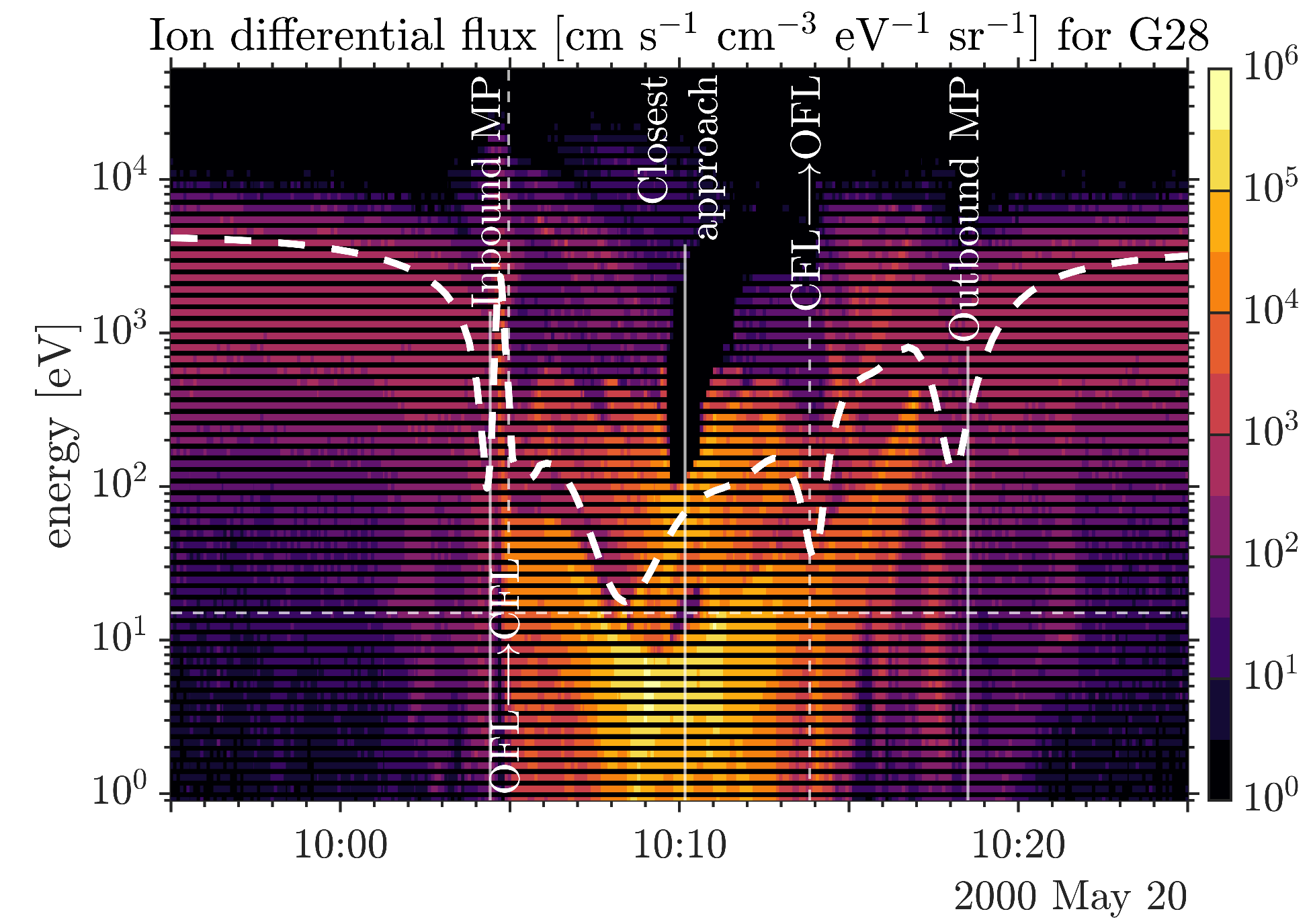}\\
\includegraphics[width=\linewidth]{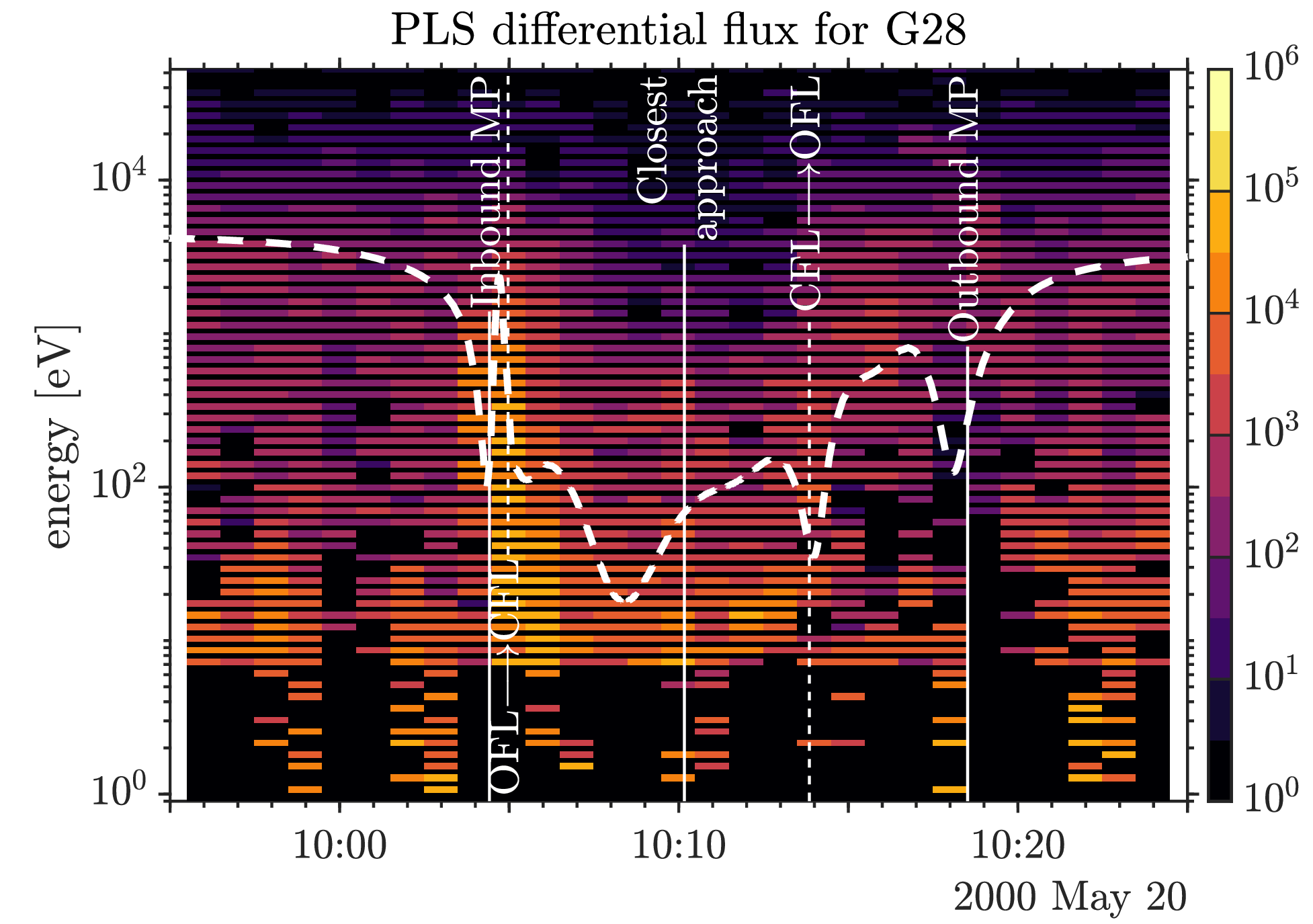}}}
\caption{Within the CFL region: same as Fig.~\ref{Fig7} for G08 (left) and G28 (right panel).\label{Fig11}}
\end{figure*}

Fig.~\ref{Fig11} shows the simulated (top row) and PLS (bottom row) ion energy spectra for G08 (left panel) and G28 (right panel). During G08, the ionospheric ion signature in the PLS data is not as strong, while our simulation predicts a clear one within the closed field line region between 20 and 200~eV. In addition, in the same region, PLS data shows a reduction in the ion flux above $\sim$200~eV coinciding with the absence of ions at energies between 80 and 3000~eV in our simulated spectrum. The lack of significant ion flux within the closed-field line region in PLS data, consistent with PWS (see Fig.~\ref{Fig9}), may indicate that we overestimate electron-impact ionisation frequency.

During G28, there are two main features in PLS data. Firstly, there is a large ion flux spreading from 10~eV to 3~keV during the inbound MP crossing at 10:05, though it is not as pronounced in our test particle simulation, being weaker in terms of flux and less spread in time. The second feature in the PLS data is a faint band between 10:10 and 10:14 starting from 7~eV and reaching 20~eV. Our simulation shows that this signature seems to be associated with H$_2^+$. PLS data did not observe ion fluxes at higher energy within the closed field line region suggesting a lack of O$_2^+$ which is observed in the simulation. So the neutral composition background may be different from what was assumed.

\section{Discussion}\label{section4}
\subsection{Within the Alfvén wings}\label{section41}

During G01, G02, G07, and G29, Galileo spacecraft gave us the opportunity to probe the close ionosphere of Ganymede within the Alfvén wings (see Table~\ref{table1}, Fig.~\ref{Fig1}). Electron number density could be mostly derived from PWS data within (part of) Ganymede's magnetosphere during these flybys. When compared with PWS observations, results from our test particle simulations show extremely good agreement in terms of plasma density $n_i$ for G01 and G07, underestimated by a factor of 2-5 for G29, and down to a factor of 10 for G02 (cf. Fig~\ref{Fig5}). Nevertheless, the trends and variations are consistent in the latter case. We have shown that the ion composition appears to be dominated by O$_2^+$ and H$_2^+$, with occasionally the contribution of H$_2$O$^+$ (excluding G29 flyby within Jupiter's shadow) to a lesser extent (cf. Fig.~\ref{Fig5}) depending on whether Galileo was on the dayside (cf. G01 and G02) or crossing the magnetopause on the anti-Jovian flank, where ions tend to escape from Ganymede's magnetosphere, on the nightside while Ganymede was still illuminated (cf. G07). The ion composition is consistent with PLS spectra exhibiting a parallel triple band structure (cf. Fig.~\ref{Fig7}, one associated with H$^+$/H$_2^+$, the second one with O$^+$/HO$^+$/H$_2$O$^+$, and the third one with O$_2^+$) with ions drifting at similar speeds but having different mass-to-charge ratios (resp. $m/z=1-2$, $m/z=16-18$, and $m/z=32$ u\,q$^{-1}$, which allows being discernible by PLS in terms of kinetic energy).


The energy at which ions are detected at Ganymede is often higher than predicted by the simulation, with larger discrepancies at CA (cf. Fig.~\ref{Fig7}). We suggest several causes, not mutually exclusive, that may help to reconcile the model and data in terms of plasma number density and ion energy spectra. Furthermore, the disagreement between G01 and G02 is very puzzling as both flybys have a very similar neutral and electromagnetic environment: similar local time (hence, the same neutral exosphere is used as input) and Ganymede above the plasma sheet (cf. Table~\ref{table1}). 

The first cause for discrepancies (at least in terms of ion number density) is the limited knowledge of the exospheric neutral densities. The modelled neutral densities from \citet{Leblanc2023} have been compared and constrained by remote-sensing observations of auroral emissions. The column number densities can only be estimated or retrieved by assuming an electron number density and temperature that is poorly constrained and unknown. In addition, not only the column number densities can be only derived on the dayside for Earth-based observations but also only those of H, O, O$_2$, H$_2$O, and O$_2$ have been reported \citep{Leblanc2023,Vorburger2024}. H$_2$ number density is less constrained and may be adjusted if needed (for example on the nightside during the outbound leg of G01). For adequate comparisons with the current results, future authors should use a more adequate analytical neutral density profile as shown in Appendix~\ref{AppExo}, especially for H$_2$. The recent discovery of a patchy CO$_2$ exosphere at Ganymede on the leading side of the North polar cap \citep{Bockelee2024} may be an additional source of ions, once ionised, to Ganymede's ionosphere. However, the aforementioned effects would not affect the shape of the simulated energy spectra, only their intensity, except if heavier ion species are included such as CO$_2^+$. Nevertheless, the signature in the ion energy spectrum would be predictable as we showed that it primarily depends on the mass of the considered ion species (the ion velocity being primarily $\sim\vec{\varv}_{\vec{E}\times\vec{B}}$, see Section~\ref{section312}).

The second cause we have identified is the spacecraft potential. Like any spacecraft, Galileo experienced charging of its surface to positive or negative values. This effect may, in general, be negligible as the relative speed between Galileo and the ion plasma flow was large. However, as seen at closest approach (see Fig.~\ref{Fig6}), the flyby speed of Galileo ($\sim 10$~km\,s$^{-1}$, see Table~\ref{table1}) is around the same value as the mean ion drift speed (see Fig.~\ref{Fig8}). In addition, the kinetic energy of ions in the spacecraft frame is of the order of ``typical'' spacecraft potential values or even lower \citep[for example, JUICE spacecraft potential within the solar wind has its potential modelled around 6\,V, with parts spanning from -36 to 8\,V,][]{Holmberg2024} meaning that low energy ions can be strongly deflected around the spacecraft prior reaching the PLS instrument. This is one of the reasons why there is a lack of confidence in PLS data around or below 15~eV, alongside the sparse data and the lack of consistency with the data at higher energies. Without the presence of a Langmuir probe, it is not possible to estimate Galileo spacecraft potential as it varied depending on the local plasma properties in the Jovian and Ganymede's magnetospheres and the illumination.

\begin{figure*}
\centering
\fcolorbox{G01}{white}{\parbox{\columnwidth}{\centering
\includegraphics[width=\linewidth]{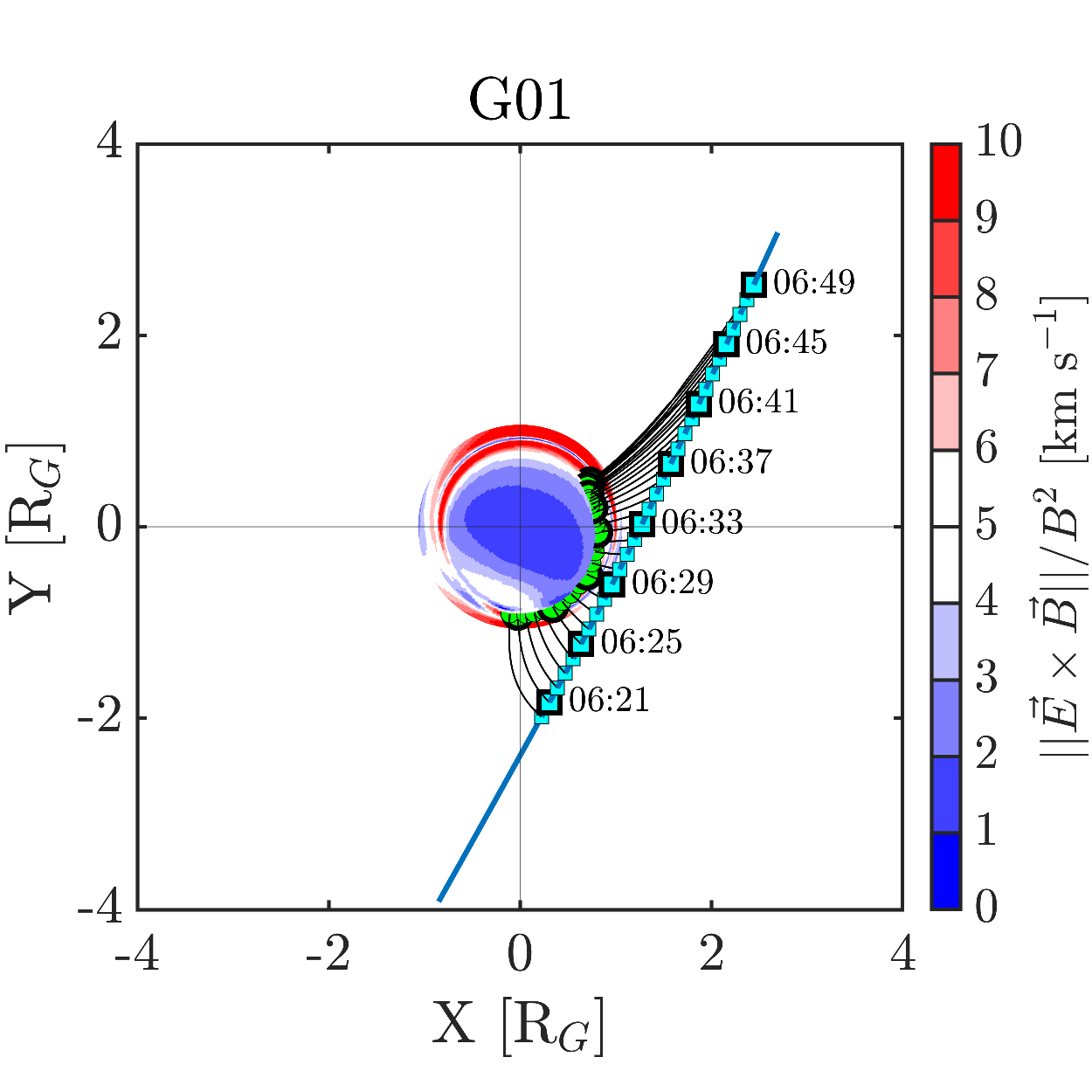}}}\!
\fcolorbox{G02}{white}{\parbox{\columnwidth}{\centering
\includegraphics[width=\linewidth]{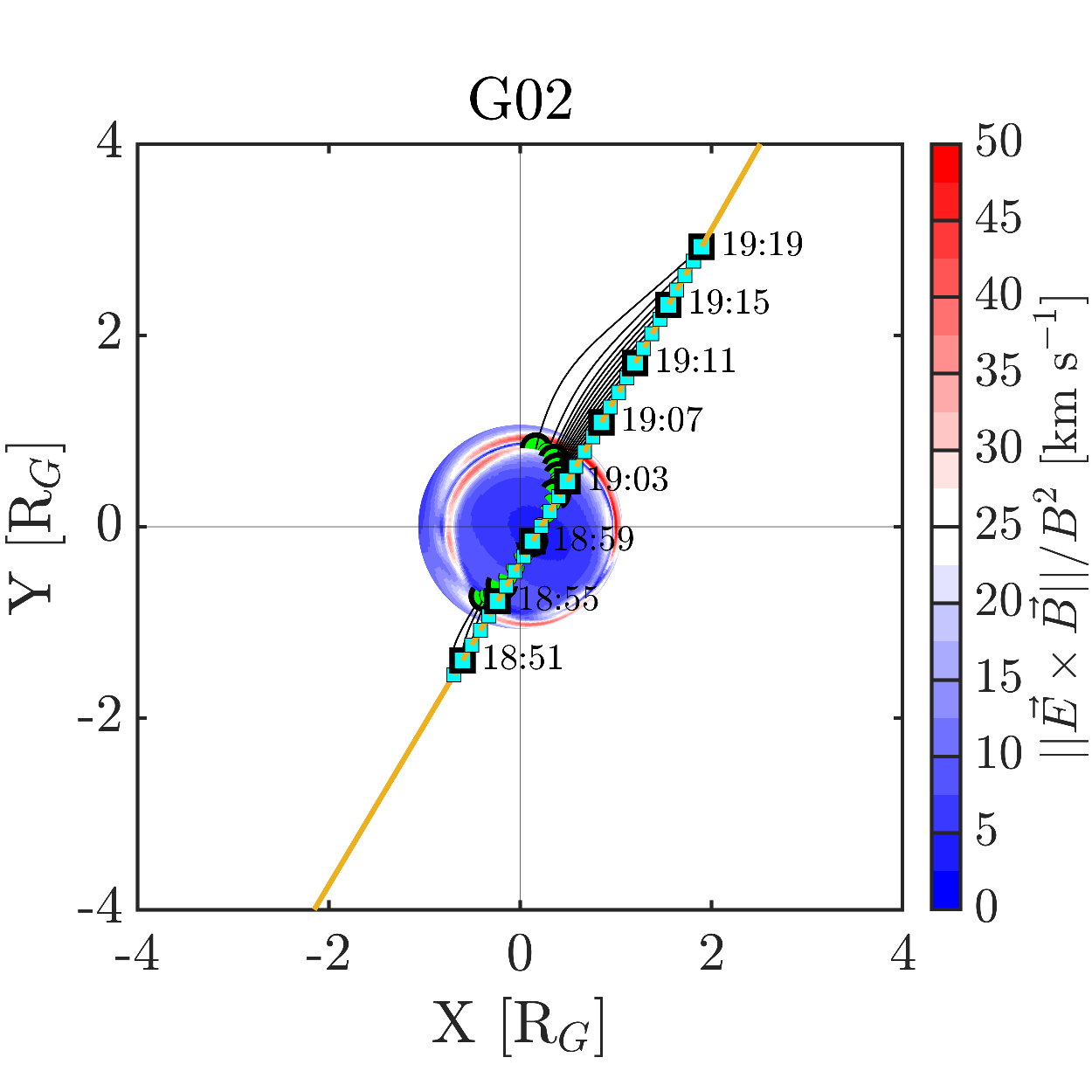}}}\\
\fcolorbox{G07}{white}{\parbox{\columnwidth}{\centering
\includegraphics[width=\linewidth]{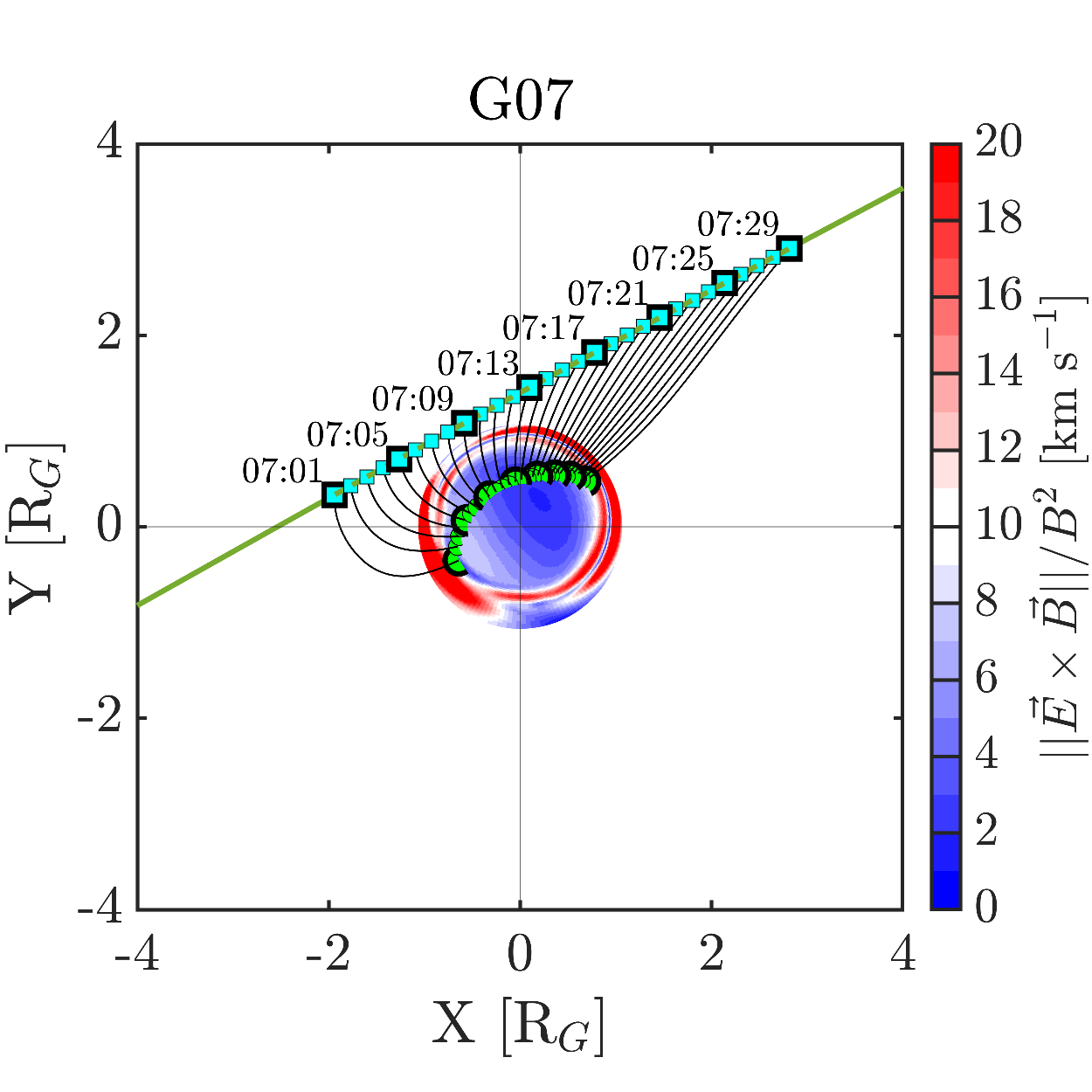}}}\!
\fcolorbox{G29}{white}{\parbox{\columnwidth}{\centering
\includegraphics[width=\linewidth]{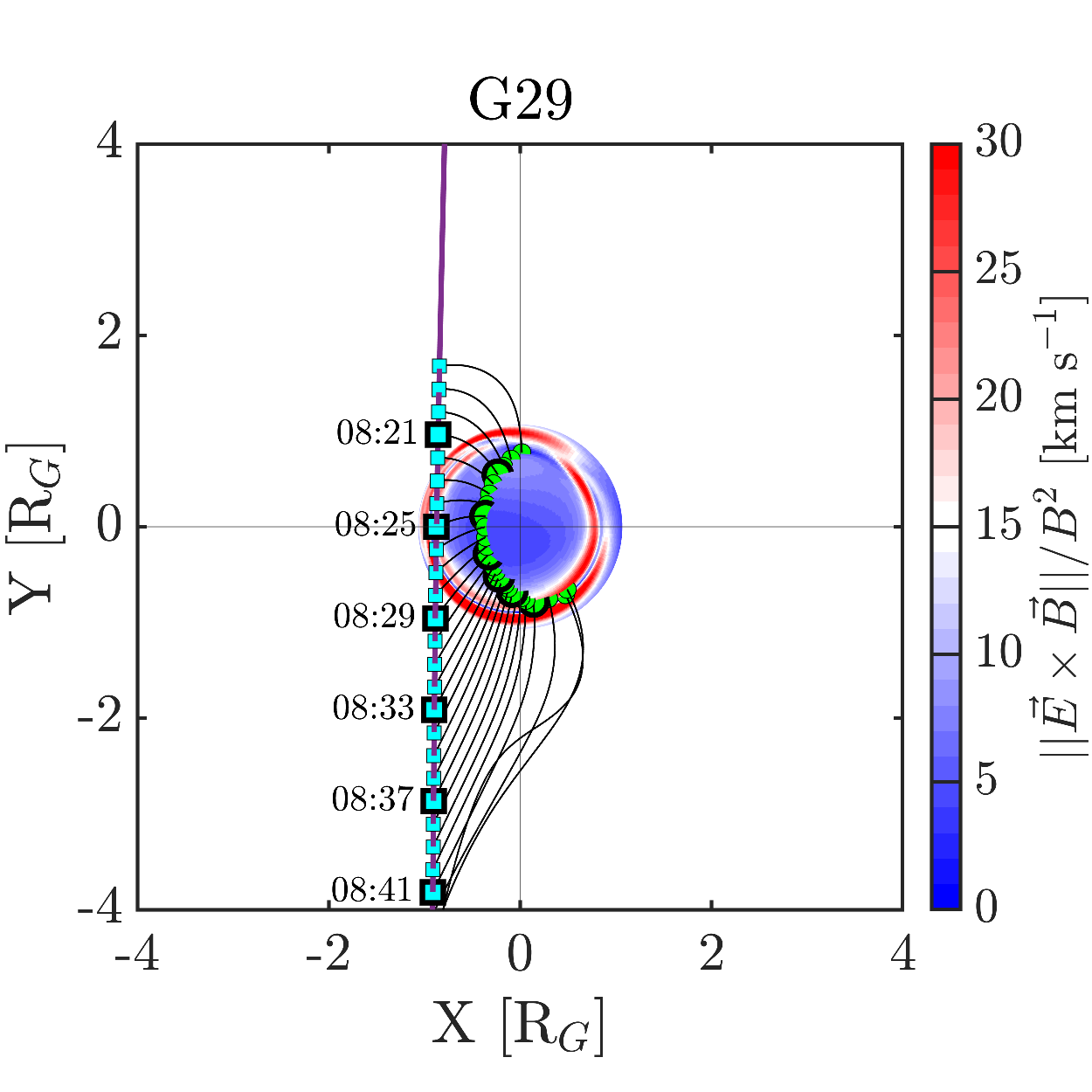}}}\\
\caption{View from the North Pole of Ganymede of the G01 (top left), G02 (top right), G07 (bottom left), and G29 (bottom right) flybys (solid red line) where $||\vec{\varv}_{\vec{E}\times\vec{B}}||$ at the lowest altitude, that said at the lower boundary of the MHD simulations around 1.05~$R_G$, is projected onto the surface of Ganymede. In addition, along the flybys, we plot the magnetic field lines that connect Galileo position (cyan squares) to the surface (green circles) during its crossing of the open field line region. Time is given for thicker markers. The figure is rotated with respect to the GPhiO-coordinate system around the $Z$-axis such that $-Y$ points towards the Sun. Note that the colourscale is different for each plot.\label{Fig12}}
\end{figure*}

The third cause we invoke is the acceleration along magnetic field lines, not considered by the MHD simulation used here. To support this idea, Fig.~\ref{Fig12} shows the configuration of the flybys with respect to the flux tube/Alfvén wings during the flybys. The axes are orientated such that the $-Y$ half-plane is the dayside. The green markers represent the projection of Galileo's trajectory on Ganymede's surface along the field lines giving us insights on Galileo's relative location within the flux tubes.  Though it is difficult to identify the exact 3D location of the Alfvén wings near the moon over the polar regions, the Alfvén wings and the flux tubes connecting Ganymede to Jupiter coincide well \citep{Neubauer1980}. Interestingly, the flybys with the best agreement in terms of ion number density, that is, G01 and G07, display the lowest $||\vec{\varv}_{\vec{E}\times\vec{B}}||$ at the baseline of the flux tube. In contrast, G02, with the largest discrepancy between simulation and observation, is also that with the largest $||\vec{\varv}_{\vec{E}\times\vec{B}}||$ at the surface. As the configuration of the flybys and the exospheric input for G01 and G02 are very similar, the ion production around Ganymede is similar, and hence the ion flux. However, the electromagnetic fields between G01 and G02 are very different. Larger ion speeds simulated at G02 cause lower ion number densities based on flux conservation, in contrast with G01. The stronger electric field used at G02 compared with G01, as a result of the MHD outcome, may be the reason why our simulated ion density is underestimated at G02 in comparison with our simulation for G01 and the PWS observations. Although we do not have direct evidence of field-aligned electric fields at Ganymede, simulations and observations done at Earth show the appearance of field-aligned currents (hence electric fields) when the Earth is hit by a sub-Alfvénic flow associated with a strong interplanetary magnetic field and develops Alfvén wings \citep[e.g.][]{Ridley2007,Beedle2024,Chen2024}. In that respect, Ganymede is a formidable plasma laboratory to investigate the interaction of a sub-Alfvénic and sub-magnetosonic flow with a magnetised body. Finally, the model ignores not only field-aligned electric fields but also Hall term in Ohm's law \citep{Jia2008,Jia2009}, and it does not take into account the mass-loading of the ionised Ganymede's exosphere at all (see Section~\ref{section23}). The relative importance of Hall term was assessed by \citet{Dorelli2015}, \citet{Zhou2019,Zhou2020}, and by \citet{Fatemi2022}. The latter showed that this term dominates the electric field within Ganymede's ionosphere. Refining by including this term and the ambipolar electric field would improve our current results. 

The fourth cause is the spatial dependence of electron-impact ionisation frequency. In our simulation, it was kept constant throughout the simulation domain. On the one hand, this means that a change in the overall electron-impact ionisation frequency impacts uniformly and similarly all the results presented. On the other hand, the ionisation frequency is unlikely uniform throughout Ganymede's magnetosphere. \citet{Carnielli2019} introduced an artificial spatial dependence by assuming no electron impact within the closed field line regions. While most likely the ionisation frequency is smaller in the closed field line regions, it is not zero. However, there are no observational constraints to date available in this region. Electron-impact ionisation frequency used in these simulations is based on the combination of electron differential flux from observations made by Voyager and Galileo \citep{Scudder1981,Paranicas1999,Cooper2001} that should be representative of the electron population at Ganymede's orbit. Based on Juno observations, \citet{Pelcener2024} showed that electron differential fluxes were found to be larger than those previously reported \citep{Paranicas1999,Carnielli2019}, suggesting higher electron-impact ionisation frequencies at Ganymede's orbit. In contrast, during the PJ34 by Juno across the Northern Alfvén wings \citep{Allegrini2022,Duling2022,Pelcener2024}, the electron differential flux was observed to drop below mainly 1~keV in Ganymede's magnetosphere. This is also highlighted in \citet{Vorburger2024}: depending on the electron population observed during Juno's flyby (referred to as polar and auroral), the electron-impact ionisation may vary by one order of magnitude for a given neutral species. Therefore, the knowledge of the electron energy distribution around Ganymede is critical for the electron-impact ionisation frequency as it depends on the energy distribution below 1~keV where most of the relevant electron-impact ionisation cross-sections peak \citep[both estimated experimentally or theoretically, e.g. ][]{Kahre1970,Itikawa2002,Itikawa2005,Deutsch2000,Deutsch2004}. In fact, higher electron-impact ionisation frequency and field-aligned acceleration may be competing effects on the ion number density: the former would naturally affect $n_i$ and the flux in the ion energy spectrum without affecting its shape, whereas the latter would affect the ion flux, hence $n_i$, and modify both flux and shape of the ion spectrum. The two phenomena are likely related. The presence of a field-aligned electric field would: 1) accelerate ionospheric ions and increase their energy as they move away from Ganymede (in the case of an electric field directed outwards), hence shifting their energy spectrum to higher energies 2) accelerate electrons moving towards Ganymede, modifying their energy distribution, their ability to ionise neutral species, therefore affecting the ion production, the intensity of the ion energy spectrum, and hence the simulated ion number density. However, as the electron-impact cross sections peak around 100~eV, it is difficult to anticipate whether electrons would be more or less efficient at ionising.

\subsection{Within the closed field line regions}\label{section42}

Regarding the closed magnetic field line region, we could only rely on PLS as there was no possibility of deriving electron number density from PWS data within Ganymede's magnetosphere for G08 and G28 flybys. Even if the electron number density was low, there was no evidence for resonance near the cyclotron frequency  (as $\omega_\text{uh}\approx \omega_\text{ce}$ for $\omega_\text{pe}\longrightarrow 0$). More recently, \citet{Buccino2022} used radio occultations between the Juno spacecraft and Earth to measure the total electron content. \citet{Buccino2022} reported electron number densities up to $10^{3}\,\text{cm}^{-3}$ at the surface in the closed field line region (egress), though these values are associated with large uncertainties ($1\sigma\approx 400\,\text{cm}^{-3}$). Nevertheless, there is a clear depletion compared with ingress located in the open/closed transition region in the southern hemisphere where electron number densities were observed up to $2\times10^{3}\,\text{cm}^{-3}$ at the surface. These results are qualitatively consistent with those obtained by \citet{Yasuda2024} performing radio occultations using Jupiter as a radio source during the G01 flyby. \citet{Yasuda2024} reported electron number densities between 5 and 20~cm$^{-3}$ in the closed field line regions and between 80 and 290~cm$^{-3}$ in the open field line regions. Plasma number densities are consistently lower within the closed field line regions. 

Regarding ion energy spectra, our test particle simulation helped us to identify where and when features are present and expected in the corresponding PLS data. For instance, the MP crossing, a lack of ion flux above 40~eV during G08 within the closed field line region, and the faint stripe between 10 and 20~eV from 10:11 to 10:14 during G28 (see Fig.~\ref{Fig11}) are captured showing that not only the Jovian plasma population is missing or weak in this region but also an ionospheric population is present though overestimated by our model. 

The electron number density could not be derived from PWS observations owing to the absence of clear upper-hybrid resonance. Both the lack of resonance in PWS data and of ion flux signature in the PLS data may be caused by a much lower electron-impact ionisation, while we have assumed a constant electron-impact everywhere for our simulations (not having a better constraint within the closed magnetic field lines). Indeed, in the closed magnetic field line region, ionising electrons (50-500~eV) are primarily from Ganymede's ionosphere and unlikely to come from the Jovian magnetosphere. This is confirmed by the fact that even thermal Jovian ions experience difficulty in accessing this region. Hybrid simulations performed by \citet{Fatemi2022} for G08 and G28 showed that thermal Jovian ion number density is reduced by a factor of 10 for G08, and they even disappear during G28 inside the closed magnetic field line region compared with the density within the Jovian magnetosphere. We observe similar effects with the test particle simulations applied to thermal Jovian ions, though more pronounced for O$^+$ than for H$^+$.

Besides the ionisation frequency, the electric field is another large uncertainty within the closed magnetic field line region. Except for the convective electric field, every term in Ohm's Law, such as the ambipolar, Hall, and Ohmic terms, is a function of $1/n_e$. \citet{Fatemi2022} found that the Ohmic term from Ohm's law, that is $\eta \vec{J}$, dominates in this `ion cavity' in their simulation. That is not surprising. In theory, $\eta$ is a function of $n_e$ but is usually kept constant or tuned for numerical reasons and stability, for instance, to damp numerical waves \citep{Jia2009,Fatemi2022}. Therefore, the electric field remains the largest unknown in this region (as either it is treated in the vacuum limit or by only considering convective and Ohmic terms), unlike magnetic fields that can be checked against MAG data. In addition, by means of coupling MHD and PIC simulations, \citet{Toth2016} and later \citet{Zhou2019} revealed the extreme complexity of the electromagnetic fields, how magnetic fields can be twisted and entangled, and suggested that a flux transfer event may have even occurred during G08, showing the need to include kinetic physics down to electron scales, for capturing the correct magnetic and electric fields upstream of Ganymede near the magnetopause, with a non-steady state approach. 

\subsection{The magnetopause crossing}\label{section43}

Some of our test particle simulations (see Figs.~\ref{Fig7} and \ref{Fig11}) showed that the magnetopause crossing must be associated with characteristic signatures, such as an increase in the plasma number density (e.g. during G01) and larger ion flux up to 50~keV. Although there is no evidence in the PWS data of such an increase in the plasma number density, the signature of larger ion fluxes is seen in the PLS data for G01, G28, and, to a lesser extent, G08. Regarding the detection, PWS may not be able to capture the magnetopause crossing due to the narrowness of the peak in space and, therefore, time ($\sim 1-2$~min) compared with PWS low time resolution ($\sim 40$\,s). Regarding the ion energy spectrum, \citet{Collinson2018} for G01 and \citet{Fatemi2022} for G08 and G28 suggested that such energisation was attributed to reconnection and/or energisation of the thermal Jovian plasma interacting with the magnetopause. Although \citet{Fatemi2022} argued that the use of a test-particle simulation should be avoided to take into account intermittent reconnection, we have shown that such an energisation occurs using electromagnetic fields from a snapshot of an MHD simulation and is seen in ionospheric O$_2^+$ (see Figs.~\ref{Fig6} and \ref{Fig11}) due to finite Larmor radius effects. In addition, the strength of our model is that we can simulate both ionospheric ions and Jovian ions, whereas, in general, hybrid and MHD simulations either consider the thermal Jovian plasma as the only source of plasma \citep[e.g.][]{Dorelli2015,Fatemi2022}, or adopted simplified approach based on boundary conditions to mimic an ionspheric source \citep[e.g][]{Jia2008,Jia2009}, or modelled the ionospheric contribution based on unsuitable background exosphere \citep[e.g][]{Duling2022,Stahl2023} as explained in Section~\ref{section22} and Appendix\,\ref{AppExo}.

\section{Conclusions}\label{section5}

In this paper, using a test particle model, that is driven by DSMC simulations of Ganymede's exosphere and MHD simulations of its electromagnetic environment, we simulate the ionospheric environment of Ganymede for all close flybys by the Galileo spacecraft near the moon.

Amongst the six close flybys we analysed, four of them (G01, G02, G07, and G29) were over the North hemisphere crossing through the Alfvén wing. Except for G02, our simulations of the total plasma density reproduce well the in-situ measurement made by Galileo and PWS. Moreover, by increasing the spatial resolution of MHD simulations and of the test particle \citep[compared to][]{Carnielli2019}, we were able to capture thinner structures such as the magnetopause where ion number densities locally peak and heavy ionospheric ions reach energies above 10~keV. In addition, our simulations provide further constraints and insights regarding the ion composition: Ganymede's ionosphere is dominated by $\mathrm{O_2^+}$ and $\mathrm{H_2^+}$.
Although our model reproduces well the trend seen in the PLS energy spectra, ion energy is often underestimated suggesting the presence of additional ion acceleration mechanisms such as field-aligned electric fields. Regarding the two remaining flybys, G08 and G28, Galileo flew upstream through the closed magnetic field line regions. Our simulations for these two flybys show that we should have expected ion number densities to be above 8~cm$^{-3}$, but PWS could never probe any electron density in this region. Similarly, there is clearly a lack of ion signatures in PLS energy spectra. Both lack of detections may suggest lower ionisation frequency in this shielded part of the magnetosphere than considered in the simulations and need additional constraints. The closed-field-line region remains a mystery.

The recent detection of H$_3^+$, along with ionospheric H$^+$, H$_2^+$, water-group ions, and O$_2^+$, by Juno \citep{Valek2022} has highlighted that even a few collisions do occur at Ganymede between ionospheric H$_2^+$ and exospheric H$_2$. Therefore, we may expect the ion composition to be more diverse than thought, which is the scope of an upcoming study.

Thanks to the upcoming JUpiter ICy Moons Explorer (JUICE) mission, launched in April 2023, we will have the opportunity to not only get closer to Ganymede's surface within the dense part of the ionosphere but also to orbit over a large range of latitude and longitude, including through the Alfvén wings and the closed field line regions. It will give us the unique opportunity to investigate the interaction of a sub-Alfvénic and sub-magnetosonic flow with a magnetised body. A series of plasma and particle observations onboard such as Radio and Plasma Wave Investigation from Radio and Plasma Wave Investigation \citep[RPWI,][]{Wahlund2024}, Particle Environmental Package (PEP, Barabash et al., in prep.), Gravity and Geophysics of Jupiter and Galilean Moons (3GM, Iess et al., in prep.), and the Sub-millimetre Wave Instrument (SWI, Hartogh et al., in prep.) may help us to lift the veil on the cause of ion energisation in the Alfvén wings as well as the low ionospheric densities within the closed-field line regions \citep{Galand2025}. In the meantime, 3D kinetic models, such as the one presented here, remain the only way to assess the full plasma environment of the moon.

\section{Acknowledgements}
Work at Imperial College London was supported by the Science and Technology Facilities Council (STFC) of the UK under ST/S000364/1 and ST/W001071/1. The research performed by F.L. was funded by the project FACOM
(ANR-22-CE49-0005-01 ACT). X.J. acknowledges support by NASA through Early Career Fellow Startup Grant \#80NSSC20K1286 and JUICE/PEP contract \#183512 with JHU/APL. H.L.F.H.'s work at DIAS was supported by a DIAS Research Fellowship in Astrophysics and by Taighde Éireann - Research Ireland award 22/FFP-P/11545. 

\section{Data availability}
The data underlying this article will be shared on reasonable request to the corresponding author. MAG \citep{KivelsonG} and PWS \citep{KurthPWS} data are available on the PDS.

\appendix
\section{Appendix: Exosphere}
\label{AppExo}
One unphysical aspect of Eq.~\ref{Eq2} is that the number density does not tend towards $0$ at infinity. To correct from it and describe exospheric number densities more accurately, \citet{Chamberlain1963} proposed to remove parts of the velocity space that cannot be populated in the absence of collisions (or local thermodynamic equilibrium). The correction is given by: 
\begin{align*}
    n_n(r)=&n_n(r_c)\exp(\lambda(r)-\lambda_c)\\
    &\times(\zeta_\text{bal}(\lambda,\lambda_c)+\zeta_\text{sat}(\lambda,\lambda_c)+\zeta_\text{esc}(\lambda,\lambda_c))\\
    \zeta_\text{bal}(\lambda,\lambda_c)=&\dfrac{2}{\sqrt{\pi}}\left[\gamma\left(\dfrac{3}{2},\lambda\right)-\sqrt{1-\dfrac{\lambda^2}{\lambda_c^2}}\exp\left(-\dfrac{\lambda^2}{\lambda+\lambda_c}\right)\gamma\left(\dfrac{3}{2},\dfrac{\lambda\lambda_c}{\lambda+\lambda_c}\right)\right]\\
    \zeta_\text{sat}(\lambda,\lambda_c)=&\dfrac{2}{\sqrt{\pi}}\sqrt{1-\dfrac{\lambda^2}{\lambda_c^2}}\exp\left(-\dfrac{\lambda^2}{\lambda+\lambda_c}\right)\gamma\left(\dfrac{3}{2},\dfrac{\lambda\lambda_c}{\lambda+\lambda_c}\right)\\
        \zeta_\text{esc}(\lambda,\lambda_c)=&\dfrac{1}{\sqrt{\pi}}\left[\Gamma\left(\dfrac{3}{2},\lambda\right)-\sqrt{1-\dfrac{\lambda^2}{\lambda_c^2}}\exp\left(-\dfrac{\lambda^2}{\lambda+\lambda_c}\right)\Gamma\left(\dfrac{3}{2},\dfrac{\lambda\lambda_c}{\lambda+\lambda_c}\right)\right]
\end{align*}
where $\zeta_\text{bal}$ is the partition function for ballistic particles (that said, flying across the exosphere along elliptic trajectories and crossing $r_c$), $\zeta_\text{sat}$ is that of satellite particles (flying across the exosphere along elliptic trajectories though not crossing $r_c$), $\zeta_\text{esc}$ is that of escaping particles (that said flying from $r_c$ along hyperbolic trajectories such that they escape from the gravitational potential well), and $\Gamma$ (resp. $\gamma$) is the upper (resp. lower) incomplete gamma function. Satellite particles have been always a subject of debate as they are not coming from $r_c$ but trapped within the potential well. If there are no collisions at all, satellite particles cannot be placed on elliptic orbits with periapsis above $r_c$.

\begin{figure*}
\centering
\includegraphics[width=\columnwidth,trim=1cm 0cm 4cm 0cm,clip]{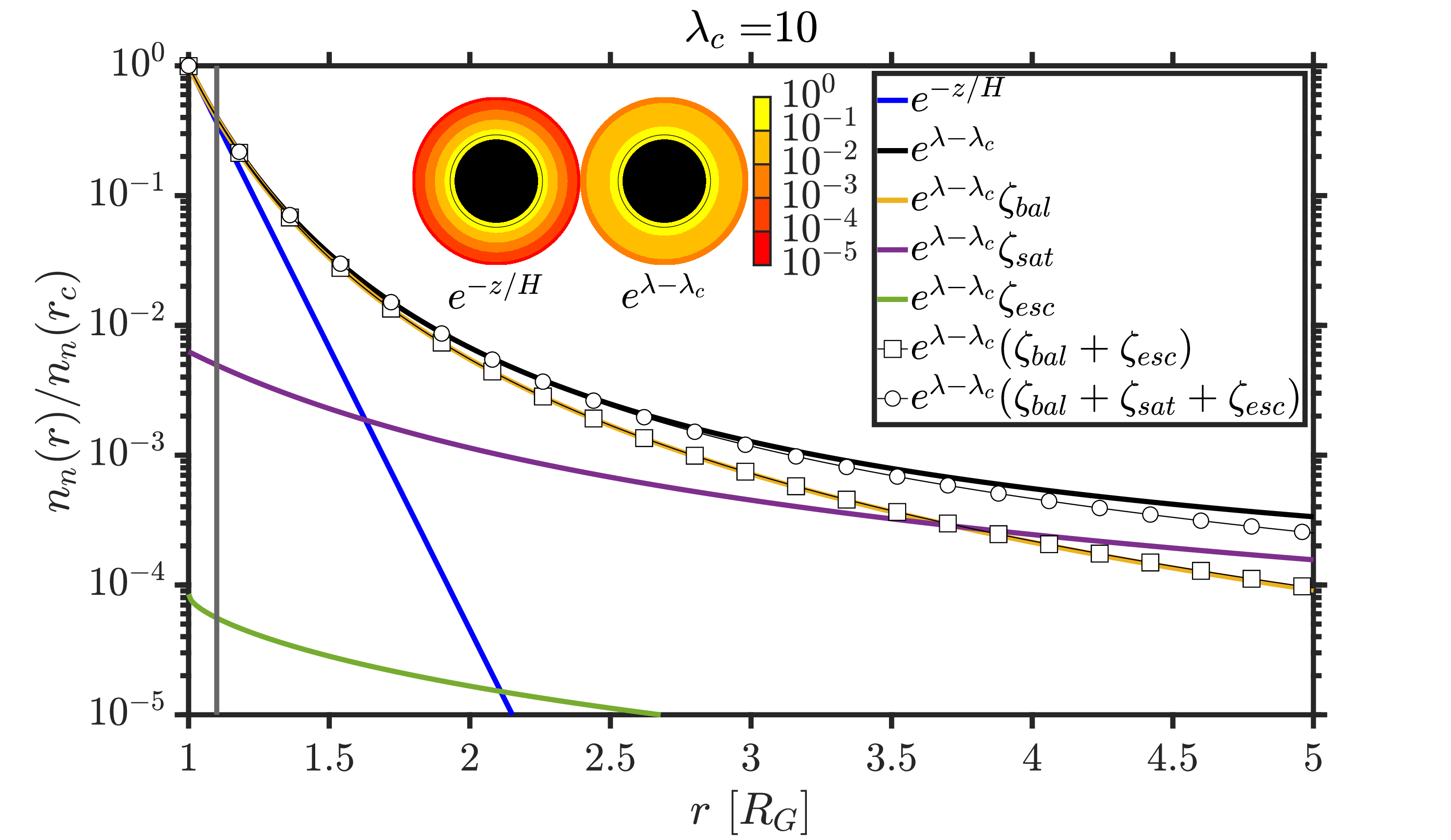}
\includegraphics[width=\columnwidth,trim=1cm 0cm 4cm 0cm,clip]{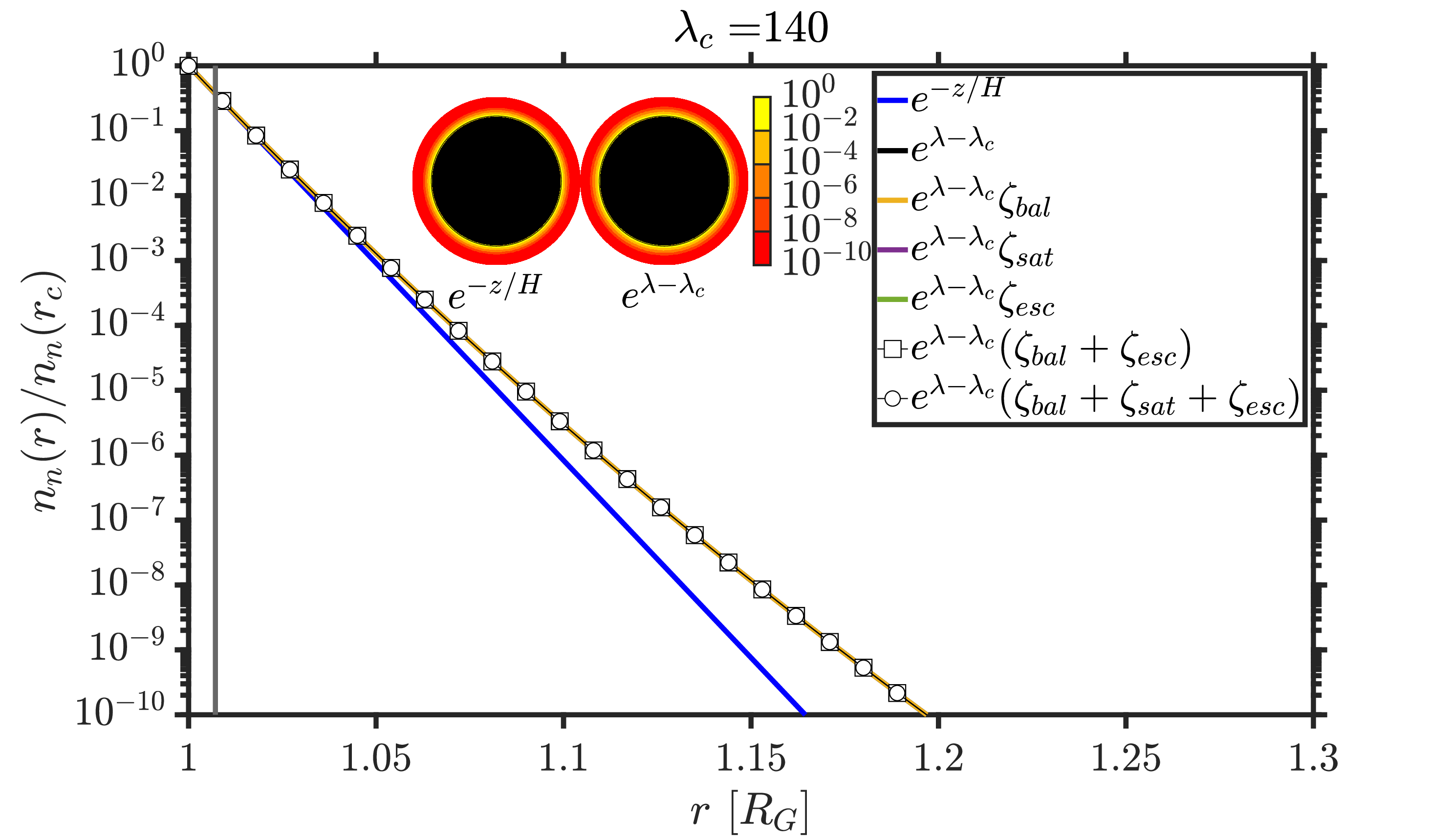}
\caption{Analytical neutral density profiles for different assumptions applied to Ganymede for $\lambda_c=10$ and $150$. The vertical line is placed at $H$ above the surface. In the inset, the sphere represents Ganymede surrounded by an atmosphere, colourscaled in logscale for both approximations, encircled by a black circle placed at the altitude $H$ above the surface. Jupiter's gravity is neglected.\label{Fig1app}}
\end{figure*}

Fig.~\ref{Fig1app} shows the analytical density profile assuming a hydrostatic equilibrium in plane parallel approximation, same for spherical symmetry, and for different corrections using Chamberlain's model \citet{Chamberlain1963}. To check this analytical solution against the simulated profiles from \citet{Leblanc2023}, we perform the following transformation to the hydrostatic equilibrium in spherical symmetry (Eq.~\ref{Eq2}):
\begin{align}
    \log\left[n_n(r_0)e^{\lambda(r)-\lambda(r_0)}\right]\times \dfrac{r}{r_0}=&\left(\log(n_0)-\lambda(r_0)\right)\dfrac{r}{r_0}+\lambda(r_0)\nonumber\\
    =&\left(\log(n_n(r_0))-\dfrac{GMm}{k_BTr_0}\right)\dfrac{r}{r_0}\nonumber\\
    &+\dfrac{GMm}{k_BTr_0}    
    \label{Eq4}
\end{align}
that is linear in $r$. In contrast, if the plane parallel approximation holds, one gets
\begin{align}
    \log\left[n_n(r_0)\exp\left(\dfrac{r-r_0}{H_n}\right)\right]\times \dfrac{r}{r_0}=&\left(\log(n_n(r_0))-\dfrac{r-r_0}{H_n}\right)\left(\dfrac{r}{r_0}\right)\nonumber\\
    =&\left(\log(n_n(r_0))+\dfrac{r_0}{H_n}\right)\dfrac{r}{r_0}\nonumber\\
    &-\dfrac{r_0}{H_n}\left(\dfrac{r}{r_0}\right)^2
    \label{Eq5}
\end{align}
that is quadratic in $r$.

\begin{figure*}
\centering
\stackinset{c}{0.5em}{t}{-1em}{\huge$\log(n_n)\times (r/r_0)$}{%
\includegraphics[width=2\columnwidth,clip,trim=3cm 0cm 4cm 0cm]{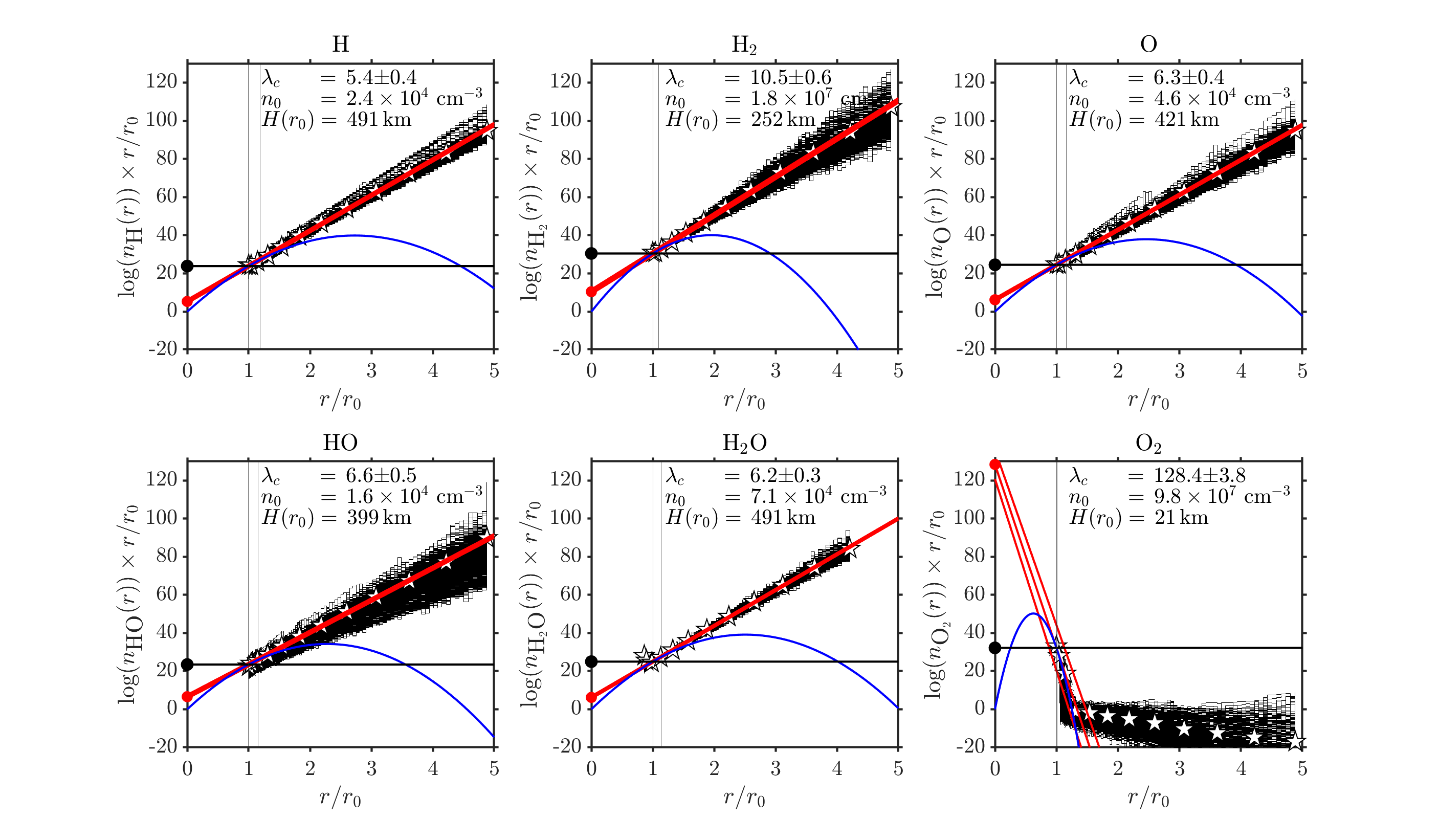}}
\caption{Exospheric profiles for all 6 species (H, H$_2$, O, HO, H$_2$O, O$_2$) used for \citet{Leblanc2023} with a Sun-Jupiter-Ganymede angle between $352^\circ$ and $358^\circ$ used for G01 and G02 simulations. Each black line represents an individual profile for a given longitude and latitude on Ganymede. $\bigstar$ corresponds to the profile averaged over latitude and longitude as a function of the altitude while the red lines are the associated linear fits. From these fits, the number density $n_n(r_0)$, the scale height $H(r_0)$, and Jeans' parameters $\lambda_c$ are derived assuming hydrostatic equilibrium in spherical coordinates. For all species, we assume $r_0=r_G$ except for H$_2$O. The horizontal black line corresponds to the fitted $\log(n_0 [\text{m}^{-3}])=(\log_{10}(n_0 [\text{cm}^{-3}])+6)\log(10)$ and the red dot to $\lambda_c$. The vertical lines are located at the Ganymede's surface and at an altitude of $r_0+H(r_0)$. The blue line is the hydrostatic equilibrium profile in the plane parallel case. Additional markers ({\large$\bullet$} and $\blacksquare$) have the same meaning than in Fig.~\ref{Fig1app}.\label{Fig2app}}
\end{figure*}

\begin{figure*}
\centering
\stackinset{c}{0.5em}{t}{-1em}{\huge$\vphantom{\log\times(r/r_0)}n_n$}{%
\includegraphics[width=2\columnwidth,clip,trim=3cm 0cm 4cm 0cm]{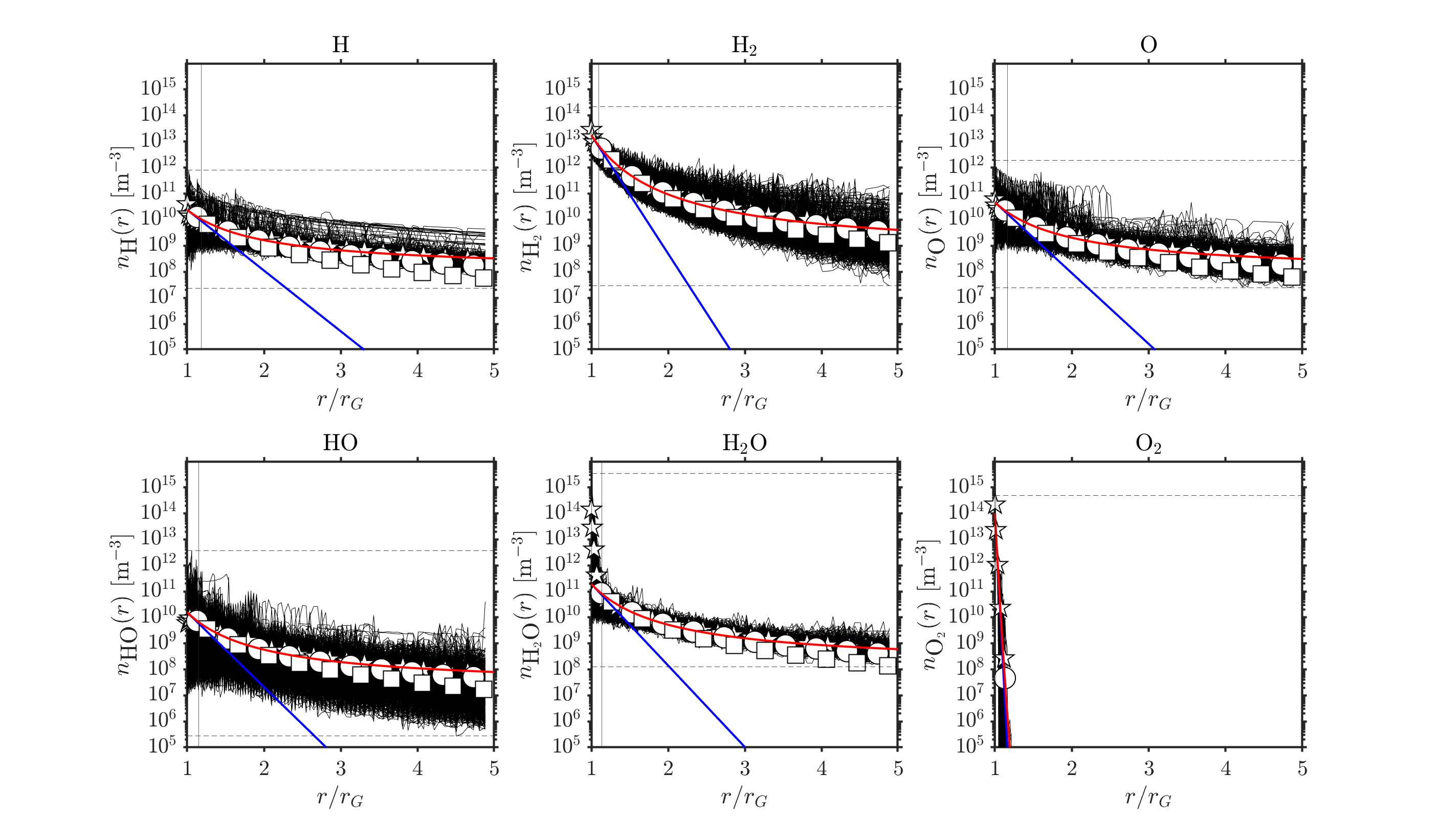}}
\caption{Same as Fig.~\ref{Fig2app} except the $y$-axis is the number density. Additional markers ({\large$\bullet$} and $\blacksquare$) have the same meaning than in Fig.~\ref{Fig1app}.\label{Fig3app}}
\end{figure*}

Fig.~\ref{Fig2app} shows the natural logarithm of the number density multiplied by $r/r_0$ for each exospheric species as a function of $r/r_0$. $r_0$ is the lower bound for our fit and optimally the exobase distance if possible. Except for H$_2$O where $r_0=1.16\,R_G$, all other species have $r_0=R_G$. That is due to the fact that H$_2$O remains collisional close to the surface on the dayside owing to a larger number densities caused by ice sublimation.  Whether these profiles follow a hydrostatic equilibrium in the plane parallel (blue lines) or spherical (red lines) approximations, they must be straight lines (Eq.~\ref{Eq4}, linear with $r$) or parabolas (Eq.~\ref{Eq5}, quadratic with $r$). In average, exospheric neutral densities (stars) tend to closely follow the spherical hydrostatic equilibrium (red lines). Near the surface, the analytical model differs from the simulation as collisions are expected to be significant. In addition, H$_2$O departs from the spherical hydrostatic equilibrium model near the surface as there is a strong localised enhancement on the dayside due to sublimation (see Fig.~\ref{Fig2}, right panel). H$_2$O is produced through two different mechanisms, resulting in two populations with different scale heights \citep{Vorburger2024}. H$_2$O is collisional near the subsolar point, where sublimation dominates H$_2$O production, while it is collisionless elsewhere \citep{Marconi2007,Shematovich2016,Leblanc2017}. However, the asymmetry driven by sublimation reduces (following an angular dependence in $\cos^\gamma\Theta$ where $\Theta$ is the angle between the local vertical and the Sun direction, 0 being the subsolar point, and $\gamma\sim5$) with altitude and almost disappears above 1200 km altitude, altitude at which the scale height associated with sublimation is tens of kilometres, consistent with the findings of \citet{Vorburger2022}. In that respect, fitting H$_2$O number density from 900 km altitude up to the upper boundary means that we may characterise the population with the largest scale height associated with sputtering. Another fit near the subsolar point and the surface shows that H$_2$O is better represented by a second population $\sim10000$ times denser with a much larger $\lambda_c\sim 80-100$ (i.e. with a smaller scale height, see Eq.~\ref{Eq2p}). H$_2$O column density is dominated by the sublimated population while H$_2$O number density is dominated by the sputtered population above 800-900 km. O$_2$ exhibits a diurnal asymmetry. Its number density profiles are only fitted between the surface and 200 km altitude, as the simulations become noisier with a stiff drop beyond that altitude. Fig.~\ref{Fig3app} shows the fitted number density profiles.

\begin{table*}
    \centering
    \caption{Fitted averaged Jeans' parameters $\lambda_c$ for Eq.~\ref{Eq2} for each exospheric species from DSMC simulation for different orbital phases. We provide the corresponding flybys for which the exosphere is used as well as the availability of remote sensing for the same configuration (though maybe not the same plasma conditions) (*) Only the sputtered component with the largest scale height was fitted. In addition, values calculated from $r_0=1.16 R_G$ extrapolated down to Ganymede's surface. $^\dag$\citet{Roth2021} $^\ddagger$\citet{McGrath2013} $^*$\citet{Leblanc2023} \label{tableA1}}
    \begin{tabular}{clccccccc}
    &&&\multicolumn{6}{c}{Species}\\
    \cline{4-9}
    $\angle$ Sun-Jup.-Gan.&Flybys &HST&H&H$_2$&O&HO&H$_2$O$^*$&O$_2$\\
    \hline
    $[027-033]$&&&5.5&10.0&6.2&6.1&7.4&131.5\\
    $[109-120]$&G07&1998-10-30$^\dag$&5.4&10.1&6.4&6.6&7.3&132.5\\
    $[157-163]$&&2003-11-30$^\ddagger$&5.4&10.2&6.1&6.5&7.6&133.6\\
    $[172-176]$&&2018-04-04$^\dag$&5.5&10.1&5.9&5.7&7.3&134.7\\
    $[178-182]$&G29&&5.1&10.4&5.8&6.1&7.0&155.0\\
    $[184-188]$&G28&&5.5&10.7&5.9&5.5&7.0&139.3\\
    $[278-291]$&G08&2010-11-19$^\dag$&5.4&10.3&6.2&6.6&7.4&131.5\\
    $[352-358]$&G01, G02&2017-02-02$^*$&5.4&10.5&6.3&6.6&7.2&128.4\\
    \end{tabular}
\end{table*}

\begin{table*}
    \centering
    \caption{Fitted averaged number density in cm$^{-3}$ at Ganymede's surface from exospheric simulations for Eq.\ref{Eq2}. (*) Only the sputtered component with the largest scale height was fitted.    \label{tableA2}}
    \begin{tabular}{ccccccc}
    & \multicolumn{6}{c}{Species}\\
    \cline{2-7}
    $\angle$ Sun-Jup.-Gan. &H&H$_2$&O&HO&H$_2$O$^*$&O$_2$\\
    \hline
    $[027-033]$&2.3$\times 10^{4}$&1.4$\times 10^{7}$&3.9$\times 10^{4}$&1.0$\times 10^{4}$&2.0$\times 10^{5}$&1.0$\times 10^{8}$\\
    $[109-120]$&2.0$\times 10^{4}$&1.2$\times 10^{7}$&4.4$\times 10^{4}$&1.4$\times 10^{4}$&1.7$\times 10^{5}$&9.2$\times 10^{7}$\\
    $[157-163]$&2.8$\times 10^{4}$&1.4$\times 10^{7}$&4.1$\times 10^{4}$&1.3$\times 10^{4}$&3.0$\times 10^{5}$&9.6$\times 10^{7}$\\
    $[172-176]$&2.2$\times 10^{4}$&1.2$\times 10^{7}$&3.2$\times 10^{4}$&8.5$\times 10^{3}$&2.0$\times 10^{5}$&9.8$\times 10^{7}$\\
    $[178-182]$&1.7$\times 10^{4}$&1.3$\times 10^{7}$&2.7$\times 10^{4}$&9.3$\times 10^{3}$&1.5$\times 10^{5}$&1.1$\times 10^{8}$\\
    $[184-188]$&2.3$\times 10^{4}$&1.5$\times 10^{7}$&3.1$\times 10^{4}$&7.3$\times 10^{3}$&1.8$\times 10^{5}$&9.9$\times 10^{7}$\\
    $[279-291]$&2.6$\times 10^{4}$&1.7$\times 10^{7}$&4.3$\times 10^{4}$&1.7$\times 10^{4}$&2.3$\times 10^{5}$&1.0$\times 10^{8}$\\
    $[352-358]$&2.4$\times 10^{4}$&1.8$\times 10^{7}$&4.6$\times 10^{4}$&1.6$\times 10^{4}$&1.9$\times 10^{5}$&1.0$\times 10^{8}$\\
    \end{tabular}
\end{table*}

Tables~\ref{tableA1} and \ref{tableA2} contain fitting parameters for all species and Ganymede configuration along the orbit. The scope of this section is not to provide a quantitative description of Ganymede exosphere. Instead, it provides a more physical, qualitative, and accurate one; in particular for those aiming at modelling the background neutral environment with a simple analytical formula. This might help and support comparisons with other authors from now for reproducibility. The fact that the profiles from \citet{Leblanc2023} can be well fitted with a \citet{Chamberlain1963}-like model is consistent with the former being a collisionless model.

\newpage
\bibliographystyle{mnras}
\bibliography{example}
\bsp	
\label{lastpage}
\end{document}